\newif\ifpreprint

\preprinttrue 

\ifpreprint
\documentclass[aip,jcp,amsmath,amssymb,preprint]{revtex4-1}
\else
\documentclass[aip,jcp,amsmath,amssymb,reprint]{revtex4-1}
\fi

\usepackage{xspace}
\usepackage{graphicx}
\usepackage{dcolumn}
\usepackage{bm}
\usepackage{amsfonts}
\usepackage[version=4]{mhchem}
\usepackage{pbox}
\usepackage{braket}
\usepackage{multirow}
\usepackage{threeparttable}
\usepackage{float}
\usepackage{xspace}
\usepackage{color}
\usepackage{fancyvrb}
\usepackage{adjustbox}
\usepackage{standalone}
\usepackage{tikz}
\usepackage{marvosym}
\usetikzlibrary{shapes.geometric, arrows}
\usepackage[note-name=,use-sort-key=false]{notes2bib}
\usepackage[colorlinks = true,
            linkcolor = red,
            urlcolor  = black,
            citecolor = red,
            anchorcolor = black]{hyperref}

\newcommand*{\kcal}{kcal mol$^{-1}$\xspace}
\newcommand*{\sunit}{$E_{\rm h}^{-2}$\xspace}
\newcommand*{\Eh}{$E_{\rm h}$\xspace}
\newcommand*{\molpro}{{\scshape Molpro}\xspace}

\newcommand*{\PSI}{{\scshape Psi4}\xspace}
\newcommand*{\ambit}{{\scshape Ambit}\xspace}
\newcommand*{\forte}{{\scshape Forte}\xspace}

\newcommand{\mref}[0]{\Psi_0}
\newcommand{\tens}[3]{{#1}_{#2}^{#3}}
\newcommand{\dfock}[1]{\epsilon_{#1}}
\newcommand{\cop}[1]{\hat{a}^\dag_{#1}}
\newcommand{\aop}[1]{\hat{a}_{#1}}
\newcommand{\sqop}[2]{\hat{a}_{#2}^{#1}}

\newcommand{\density}[2]{\gamma_{#2}^{#1}}

\newcommand{\cumulant}[2]{\lambda_{#2}^{#1}}
\newcommand{\no}[1]{ \{ {#1} \}}

\makeatletter

\newbox\swb@xone
\newbox\swb@xtwo
\newbox\swb@xthree
\newbox\swb@xfour
\newdimen\swdimen@ne
\newdimen\swdimentw@

\newcommand{\acontraction}[5][1ex]{%
  \mathchoice
    {\acontraction@\displaystyle{#2}{#3}{#4}{#5}{#1}}%
    {\acontraction@\textstyle{#2}{#3}{#4}{#5}{#1}}%
    {\acontraction@\scriptstyle{#2}{#3}{#4}{#5}{#1}}%
    {\acontraction@\scriptscriptstyle{#2}{#3}{#4}{#5}{#1}}}%
\newcommand{\acontraction@}[6]{%
  \setbox\swb@xone=\hbox{${}#1{}#2{}$}%
  \setbox\swb@xtwo=\hbox{${}#1{}#3{}$}%
  \setbox\swb@xthree=\hbox{${}#1{}#4{}$}%
  \setbox\swb@xfour=\hbox{${}#1{}#5{}$}%
  \swdimen@ne=\wd\swb@xtwo%
  \advance\swdimen@ne by \wd\swb@xfour%
  \divide\swdimen@ne by 2%
  \advance\swdimen@ne by \wd\swb@xthree%
  \vbox{%
    \hbox to 0pt{%
      \kern \wd\swb@xone%
      \kern 0.5\wd\swb@xtwo%
      \acontraction@@{\swdimen@ne}{#6}%
      \hss}%
    \vskip 0.5ex
    \vskip\ht\swb@xtwo}}

\newcommand{\acontraction@@}[3][0.05em]{%
  \hbox{%
    \vrule width #1 height 0pt depth #3%
    \vrule width #2 height 0pt depth #1%
    \vrule width #1 height 0pt depth #3%
    \relax}}

\newcommand{\tcontraction}[5][1ex]{%
  \mathchoice
    {\tcontraction@\displaystyle{#2}{#3}{#4}{#5}{#1}}%
    {\tcontraction@\textstyle{#2}{#3}{#4}{#5}{#1}}%
    {\tcontraction@\scriptstyle{#2}{#3}{#4}{#5}{#1}}%
    {\tcontraction@\scriptscriptstyle{#2}{#3}{#4}{#5}{#1}}}%
\newcommand{\tcontraction@}[6]{%
  \setbox\swb@xone=\hbox{${}#1{}#2{}$}%
  \setbox\swb@xtwo=\hbox{${}#1{}#3{}$}%
  \setbox\swb@xthree=\hbox{${}#1{}#4{}$}%
  \setbox\swb@xfour=\hbox{${}#1{}#5{}$}%
  \swdimen@ne=\wd\swb@xtwo%
  \advance\swdimen@ne by \wd\swb@xfour%
  \divide\swdimen@ne by 2%
  \advance\swdimen@ne by \wd\swb@xthree%
  \vbox{%
    \hbox to 0pt{%
      \kern \wd\swb@xone%
      \kern 0.5\wd\swb@xtwo%
      \tcontraction@@{\swdimen@ne}{#6}%
      \hss}%
    \vskip 0.5ex
    \vskip\ht\swb@xtwo}}

\newcommand{\tcontraction@@}[3][0.075em]{%
  \hbox{%
    \vrule width #1 height 0pt depth #3%
    \vrule width #2 height 0pt depth #1%
    \vrule width #1 height 0pt depth #3%
    \relax}}

\makeatother

\usepackage[normalem]{ulem}

\definecolor{goodorange}{RGB}{225,125,0}
\definecolor{goodgreen}{RGB}{0,125,0}
\definecolor{goodred}{RGB}{220,50,25}
\definecolor{goodblue}{RGB}{25,25,150}

\newcommand{\note}[2]{
\ifthenelse{\equal{#1}{F}}{
\colorbox{goodorange}{\textcolor{white}{\footnotesize \fontfamily{phv}\selectfont #1}}
    \textcolor{goodorange}{{\footnotesize \fontfamily{phv}\selectfont #2}}\xspace
}{}
\ifthenelse{\equal{#1}{Y}}{
\colorbox{goodred}{\textcolor{white}{\footnotesize \fontfamily{phv}\selectfont #1}}
    \textcolor{goodred}{{\footnotesize \fontfamily{phv}\selectfont #2}}\xspace
}{}
}

\begin{document}

\title{Driven similarity renormalization group: Third-order multireference perturbation theory}

\author{Chenyang Li}
\email{cli62@emory.edu}
\author{Francesco A. Evangelista}
\email{francesco.evangelista@emory.edu}
\affiliation{Department of Chemistry and Cherry Emerson Center for Scientific Computation, Emory University, Atlanta, GA, 30322}

\date{\today}

\begin{abstract}

A third-order multireference perturbation theory based on the driven similarity renormalization group approach (DSRG-MRPT3) is presented.
The DSRG-MRPT3 method has several appealing features: a) it is intruder free, b) it is size consistent, c) it leads to a non-iterative algorithm with ${\cal O}(N^6)$ scaling, and d) it includes reference relaxation effects.
The DSRG-MRPT3 scheme is benchmarked on the potential energy curves of \ce{F2}, \ce{H2O2}, \ce{C2H6}, and \ce{N2} along the \ce{F-F}, \ce{O-O}, \ce{C-C}, and \ce{N-N} bond dissociation coordinates, respectively.
The nonparallelism errors of DSRG-MRPT3 are consistent to those of CASPT3 and MRCISD, and show significant improvements over those obtained from DSRG second-order multireference perturbation theory.
Our efficient implementation of the DSRG-MRPT3 based on factorized electron repulsion integrals enables studies of medium-sized open-shell organic compounds.
This point is demonstrated with computations of the singlet-triplet splitting ($\Delta_{\rm ST} = E_{\rm T} - E_{\rm S}$) of 9,10-anthracyne.
At the DSRG-MRPT3 level of theory, our best estimate of the adiabatic $\Delta_{\rm ST}$ is 3.9 \kcal, a value that is within 0.1 \kcal from multireference coupled cluster results.
\end{abstract}

\maketitle

\section{Introduction}
Multireference perturbation theory (MRPT) based on a complete active space (CAS)\cite{Roos:1980fd,Ruedenberg:1982cv,Werner:1985dh} wave function is one of the simplest and most popular quantum chemistry approaches for studying near-degenerate electronic states.
A number of MRPTs have been proposed, among which the second-order complete active space perturbation theory (CASPT2) by Roos and co-workers is perhaps the most successful one.\cite{Andersson:1990jz,Andersson:1992cq}
The partially- and strongly-contracted variants of second-order $n$-electron perturbation theory (pc- and sc-NEVPT2), introduced more recently, are also growing in popularity.\cite{Dyall:1995ct,Angeli:2001bg,Angeli:2002ik,Angeli:2007by}
One of the advantages of CASPT2, pc- and sc-NEVPT2 is that they employ an internally contracted formalism, whereby the first-order correction to the wave function is generated by excitation operators acting on the entire reference wave function.
Therefore, the computational cost of these methods scales only polynomially with the number of active space orbitals---a tremendous reduction in cost compared to exponentially-scaling uncontracted multireference formalisms.

Despite their success, CASPT2 and NEVPT2 have some crucial limitations.
Perhaps the most unsettling feature of CASPT2---and most other MRPT methods---is the intruder-state problem.\cite{Evangelisti:1987fw,Paldus:1993dx,Kowalski:2000cj,Roos:1995jz} Intruder states are encountered when the zeroth-order energy of the reference and excited configurations are near degenerate.
Intruder states introduce singularities in the energy denominators and yield excitation amplitudes with unphysically large values.
Several approaches have been proposed to address intruders,\cite{Dyall:1995ct,Andersson:1995vv,Ghigo:2004fd,Witek:2002jd,Taube:2009jz,Camacho:2009cr} yet the most straightforward is level (denominator) shifting.\cite{Roos:1995jz,Forsberg:1997ke}
When intruder states are weakly coupled to the reference, level shifting works well for both ground and excited states.
However, in computations on multiple excited states, finding a ubiquitous level shift that removes intruders for all states may be challenging.\cite{Roos:1996ea}
In addition, level shifting introduces some arbitrariness in CASPT2 results.\cite{Camacho:2009cr,Vancoillie:2016gp}
Another less severe problem is the small size-consistency error carried by CASPT2 due to the use of projectors in the zeroth-order Hamiltonian [$\hat{H}^{(0)}$].\cite{Rintelman:2005fj,Helgaker2000book}
These projectors are introduced because the reference is not an eigenfunction of the average Fock operator.

In NEVPT2 the above two issues are neatly solved by employing Dyall's zeroth-order Hamiltonian that includes two-electron interactions within the active orbitals.\cite{Dyall:1995ct}
However, both CASPT2 and NEVPT2 have computational bottlenecks that prevent computations with large active spaces.
For a complete active space, the energy expressions of CASPT2 and NEVPT2 demand the four-particle reduced density matrix (4RDM) of the reference. 
The memory cost of storing this quantity grows as the eighth power of the number of active orbitals ($N_{\bf A}$) and quickly becomes the Achilles' heel of these methods when $N_{\bf A} > 16$.
The computational cost of CASTP2 and NEVPT2 is then dominated by tensor contractions that scale as ${\cal O}(N_{\bf A}^{8})$ and ${\cal O}(N_{\bf A}^{9})$, respectively.
Several attempts have been made to avoid the 4RDM via cumulant decompositions,\cite{Kutzelnigg:1997ut,Kutzelnigg:1999cj,Yanai:2007ix} and encouraging results have been obtained for CASPT2.\cite{Kurashige:2011ck}
Further approximations to the three-particle density matrix are less promising as ``false intruders'' appear on the potential energy curves.\cite{Kurashige:2011ck,Zgid:2009fu}
Unfortunately, even when the 4RDM is neglected, CASPT2 and pc-NEVPT2 still require removing linear dependencies in the excitation manifold by diagonalizing the overlap metric.\cite{Andersson:1992cq,Angeli:2001bg}
This step also has a computational costs proportional to ${\cal O}(N_{\bf A}^{9})$ and restricts state-of-the-art CASPT2 computations to $N_{\bf A} \approx 30$.\cite{Kurashige:2011ck}

Multireference perturbation theories based on the driven similarity renormalization group\cite{Evangelista:2014kt} (DSRG) provide a solution to both the intruder-state problem and the computational scaling limitations of CASPT2 and NEVPT2.
The DSRG is a many-body approach closely related to the in-medium similarity renormalization group (IM-SRG),\cite{Giazek:1993ik,Wegner:2000kq,Tsukiyama:2011eo,Hergert:2013kf,Hergert:2016fd} coupled cluster,\cite{Cizek:1966cy,Bartlett:1981hg,Crawford:2000by,Bartlett:2007kv} and canonical transformation theories.\cite{Yanai:2006gi,Yanai:2007ix,Neuscamman:2010ej}
Like in the IM-SRG, the DSRG separates excitation energy scales via a continuous unitary transformation of the Hamiltonian controlled by a flow parameter $s$.
A perturbative analysis of the DSRG shows that this transformation folds in correlation effects from excited configurations that correspond to energy denominators larger than a cutoff $\Lambda = s^{-1/2}$, while it leaves untouched those excitations for which the denominators are smaller than $\Lambda$.\cite{Kehrein2010book,Hergert:2016un}
As such, the DSRG avoids intruders for finite values of $s$ and yields a transformed (renormalized) Hamiltonian with modified many-body interactions.
Nonetheless, the DSRG is distinct from the IM-SRG.
While the IM-SRG directly determines the renormalized Hamiltonian by solving a collection of ordinary differential equations, the DSRG obtains it from a set of coupled nonlinear equations.
Another critical ingredient of the DSRG is the use of Fock-space many-body conditions,\cite{Kutzelnigg:1982kr,*Kutzelnigg:1983dr,*Kutzelnigg:1984eg,*Kutzelnigg:1985fj,Stolarczyk:1985fk,*Stolarczyk:1985ct,*Stolarczyk:1988ci,*Stolarczyk:1988cv,Lindgren:1978io,Nooijen:1996hp} which lead to equations in terms of normal-ordered second-quantized operators.
For multireference theories, the many-body approach effectively avoids the need to orthogonalize the excitation manifold.\cite{Datta:2011ca,Datta:2012hu,Li:2015iz,Li:2016hb}

In our previous work, we have explored the multireference DSRG\cite{Li:2015iz,Li:2016hb} (MR-DSRG) and its second-order perturbation theory (DSRG-MRPT2).\cite{Li:2015iz,Hannon:2016bh}
The MR-DSRG formalism is built upon the algebra of Mukherjee and Kutzelnigg's generalized normal ordering and Wick's theorem,\cite{Kutzelnigg:1997ut,Mukherjee:1997tk,Mahapatra:1998cp,Shamasundar:2009ee,Kong:2010kg,Kutzelnigg:2010iu,Sinha:2013dx} where operator contractions lead to density cumulants.\cite{Kutzelnigg:1997ut,Kutzelnigg:2010iu,Hanauer:2012bd}
The advantage of this scheme is that the DSRG-MRPT2 energy requires at most the three-particle density cumulant.
Consequently, the DSRG-MRPT2 approach has a scaling that is proportional to $\mathcal{O}(N_{\bf A}^6)$, and could potentially be applied to systems with large active spaces.

Our previous work has shown that the accuracy of the DSRG-MRPT2 is similar to that of other second-order MRPTs.\cite{Li:2015iz}
Furthermore, the linearized MR-DSRG with one- and two-body operators [MR-LDSRG(2)] greatly improves upon the accuracy of DSRG-MRPT2, but requires a recursive evaluation of the Hamiltonian and an iterative update of the cluster amplitudes.\cite{Li:2016hb}
The cost of MR-LDSRG(2) computations currently limits applications of this method to systems with 200--300 orbitals.
In this work we propose to overcome this limitation by developing a third-order multireference perturbation theory (MRPT3) based on the DSRG.  A noniterative DSRG-MRPT3 would be less expensive than the MR-LDSRG(2) scheme and likely to be more accurate than second-order perturbation theory.
Indeed, several third-order MRPTs have been formulated and they were found to be superior with respect to the corresponding second-order MRPTs.\cite{Wolinski:1987jb,Werner:1996in,Angeli:2002ip,Angeli:2006gf,Jiang:2006cn,Khait:2009ew}
For example, the CASPT3 implementation of Werner provides geometries and harmonic frequencies of small molecules that are as accurate as those of multireference configuration interaction with singles and doubles (MRCISD), but only costs as one iteration of MRCISD.\cite{Werner:1996in}

In this work, we derive and implement a third-order DSRG-MRPT (DSRG-MRPT3).
The zeroth-order Hamiltonian is chosen to contain only the diagonal blocks of the Fock operator, an identical choice made in the CASPT2D approach.\cite{Andersson:1990jz}
When applied to the DSRG-MRPT3, this choice of $\hat{H}^{(0)}$ leads to an efficient non-iterative formalism that is free from the intruder-state problem,\cite{Li:2015iz} is rigorously size extensive,\cite{Bartlett:1981hg}
and may be interfaced with any reference wave function for which the one-, two-, and three-body density cumulants are computed.
In this study we also develop relaxed DSRG-MRPT2 and MRPT3 approaches in which the reference wave function is optimized under the effects of dynamic electron correlation.
Most state-specific MRPT2 approaches, including CASPT2 and NEVPT2, do not account for reference relaxation effects, with the notable exception of  Mukherjee's state-specific MRPT2,\cite{Mahapatra:1999gh,Mahapatra:1998kj,SinhaMahapatra:1999bp,Chaudhuri:2005dt,Chattopadhyay:2016cc} generalized Van Vleck PT2,\cite{Shavitt:1980jc,Kirtman:1981kf,Khait:2002ca,Hoffmann:2009ih} and multiconfigurational PT2.\cite{Rolik:2003be,Szabados:2005jk,Chaudhuri:2005dt}

We begin our discussion of DSRG-MRPT3 by providing a brief overview of the general MR-DSRG ansatz in Sec.~\ref{sec:mrdsrg}, followed in Sec.~\ref{sec:mrpt3} by a detailed perturbative analysis.
Then, in Sec.~\ref{sec:implementation}, we provide details of the DSRG-MRPT3 implementation in our open-source code \forte.\cite{FORTE2016}
Section \ref{sec:results} consists of two parts.
In the first one we study the potential energy curves of \ce{F2}, \ce{HO-OH}, \ce{H3C-CH3}, and \ce{N2} to assess the accuracy of the DSRG-MRPT3 method and compare its performance with that of the DSRG-MRPT2 and the MR-LDSRG(2) approaches.
In the second part we compute the singlet-triplet gap of 9,10-anthracyne and compare DSRG-MRPT3 results with other multireference methods including CASPT2,\cite{Celani:2000cf} MRCISD,\cite{Knowles:1988hv,Werner:1988ku,Shamasundar:2011ew} and Mukherjee multireference coupled cluster theory.\cite{Mahapatra:1998cp,Mahapatra:1999ev,Evangelista:2006gf,Mahapatra:2010kj,Evangelista:2010cq}
Finally, in Sec.~\ref{sec:conclusions} we conclude and discuss future applications of the DSRG-MRPT3 scheme.

\section{Theory}
\label{sec:theory}

\subsection{An overview of MR-DSRG}
\label{sec:mrdsrg}

In this section we briefly review the MR-DSRG formalism.
Readers who are interested in the details regarding the operator parameterization should refer to the original DSRG (Ref.~\citenum{Evangelista:2014kt}) and MR-DSRG (Refs.~\citenum{Li:2015iz} and \citenum{Li:2016hb}) papers.
In MR-DSRG theory, we employ the generalized normal ordering of Mukherjee and Kutzelnigg (MK-GNO) to deal with the algebra of second-quantized operators.\cite{Kutzelnigg:1997ut,Mukherjee:1997tk,Mahapatra:1998cp,Shamasundar:2009ee,Kong:2010kg,Kutzelnigg:2010iu,Sinha:2013dx}
Under MK-GNO, operator contractions are associated with density cumulants,\cite{Kutzelnigg:1997ut,Kutzelnigg:2010iu,Hanauer:2012bd} which embody all information of the reference wave function.
The reference wave function used to define the MK-GNO Fermi vacuum is a multideterminantal wave function:
\begin{align}
\label{eq:mref}
    \ket{\mref} = \sum_{\mu = 1}^{d} c_{\mu} \ket{\Phi^{\mu}},
\end{align}
where each determinant $\Phi^{\mu}$ is weighted by the coefficient $c_{\mu}$.
In this work we further assume that the set of determinants $\{\Phi^{\mu}, \mu = 1, 2, \dots, d \}$ forms a complete active space (CAS), although this is not generally required by the MR-DSRG formalism.
The coefficients $c_{\mu}$ and the molecular orbitals are determined by the CAS self-consistent field (CASSCF) procedure.\cite{Roos:1980fd,Ruedenberg:1982cv}
The set of spin orbitals $\{\phi^{p}, p = 1,2,\dots,N \}$ then falls into three subsets: core ($\bf C$, with indices $m, n$), active ($\bf A$, with indices $u, v, w, x, y, z$), and virtual ($\bf V$, with indices $e, f, g, h$) of dimension $N_{\bf C}$, $N_{\bf A}$, and $N_{\bf V}$, respectively.
Two composite orbital sets are introduced: hole ($\bf H = C \cup A$, with indices $i,j,k,l$) and particle ($\bf P = A \cup V$, with indices $a,b,c,d$) of size $N_{\bf H} = N_{\bf C} + N_{\bf A}$ and $N_{\bf P} = N_{\bf A} + N_{\bf V}$, respectively.
General orbitals ($\bf G$) are denoted by indices $p,q,r,s$.

The philosophy of the DSRG ansatz is to define a continuous ($s$-dependent) unitary operator $\hat{U}(s)$ that transforms the bare Hamiltonian $\hat{H}$ to a band-diagonal operator $\bar{H}(s)$, namely:
\begin{equation}
\label{eq:dsrg_trans}
  \hat{H} \rightarrow \bar{H}(s) = \hat{U}^{\dag} (s) \hat{H} \hat{U}(s), \quad s \geq 0.
\end{equation}
When the flow variable $s$ approaches infinity, we require this transformation to exactly zero the couplings between the reference state and its internally-contracted excited configurations.
In the many-body formalism,\cite{Lindgren:1978io,Nooijen:1996hp,Datta:2011ca,Datta:2012hu,Demel:2013kz} this coupling is conveniently represented by the non-diagonal terms of the Hamiltonian $[\bar{H}^{\rm N}(s)]$.
The DSRG assumes that the unitary transformation is controlled by a Hermitian source operator $\hat{R}(s)$.  The corresponding many-body DSRG flow equation\cite{Evangelista:2014kt,Li:2015iz,Li:2016hb} realizes this idea by equating the non-diagonal terms of the Hamiltonian to the source operator:
\begin{align}
\label{eq:dsrg_flow}
    \bar{H}^{\rm N}(s) = \hat{R}(s).
\end{align}
The source operator $\hat{R}(s)$ renormalizes the Hamiltonian in such a way that excited configurations that are energetically separated from the reference by at least $s^{-1/2}$ are decoupled to each other.\cite{Kehrein2010book,Evangelista:2014kt}
Note, the parametrization of $\hat{R}(s)$ that can achieve this renormalization transformation is not unique.
In this work we use a source operator (see Ref.~\citenum{Evangelista:2014kt}) that reproduces the transformation of second-order perturbation theory based on the single-reference similarity renormalization group.\cite{Tsukiyama:2011eo,Hergert:2016fd}

Once $\hat{R}(s)$ is defined, the unitary operator $\hat{U}(s)$ can be determined via Eqs.~\eqref{eq:dsrg_trans} and \eqref{eq:dsrg_flow}.
In order to set up Eq.~\eqref{eq:dsrg_flow}, we need to write the DSRG transformed Hamiltonian $\bar{H}(s)$ as a sum of second-quantized operators.
The unitary operator $\hat{U}(s)$ is expressed as the exponential of an anti-Hermitian operator $\hat{A} (s)$, and $\hat{A} (s)$ is further related to the coupled cluster excitation operator $\hat{T}(s)$,
\begin{align}
    \hat{U} (s) = \exp{[\hat{A}(s)]} = \exp{[\hat{T}(s) - \hat{T}^{\dag}(s)]}.
\end{align}
Note that internal amplitudes $\tens{t}{uv\cdots}{xy\cdots} (s)$ with $u, v, \dots, x, y, \dots \in \bf A$ are redundant and hence set to zero.
Using the Baker--Campbell--Hausdorff (BCH) expansion, we write the renormalized Hamiltonian $\bar{H}(s)$ as a series of commutators of $\hat{H}$ and $\hat{A}(s)$,
\begin{equation}
\label{eq:exactBCH}
    \bar{H}(s) = \hat{H} + \sum_{k=1}^{\infty} \frac{1}{k!} \underbrace{[\cdots[[\hat{H}, \hat{A}(s)], \hat{A}(s)], \cdots]}_{k\text{ nested commutators}}.
\end{equation}
The many-body expression of $\bar{H}(s)$ is obtained by evaluating the commutators using the MK-GNO Wick's theorem and subsequently collecting the same rank of normal-ordered operators.\cite{Evangelista:2014kt,Li:2015iz,Li:2016hb}
Because the BCH formula in Eq.~\eqref{eq:exactBCH} does not terminate, we truncate each commutator $[\hat{H}, \hat{A}(s)]$ to contain at most two-body operators,\cite{Yanai:2006gi,Yanai:2007ix,Li:2016hb} and consider the series converged when the Frobenius norm of the $k$-nested commutator is less than $10^{-12}$ \Eh.\cite{Evangelista:2012hz,Evangelista:2014kt,Li:2016hb}
We shall use the subscript ``1,2'' whenever the operator is truncated to contain at most two-body terms.

The MR-DSRG energy can be computed in three ways.
The first approach simply computes the expectation value of $\bar{H}(s)$ using the reference wave function $\mref$,
\begin{align}
\label{eq:energy_unrelaxed}
    E_{\rm u}(s) = \braket{\mref | \bar{H}(s) | \mref}.
\end{align}
The energy obtained via Eq.~\eqref{eq:energy_unrelaxed} is unrelaxed because the weight of each reference determinant ($c_{\mu}$) is fixed and equal to the reference CASCI/CASSCF wave function.
Alternatively, the coefficient $c_{\mu}$ may be optimized by diagonalizing $\bar{H}(s)$ within the set of reference determinants $\{\Phi^{\mu}\}$:
\begin{align}
\label{eq:energy_relaxed}
    \sum_{\mu}^{d} \braket{\Phi_{\nu} | \bar{H}(s) | \Phi^{\mu}} c_{\mu} = E(s) c_{\nu}.
\end{align}
We consider two approaches to treat reference relaxation effects.
The first one consists in the iterative solutions of Eqs.~\eqref{eq:energy_relaxed} and \eqref{eq:dsrg_flow} until self consistency is reached.
Achieving self consistency of the amplitude and eigenvalue equations is a necessary condition to guarantee that when $\hat{A}(s)$ is not truncated, the BCH expansion of $\bar{H}(s)$ is not approximated, and $s \rightarrow \infty$, then the full MR-DSRG is equivalent to full configuration interaction.
In the second approach, we approximate the relaxed energy with a single-step correction which consist in solving Eq.~\eqref{eq:energy_relaxed} only once and take the eigenvalue of the matrix $\braket{\Phi_{\nu} | \bar{H}(s) | \Phi^{\mu}}$ as the relaxed energy.
This single-step relaxation scheme is economical since it requires only one $N^{6}$ step.
Both relaxation schemes are compared and benchmarked in Sec.~\ref{sec:results}. 

\subsection{The DSRG-MRPT3 method}
\label{sec:mrpt3}

We first partition the normal-ordered bare Hamiltonian into a zeroth-order term $\hat{H}^{(0)}$ and a first-order fluctuation potential $\hat{H}^{(1)}$.
The DSRG perturbation theory is derived from an order-by-order expansion of the source operator $\hat{R} (s)$, the anti-Hermitian operator $\hat{A}(s)$, and the unrelaxed energy $E_{\rm u} (s)$, while the coefficients $c_{\mu}$ in Eq.~\eqref{eq:energy_relaxed} are not treated perturbatively.\cite{Li:2015iz}
The zeroth- till third-order DSRG transformed Hamiltonian are given by
\begin{align}
    \bar{H}^{(0)} (s) =& \hat{H}^{(0)}, \label{eq:Hbar_0th} \\
    \bar{H}^{(1)} (s) =& [\hat{H}^{(0)}, \hat{A}^{(1)} (s)] + \hat{H}^{(1)}, \label{eq:Hbar_1st} \\
    \bar{H}^{(2)} (s) =& [\hat{H}^{(0)}, \hat{A}^{(2)} (s)] + \frac{1}{2} [\tilde{H}^{(1)} (s), \hat{A}^{(1)} (s)], \label{eq:Hbar_2nd} \\
    \bar{H}^{(3)} (s) =& [\hat{H}^{(0)}, \hat{A}^{(3)} (s)] + \frac{1}{2} [\tilde{H}^{(1)} (s), \hat{A}^{(2)} (s)] \notag\\
    &+ \frac{1}{2} [\tilde{H}^{(2)} (s), \hat{A}^{(1)} (s)], \label{eq:Hbar_3rd}
\end{align}
where we have introduced the combined first-order Hamiltonian $\tilde{H}^{(1)} (s) = \hat{H}^{(1)} + \bar{H}^{(1)} (s)$ and the second-order counterpart $\tilde{H}^{(2)} (s) = \bar{H}^{(2)} (s) - \frac{1}{6} [[\hat{H}^{(0)}, \hat{A}^{(1)}(s)], \hat{A}^{(1)}(s)]$.
The first- and second-order cluster amplitudes are determined by:
\begin{align}
    \hat{R}^{(1)} (s) &= [\bar{H}^{(1)}]^{\rm N} (s), \label{eq:R_1st}\\
    \hat{R}^{(2)} (s) &= [\bar{H}^{(2)}]^{\rm N} (s). \label{eq:R_2nd}
\end{align}

To simplify the structure of Eqs.~\eqref{eq:Hbar_0th}--\eqref{eq:Hbar_3rd}, the zeroth-order Hamiltonian $\hat{H}^{(0)}$ is chosen to include the reference energy and the diagonal blocks of the Fock operator
\begin{align}
    \hat{H}^{(0)} = E_0 + \hat{F}^{(0)},
\end{align}
where
\begin{align}
\label{eq:F0th}
    \hat{F}^{(0)} &= \sum_{mn}^{\bf C} \tens{f}{m}{n} \no{\sqop{m}{n}} + \sum_{uv}^{\bf A} \tens{f}{u}{v} \no{\sqop{u}{v}} + \sum_{ef}^{\bf V} \tens{f}{e}{f} \no{\sqop{e}{f}} \notag\\
    &= \sum_{p} \dfock{p} \no{\sqop{p}{p}}.
\end{align}
Here the string of normal-ordered creation ($\cop{}{}$) and annihilation ($\aop{}{}$) operators are compactly written as $\no{\sqop{pq\dots}{rs\dots}} = \no{ \cop{p}\cop{q} \dots \aop{s}\aop{r} }$.
The generalized Fock matrix element $\tens{f}{p}{q}$ is expressed in terms of one-electron ($\tens{h}{p}{q}$) and antisymmetrized two-electron ($\tens{v}{pq}{rs}$) integrals as well as the one-particle density matrix $\density{p}{q} = \braket{\mref| \cop{p}\aop{q} |\mref}$,
\begin{align}
    \tens{f}{p}{q} = \tens{h}{p}{q} + \sum_{ij} \tens{v}{pi}{qj} \density{i}{j}.
\end{align}
As indicated by the last line of Eq.~\eqref{eq:F0th}, molecular orbitals are semicanonicalized such that the core, active and virtual blocks of the generalized Fock matrix are diagonal ($\dfock{p} = \tens{f}{p}{p}$).

With this zeroth-order Hamiltonian, the commutator $[\hat{H}^{(0)}, \hat{A}^{(n)} (s)]$ yields only one- and two-body non-diagonal terms (that is, $\no{\sqop{a}{i}}$, $\no{\sqop{ab}{ij}}$, $\no{\sqop{i}{a}}$, and $\no{\sqop{ij}{ab}}$).
As a result, expectation values of the form $\braket{\mref| [\hat{H}^{(0)}, \hat{A}^{(n)} (s)] |\mref} = 0$ for any $n$.
The unrelaxed zeroth- till third-order energies written in terms of the cluster operators are thus:
\begin{align}
    E^{(0)}_{\rm u} (s) =& E_0, \label{eq:E(0)} \\
    E^{(1)}_{\rm u} (s) =& 0, \label{eq:E(1)} \\
    E^{(2)}_{\rm u} (s) =& \braket{\mref| [\tilde{H}^{(1)} (s), \hat{T}^{(1)} (s)] |\mref}, \label{eq:E(2)} \\
    E^{(3)}_{\rm u} (s) =& \braket{\mref| [\tilde{H}^{(1)} (s), \hat{T}^{(2)} (s)] |\mref} \notag\\
    &+ \braket{\mref| [\tilde{H}^{(2)} (s), \hat{T}^{(1)} (s)] |\mref}, \label{eq:E(3)}
\end{align}
where the prefactors that enter in Eqs.~\eqref{eq:Hbar_2nd} and \eqref{eq:Hbar_3rd} cancel since $\braket{\mref| [\hat{H}, \hat{A} (s)] |\mref} = 2 \braket{\mref| [\hat{H}, \hat{T} (s)] |\mref}$.

Note that the contribution $\braket{\mref| [\tilde{H}^{(2)} (s), \hat{T}^{(1)} (s)] |\mref}$ to $E^{(3)}_{\rm u} (s)$ reported in Eq.~\eqref{eq:E(3)} contains contractions that involve the four-body cumulant of the reference.
To avoid the cost of computing and storing the four-body cumulant we approximate the third-order transformed Hamiltonian with:
\begin{equation}
\begin{split}
    \bar{H}^{(3)} (s) \approx& [\hat{H}^{(0)}, \hat{A}^{(3)} (s)] + \frac{1}{2} [\tilde{H}^{(1)} (s), \hat{A}^{(2)} (s)] \notag\\
    &+ \frac{1}{2} [\tilde{H}^{(2)}_{1,2} (s), \hat{A}^{(1)} (s)],
    \end{split}
\end{equation}
where $\tilde{H}^{(2)}_{1,2} (s)$ includes only the one- and two-body components of the operator $\tilde{H}^{(2)}(s)$ and neglects the remaining three-body components.
This approximation is consistent with the linear MR-DSRG truncated to two-body operators [MR-LDSRG(2)].\cite{Li:2016hb} By this we mean that if the perturbative series for $\bar{H}^{(n)}$ were to be resummed to infinite order, consistently truncating all commutators to one- and two-body operators, it would yield (if convergent) the unrelaxed MR-LDSRG(2) transformed Hamiltonian.

From Eq.~\eqref{eq:R_1st}, we can derive the explicit expressions of the first-order amplitudes,\cite{Li:2015iz}
\begin{align}
    \tens{t}{a}{i,(1)} (s) &= [ \tens{f}{a}{i,(1)} + \sum\limits_{ux}^{\mathbf{A}} \tens{\Delta}{u}{x} \tens{t}{ax}{iu,(1)} (s) \density{x}{u} ]\frac{1 - e^{-s(\tens{\Delta}{a}{i})^2} }{\tens{\Delta}{a}{i}}, \label{T1_1st} \\
    \tens{t}{ab}{ij,(1)} (s) &= \tens{v}{ab}{ij,(1)} \frac{1 - e^{-s(\tens{\Delta}{ab}{ij})^2}}{\tens{\Delta}{ab}{ij}}, \label{T2_1st}
\end{align}
where $\tens{\Delta}{ab\dots}{ij\dots} = \dfock{i} + \dfock{j} + \dots - \dfock{a} - \dfock{b} - \dots$ is a general M{\o}ller--Plesset denominator. 
Analogous expressions for the second-order amplitudes can be derived from Eq.~\eqref{eq:R_2nd},
\begin{align}
    \tens{t}{a}{i,(2)} (s) &= [ \tens{f}{a}{i,(2)} (s) + \sum\limits_{ux}^{\mathbf{A}} \tens{\Delta}{u}{x} \tens{t}{ax}{iu,(2)} (s) \density{x}{u} ]\frac{1 - e^{-s(\tens{\Delta}{a}{i})^2} }{\tens{\Delta}{a}{i}}, \label{T1_2nd} \\
    \tens{t}{ab}{ij,(2)} (s) &= \tens{v}{ab}{ij,(2)} (s) \frac{1 - e^{-s(\tens{\Delta}{ab}{ij})^2}}{\tens{\Delta}{ab}{ij}}. \label{T2_2nd}
\end{align}
Here $\tens{f}{a}{i,(2)} (s)$ and $\tens{v}{ab}{ij,(2)} (s)$ are one- and two-body non-diagonal elements of $\frac{1}{2} [\tilde{H}^{(1)}(s), \hat{A}^{(1)} (s)]$, respectively.

Once the first- and second-order amplitudes are known, the unrelaxed DSRG-MRPT3 energy is obtained by summing the zeroth- to third-order scalar terms of Eqs.~\eqref{eq:E(0)}--\eqref{eq:E(3)},
\begin{align}
\label{eq:E_unrelaxed}
    E^{[3]}_{\rm u} = \sum_{n=0}^{3} E^{(n)}_{\rm u}.
\end{align}
The relaxed energy is obtained by a \emph{one-step} diagonalization [Eq.~\eqref{eq:energy_relaxed}] using the summed third-order DSRG transformed Hamiltonian:
\begin{align}
\label{eq:E_relaxed}
    \sum_{\mu}^{d} \braket{\Phi_{\nu} | \bar{H}^{[3]}_{1,2} (s) | \Phi^{\mu}} c_{\mu} = E^{[3]}(s) c_{\nu},
\end{align}
where we have applied the operator truncation $\bar{H}^{[3]}_{1,2} (s) = \sum_{n=0}^{3} \bar{H}^{(n)}_{1,2} (s)$.
As such, both relaxed and unrelaxed DSRG-MRPT3 energies are computed in a non-iterative fashion.
In Appendix \ref{sec:ref_relax}, we show that solving Eq.~\eqref{eq:E_relaxed} only requires minuscule amount of work comparing to computing Eq.~\eqref{eq:E_unrelaxed}.
For brevity, we shall take the energy of Eq.~\eqref{eq:E_relaxed} as our DSRG-MRPT3 energy, otherwise as uDSRG-MRPT3 for the unrelaxed energy.

We note that other more sophisticated zeroth-order Hamiltonian such as the one proposed by Dyall\cite{Dyall:1995ct} or the retaining-excitation Hamiltonian\cite{Fink:2006gu,Fink:2009gx} can also be straightforwardly applied to the DSRG perturbation theory.
However as discussed in Appendix \ref{sec:other_H0th}, using these Hamiltonians mostly lead to unnecessary complications and bring only small advantages.

\section{Implementation}
\label{sec:implementation}

To obtain the DSRG-MRPT3 working equations, the core task is to evaluate the commutators in Eqs.~\eqref{eq:Hbar_1st}--\eqref{eq:Hbar_3rd}.
Essentially the $k$-th nested commutator [$\hat{C}_{k} (s)$] is computed from the following recursive relation:
\begin{align}
    \hat{C}_{k} (s) = \frac{1}{k} [\hat{C}_{k-1} (s), \hat{A}(s)]_{1,2}, \quad k = 1, 2, 3, \cdots,
\end{align}
where $\hat{C}_{0} (s) = \hat{H}$.
In Ref.~\citenum{Li:2016hb} we have reported all terms of $\hat{C}_{1} (s) = [\hat{H}, \hat{A}(s)]_{1,2}$ and we shall not repeat them here.
Instead, we would like to discuss those terms that are necessary to implement the DSRG-MRPT3 scheme.

Computing the scalar term of commutator $[\hat{H}, \hat{T}(s)]$ for example requires only the non-diagonal terms of $\hat{H}$ ($\tens{f}{i}{a}$ and $\tens{v}{ij}{ab}$).
This statement suggests that we need elements such as $\tens{[\tilde{H}^{(n)} (s)]}{i}{a}$ and $\tens{[\tilde{H}^{(n)} (s)]}{ij}{ab}$ for $n = 1, 2$ to compute the second- and third-order energies.
For $n = 1$ we are able to write out explicit expressions due to the simple structure of $\hat{H}^{(0)}$:
\begin{align}
    \tens{[\tilde{H}^{(1)} (s)]}{i}{a} =& \tens{f}{a}{i, (1)} + [\tens{f}{a}{i, (1)} + \sum_{ux} \tens{\Delta}{u}{x} \tens{t}{ax}{iu,(1)} (s) \density{x}{u}] e^{-s(\Delta_{a}^{i})^2}, \\
    \tens{[\tilde{H}^{(1)} (s)]}{ij}{ab} =& \tens{v}{ab}{ij, (1)} + \tens{v}{ab}{ij, (1)} e^{-s(\Delta_{ab}^{ij})^2}.
\end{align}
When $n = 2$ we call for the non-diagonal elements of $\frac{1}{2} [\tilde{H}^{(1)}(s), \hat{A}^{(1)} (s)]_{1,2}$, computing which necessitate \emph{all} elements of $\tilde{H}^{(1)}(s)$.
Fortunately except for non-diagonal terms of $\tilde{H}^{(1)}(s)$, the remainders are equivalent to those in the first-order bare Hamiltonian.
We therefore factorize the two-electron integrals using density fitting (DF) or Cholesky decomposition (CD) techniques\cite{Whitten:1973ju,Dunlap:1979gh,Werner:2003gq,Beebe:1977dp,Koch:2003go,Aquilante:2008gk,Aquilante:2011fd,Weigend:2009bh,DePrince:2013ki,Hannon:2016bh} to avoid storing the four-index tensor of size $N^{4}$.
Specifically two-electron integrals are approximated as a contraction of two three-index tensors:
\begin{align}
\label{eq:int_fact}
    (pq | rs) \approx \sum_Q^M B^Q_{pq} B^{Q}_{rs},
\end{align}
where $Q$ is the auxiliary index.
The quantity $M$ is maximum value of $Q$ and it is roughly three times as large as the regular basis set ($M \approx 3 N$).

For small active spaces ($N_{\bf A} \ll N_{\bf C}$) the asymptotic scaling of the DSRG-MRPT3 method is ${\cal O}(N_{\bf V}^4 N_{\bf C}^2)$, resulting from the contraction:
\begin{align}
\label{eq:scaling_problem}
    \tens{\bar{H}}{ef}{ij,(2)} (s) \leftarrow \sum_{gh}^{\bf V} \tens{v}{ef}{gh,(1)} \tens{t}{gh}{ij,(1)} (s).
\end{align}
This tensor contraction is also contained in the coupled cluster equations and it is a scourge because its computational cost cannot be reduced by factorizing the two-electron integrals $\tens{v}{ef}{gh,(1)}$ via DF or Cholesky decomposition,\cite{Schutz:2003jy} instead it requires the use of alternative factorizations.\cite{Parrish:2014ig}
As a compromise, we form $\tens{v}{ef}{gh,(1)}$ in batches of compound indices $gh$ according to Eq.~\eqref{eq:int_fact}, with the size of each batch automatically determined by available memory.
For large active spaces, the computational cost of the DSRG-MRPT3 is constrained by terms such as:
\begin{align}
\label{eq:E_cumulant}
    E^{(2)}_{\rm u} \leftarrow \frac{1}{4} \sum_{xyz \atop uvw}^{\bf A} \sum_{e}^{\bf V} \tens{v}{xy}{we,(1)} (s) \tens{t}{ez}{uv,(1)} (s) \cumulant{xyz}{uvw},
\end{align}
where $\cumulant{xyz}{uvw}$ is the three-particle density cumulant of the reference.
Although the term given in Eq.~\eqref{eq:E_cumulant} scales as ${\cal O}(N_{\bf A}^6 N_{\bf V})$, the effort made to compute it is still significantly less than the most expensive step in CASPT2 or NEVPT2 [${\cal O}(N_{\bf A}^9)$].

The DSRG-MRPT3 method is implemented in our open-source code \forte,\cite{FORTE2016} a plugin to the \PSI package\cite{PSI4} that specializes on multireference methods.
All tensor contractions are written using the syntax provided by the open-source tensor library \ambit.\cite{AMBIT2016}
The three-index DF integrals generated by \PSI are read using a general interface developed by Hannon \textit{et al.}\cite{Hannon:2016bh}
The DSRG-MRPT3 equations are spin integrated such that one-body terms are decomposed into $\alpha$ and $\beta$ blocks and two-body terms are divided into $\alpha\alpha$, $\alpha\beta$, and $\beta\beta$ contributions.
For convenience, the current implementation constantly stores nine tensors of size $N_{\bf H}^2 N_{\bf P}^2 / 16$ coming from the spin integration of the $t_2$ amplitudes (either first- or second-order) and two two-body intermediates.
This storage requirement limits practical applications to about 650 basis functions of 100 correlated electrons on a computer with 120 GB of memory.

\section{Computational Details}
\label{sec:comput_detail}

We tested the DSRG-MRPT3 method on the ground-state potential energy curves (PECs) of \ce{F2}, \ce{H2O2}, \ce{C2H6}, and \ce{N2} by breaking the respective F--F, O--O, C--C, and N--N bonds.
The DSRG-MRPT3 PECs were compared to those of other multireference methods including DSRG-MRPT2,\cite{Li:2015iz} MR-LDSRG(2),\cite{Li:2016hb} NEVPT2,\cite{Angeli:2001bg,Angeli:2002ik,Angeli:2007by} CASPT2,\cite{Andersson:1992cq,Werner:1996in} CASPT3,\cite{Werner:1996in} MRCISD,\cite{Knowles:1988hv,Werner:1988ku} and MRCISD with Davidson correction (MRCISD+Q).\cite{Langhoff:1974kz}
All multireference computations are based on minimal CASSCF references, specifically, CAS(2,2) for \ce{F2}, \ce{H2O2}, and \ce{C2H6}, while CAS(6,6) for \ce{N2}.
Full configuration interaction (FCI) data served as the benchmark for \ce{F2} and \ce{N2}.\cite{Bytautas:2007fa,Li:2016hb}
For \ce{H2O2} and \ce{C2H6}, we took the reference data from Ref.~\citenum{Yang:2013gq}, which were computed using coupled cluster singles, doubles and triples augmented with second-order perturbative quadruples corrections [CCSDT(2)$_\text{Q}$].\cite{Hirata:2004fm}
Dunning's correlation-consistent double-$\zeta$ (cc-pVDZ) basis set\cite{Dunning:1989bx} was used and the molecular orbitals constructed mainly from the 1$s$ orbitals of C, N, O, and F atoms were frozen in all post-CASSCF computations.

To measure the quality of the potential energy curves we use the nonparallelism error (NPE) computed with respect to a reference method [FCI or CCSDT(2)$_\mathrm{Q}$] over a range of bond lengths $\bf R$, defined as:
\begin{equation}
\text{NPE} = \max_{r \in {\bf R}} \Delta E(r) -  \min_{r \in {\bf R}} \Delta E(r),
\end{equation}
where $\Delta E(r)$ is the error with respect to the reference energy at bond length $r$.

\begin{figure}[!ht]
\centering
    \ifpreprint
    \includegraphics[width=0.55\columnwidth]{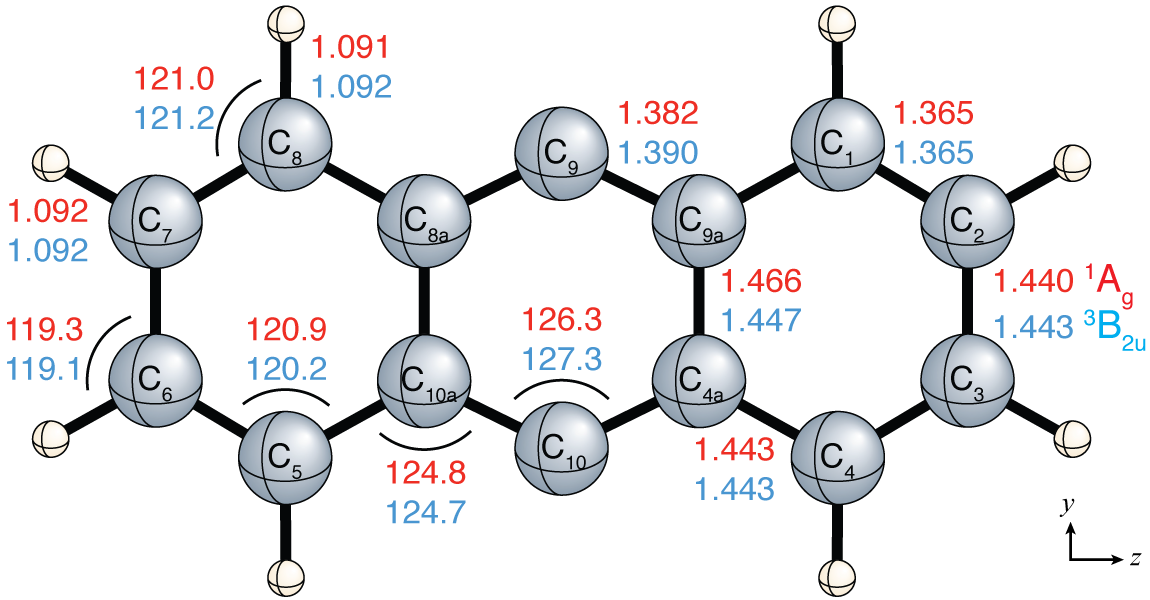}
    \else
    \includegraphics[width=0.8\columnwidth]{figure_1}
    \fi
    \caption{Optimized geometries (in {\AA}ngstr{\"o}ms and degrees) of the singlet and triplet 9,10-anthracyne using DSRG-MRPT3 ($s = 0.5$ \sunit) with the CASSCF(2,2) reference and the cc-pVDZ basis set.  This figure was made with the cheMVP package, see Ref.~\citenum{cheMVP}}
    \label{fig:anthracyne}
\end{figure}

As an application to medium-sized molecules, we studied the singlet-triplet splitting ($\Delta_{\rm ST} = E_{\rm T} - E_{\rm S}$) of 9,10-anthracyne (9,10-didehydroanthracene, see Fig.~\ref{fig:anthracyne}), which is recently found capable of retro-Bergman cyclization on a NaCl/Cu(111) surface when manipulated with the CO tip of an atomic force microscope.\cite{Schuler:2016dv}
We first optimized the geometries of singlet and triplet 9,10-anthracyne at the CASSCF(2,2)-DSRG-MRPT3/cc-pVDZ level of theory using gradients from 3-point finite-difference computations.
These geometries were characterized as minima by finite-difference harmonic vibrational analyses.
Then we computed the $\Delta_{\rm ST}$ of 9,10-anthracyne using various multireference methods including DSRG-MRPT2/3, NEVPT2, the partially contracted version of CASPT2\cite{Celani:2000cf} and MRCISD\cite{Shamasundar:2011ew} as implemented in the RS2C and CIC modules of MOLPRO,\cite{MOLPRO-WIREs}
Mukherjee multireference coupled cluster theory with singles and doubles (Mk-MRCCSD),\cite{Mahapatra:1998cp,Mahapatra:1999ev,Evangelista:2006gf,Mahapatra:2010kj} and Mk-MRCCSD with perturbative triples [Mk-MRCCSD(T)]\cite{Evangelista:2010cq} based on a CASSCF reference.
Both cc-pVDZ and cc-pVTZ basis sets were adopted and the $1s$-like orbitals on carbon atoms were excluded for dynamic correlations.
The two-electron integrals in DSRG-MRPT2/3 computations were approximated by Cholesky decomposition\cite{Beebe:1977dp,Koch:2003go,Aquilante:2008gk,Aquilante:2011fd} with a threshold of $10^{-8}$ a.u.~for geometry optimizations and $10^{-6}$ a.u.~for single points.

The NEVPT2, CASPT2, CASPT3, and MRCISD energies were computed using the \molpro 2015.1 program\cite{MOLPRO-WIREs,MOLPRO2015} and the remaining were obtained from \PSI.\cite{PSI4}
The $xyz$ coordinates of the optimized 9,10-anthracyne, as well as all energies on the PECs of \ce{F2}, \ce{H2O2}, \ce{C2H6}, and \ce{N2} are available in the supplementary material.\cite{SI}
For convenience, the supplementary material\cite{SI} also includes the internal coordinates of \ce{H2O2} and \ce{C2H6}, which are taken from Refs.~\citenum{Yang:2013gq} and \citenum{Karton:2007da}, respectively.

\section{Results}
\label{sec:results}

\subsection{Potential energy curves}
In this section we assess the accuracy of DSRG-MRPT3 by investigating four bond breaking processes.
We use a recent benchmark set by Yang, Jalan, Green, and Truhlar (YJGT) that was used to compare the performance of numerous single-reference coupled cluster and multireference methods.\cite{Yang:2013gq}
The YJGT set contains \ce{F2}, \ce{H2O2} and \ce{C2H6} scanned along the \ce{F-F}, \ce{O-O}, and \ce{C-C} bonds, respectively.
To test a multiple-bond breaking process, we consider the dissociation curve of \ce{N2}, a routine benchmark for multireference methods.\cite{Yanai:2007ix,Das:2010bj,Kurashige:2014bq,FossoTande:2016hb}

\begin{figure*}[ht]
\centering
    \ifpreprint
    \includegraphics[width=0.90\columnwidth]{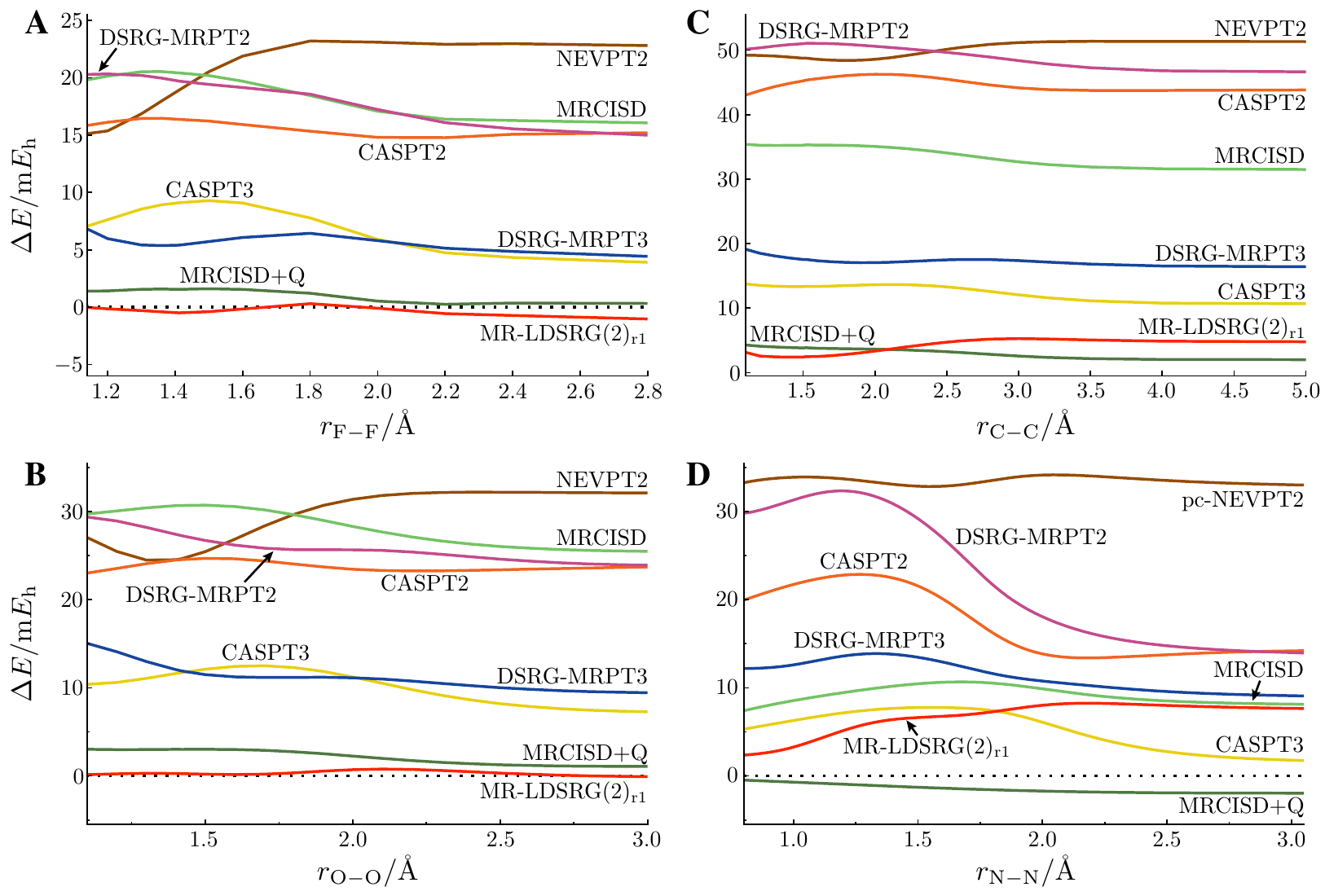}
    \else
    \includegraphics[width=1.80\columnwidth]{figure_2}
    \fi
    \caption{Energy deviations of various multireference methods for the ground-state potential energy curves of: (A) \ce{F2}, (B) \ce{H2O2}, (C) \ce{C2H6}, and (D) \ce{N2} relative to FCI, CCSDT(2)$_\text{Q}$, CCSDT(2)$_\text{Q}$, and FCI, respectively. All DSRG methods employ $s = 0.5$ \sunit. The MR-LDSRG(2)$_\text{r1}$ theory adopts the one-step relaxation scheme.}
    \label{fig:pec}
\end{figure*}

\newcolumntype{d}[1]{D{.}{.}{#1}}
\begin{table*}[ht!]
\centering
\ifpreprint
\renewcommand{\arraystretch}{0.8}
\else
\renewcommand{\arraystretch}{1.25}
\fi
\caption{Maximum error (MAX) and nonparallelism error (NPE) for the ground-state potential energy curves of \ce{F-F}, \ce{HO-OH}, \ce{H3C-CH3}, and \ce{N}$\equiv${N} computed with various methods (reported in units of m\Eh).  All DSRG methods employ a value of the flow variable $s = 0.5$ \sunit. The last column shows the average NPE.}

\label{tab:error}
\ifpreprint
\begin{tabular*}{\columnwidth}{@{\extracolsep{\stretch{1}}}l*{2}{d{2.2}}d{2.2}d{1.2}d{2.2}d{1.2}*{3}{d{2.2}}@{}}
\else
\begin{tabular*}{2.08\columnwidth}{@{\extracolsep{\stretch{1}}}l*{2}{d{2.2}}d{2.2}d{1.2}d{2.2}d{1.2}*{3}{d{2.2}}@{}}
\fi
\hline
\hline
 & \multicolumn{2}{c}{\ce{F2}} & \multicolumn{2}{c}{\ce{H2O2}} & \multicolumn{2}{c}{\ce{C2H6}} & \multicolumn{2}{c}{\ce{N2}} & \multicolumn{1}{c}{Average} \\
 \cline{2-3} \cline{4-5} \cline{6-7} \cline{8-9}
 Method & \multicolumn{1}{c}{MAX} & \multicolumn{1}{c}{NPE} & \multicolumn{1}{c}{MAX} & \multicolumn{1}{c}{NPE} & \multicolumn{1}{c}{MAX} & \multicolumn{1}{c}{NPE} & \multicolumn{1}{c}{MAX} & \multicolumn{1}{c}{NPE} & \multicolumn{1}{c}{NPE} \\
\hline
pc-NEVPT2 & 23.20 & 8.08 & 32.21 & 7.71 & 51.36 & 2.97 & 34.17 & 1.41 & 5.04 \\
CASPT2 & 16.46 & 1.69 & 24.69 & 1.66 & 46.25 & 3.22 & 22.87 & 9.49 & 4.02 \\
uDSRG-MRPT2 & 25.23 & 10.24 & 31.14 & 7.23 & 52.14 & 5.50 & 32.91 & 18.87 & 10.46 \\
DSRG-MRPT2 & 20.33 & 5.34 & 29.38 & 5.46 & 51.06 & 4.42 & 32.35 & 18.34 & 8.39 \\
CASPT3 & 9.29 & 5.38 & 12.51 & 5.23 & 13.76 & 3.08 & 7.77 & 6.31 & 5.00 \\
uDSRG-MRPT3 & 9.46 & 5.03 & 15.50 & 6.04 & 19.15 & 2.73 & 15.19 & 6.05 & 4.96 \\
DSRG-MRPT3 & 6.81 & 2.39 & 15.04 & 5.59 & 19.13 & 2.71 & 13.88 & 4.78 & 3.87 \\
MRCISD & 20.53 & 4.47 & 30.73 & 5.24 & 35.40 & 3.89 & 10.67 & 3.27 & 4.22 \\
MRCISD+Q & 1.60 & 1.35 & 3.02 & 1.94 & 4.29 & 2.27 & 1.96 & 1.50 & 1.77 \\
uMR-LDSRG(2) & 3.83 & 4.86 & 3.89 & 3.95 & 6.08 & 3.43 & 9.30 & 5.25 & 4.37 \\
MR-LDSRG(2)$_{\rm r1}$ & -1.03 & 1.34 & 0.78 & 0.86 & 5.28 & 2.87 & 8.25 & 4.98 & 2.51 \\
MR-LDSRG(2) & -5.19 & 4.24 & -3.87 & 3.75 & 4.92 & 3.12 & 7.61 & 4.81 & 3.98 \\
\hline
\hline
\end{tabular*}
\end{table*}

The errors in the computed ground-state potential energy curves of \ce{F2}, \ce{H2O2}, \ce{C2H6} and \ce{N2} are shown in Fig.~\ref{fig:pec} and Table \ref{tab:error}.
Fig.~\ref{fig:pec} shows that third-order MRPTs recover a larger fraction of electron correlation than the corresponding second-order methods.
For example, going from DSRG-MRPT2 to DSRG-MRPT3, the maximum error is reduced by an amount between one third (\ce{F2}) and one half (\ce{H2O2}).
Despite the fact that the DSRG-MRPT3 does not always yield the curve with the smallest absolute error, we find that the average NPE of the DSRG-MRPT3 (3.87 m\Eh) is smaller than that of CASPT3 (5.00 m\Eh), MRCISD (4.22 m\Eh), and the MR-LDSRG(2) (3.98 m\Eh).
Interestingly, the DSRG-MRPT3 significantly improves upon the PEC of \ce{N2} computed with DSRG-MRPT2.
For this molecule, the DSRG-MRPT3 gives a NPE equal to 4.78 m\Eh vs. 18.34 m\Eh for the DSRG-MRPT2.
Overall, a comparison of the various MR-DSRG methods considered here suggests that the accuracy of these methods follows the trend: DSRG-MRPT2 $\ll$ DSRG-MRPT3 $<$ MR-LDSRG(2).

\begin{figure*}[!ht]
\centering
    \ifpreprint
    \includegraphics[width=0.90\columnwidth]{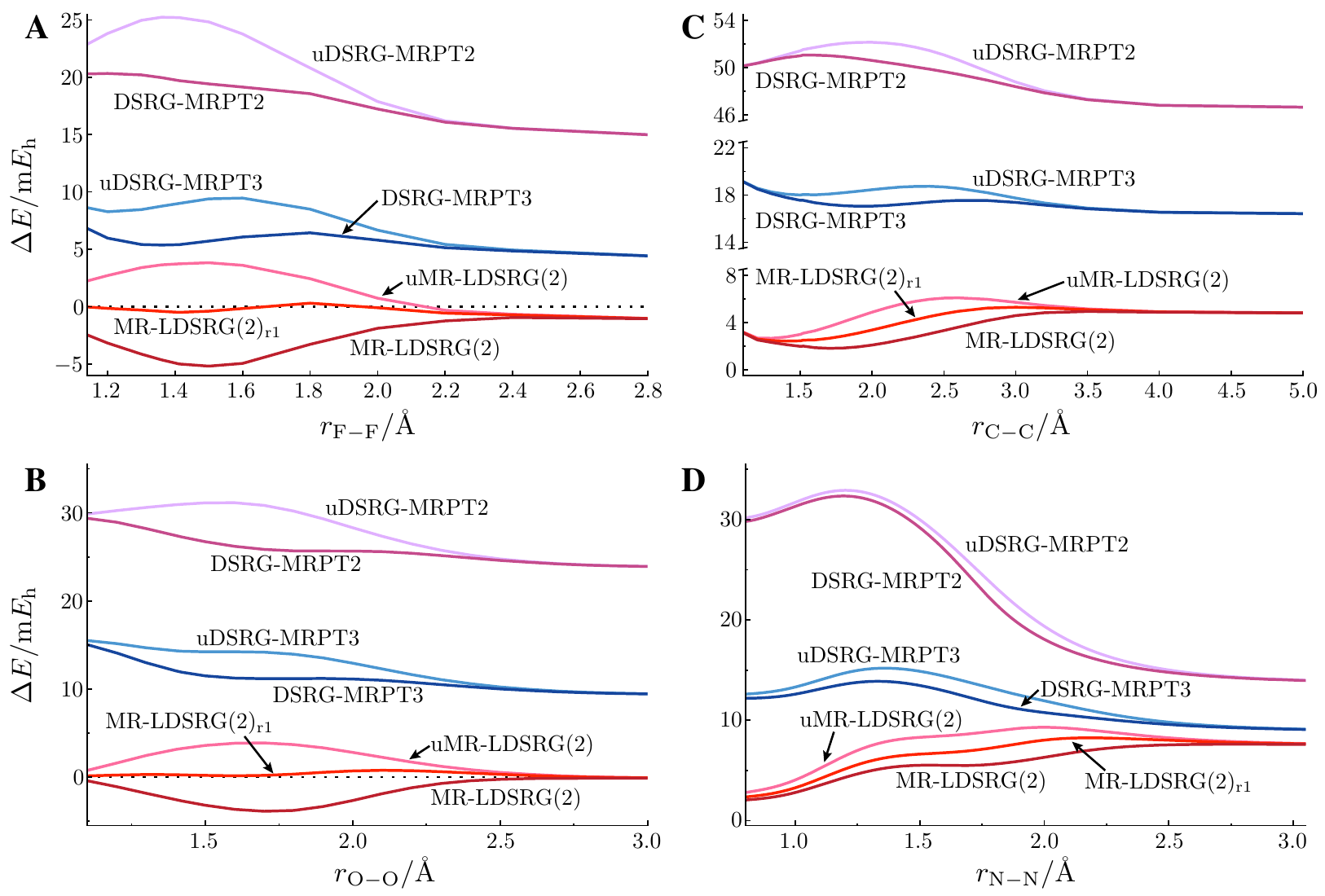}
    \else
    \includegraphics[width=1.75\columnwidth]{figure_3}
    \fi
    \caption{Energy deviations of various multireference DSRG ($s = 0.5$ \sunit) methods for the ground-state potential energy curves of: (A) \ce{F2}, (B) \ce{H2O2}, (C) \ce{C2H6} , and (D) \ce{N2} relative to FCI, CCSDT(2)$_\text{Q}$, CCSDT(2)$_\text{Q}$, and FCI, respectively.}
    \label{fig:relax}
\end{figure*}

Figure \ref{fig:relax} shows a comparison of different reference-relaxation (RR) approaches applied to perturbative and nonperturbative MR-DSRG methods.
For all combinations of molecules and methods the one-step RR curves are more accurate than the corresponding unrelaxed curves.
For example, the average NPE of the unrelaxed DSRG-MRPT3 is 1.09 m\Eh higher than the one-step RR approach.
For the MR-LDSRG(2) series, the one-step RR results (labeled with the subscript ``r1'') show energy errors that are halfway between those from unrelaxed and fully relaxed computations.
These findings suggest that at the perturbative level it is generally advantageous to introduce reference relaxation effects and that a one-step approach is a good compromise between accuracy and cost.
Beyond perturbation theory we observe that full relaxation of the reference does in certain cases give results inferior to the one-step correction (see the \ce{F2} and \ce{H2O2} curves).  However, in our opinion the fully relaxed MR-LDSRG(2) is the most rigorous formulation of the nonpertubative MR-DSRG approach and should be the default choice in applications to chemical problems.

\subsection{Singlet-triplet splittings of 9,10-anthracyne}

\begin{figure*}[!ht]
\centering
    \ifpreprint
    \includegraphics[width=1.0\columnwidth]{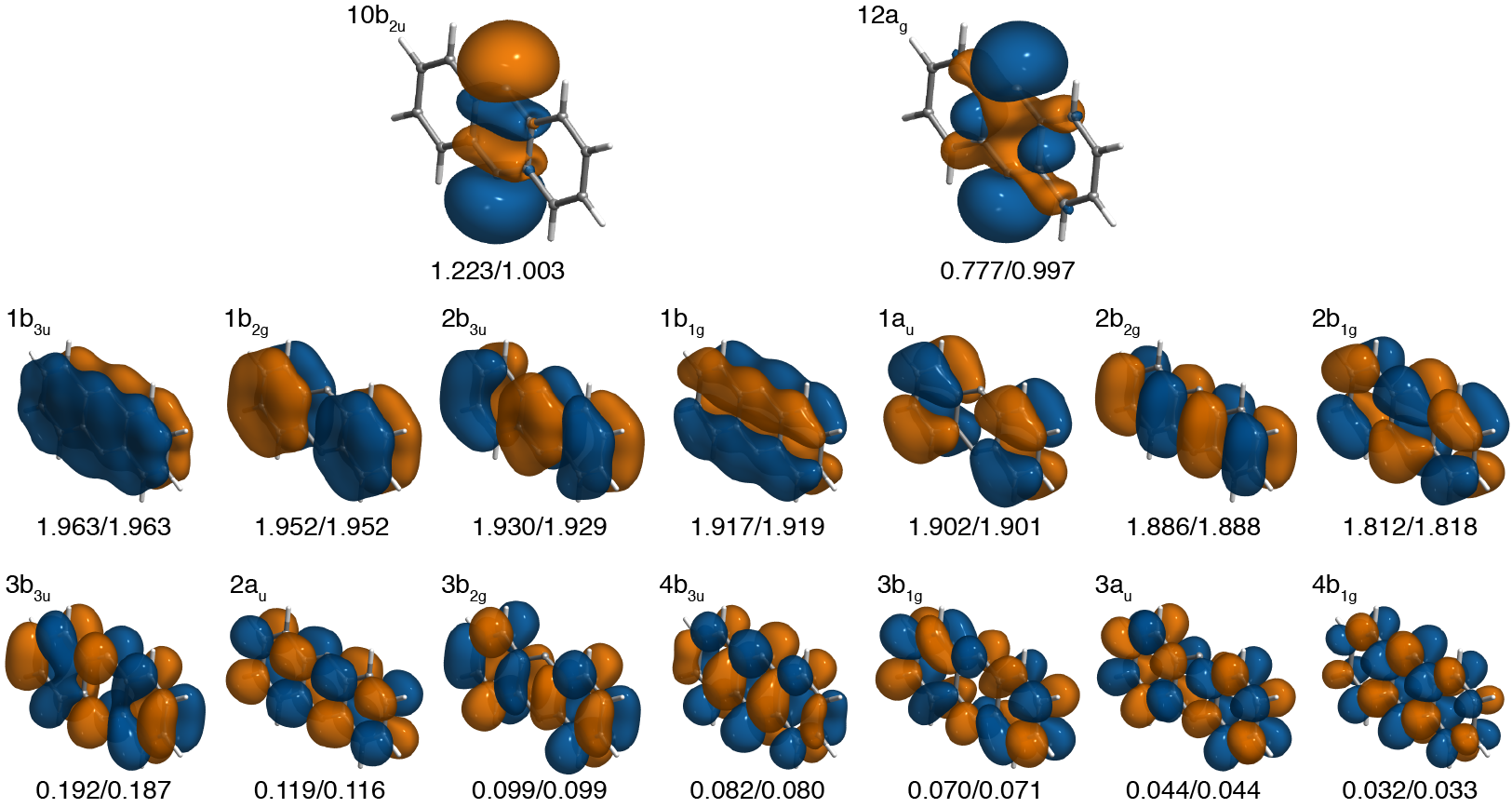}
    \else
    \includegraphics[width=1.90\columnwidth]{figure_4_png}
    \fi
    \caption{The $\sigma$ and $\pi$ active orbitals of singlet 9,10-anthracyne computed at the CASSCF(16,16)/cc-pVDZ level of theory.
    These orbitals were canonicalized by diagonalization of the active block of the active Fock matrix.  Orbital occupation numbers (ONs) are reported for both the singlet and triplet states and are given in the format: singlet ON/triplet ON.}
    \label{fig:orbs}
\end{figure*}

The simplest active space for 9,10-anthracyne is a CAS(2,2) that consists of two electrons in two $\sigma$ orbitals that belong to the dehydrogenated carbon atoms.
From our previous experience with \textit{p}-benzyne,\cite{Li:2015iz,Li:2016hb} adding six $\pi$-type orbitals to the active space increases the singlet-triplet splitting ($\Delta_{\rm ST} = E_{\rm T} - E_{\rm S}$) by up to 2 \kcal.
As noted before,\cite{Li:2016hb} this energy shift likely results from an improved treatment of static correlation effects by the larger active space, since both the CASSCF and DSRG-MRPT2 singlet-triplet splittings are shifted by a similar amount.
For 9,10-anthracyne, there are fourteen $\pi$ orbitals, so that a full treatment of the $\sigma$ and $\pi$ orbitals leads to a CAS(16,16) reference wave function.
The sixteen orbitals of the CAS(16,16) space are shown in Figure \ref{fig:orbs} along with their corresponding occupation numbers (ONs).
The ON for each of the sixteen orbitals lies in the range of 0.02--1.98, justifying the inclusion of all $\pi$ orbitals to the active space.\cite{Bofill:1989dd,Keller:2015ds}
Since---with the exception of the DSRG-MRPT2/3 approaches---the CAS(16,16) is too large for the methods considered here, we also report results for three smaller active spaces that include a subset of the $\pi$ space: CAS(4,4), CAS(8,8), and CAS(12,12).
Specifically, the CAS(4,4) active space includes the $2b_{1g}$ and $3b_{3u}$ orbitals, while the CAS(8,8) further adds $1a_{u}, 2b_{2g}, 2a_{u}$, and $3b_{2g}$ orbitals.
The CAS(12,12) active space is obtained by augmenting the CAS(8,8) reference with the $2b_{3u}, 1b_{1g}, 4b_{3u}$, and $3b_{1g}$ orbitals.

The CASSCF(2,2)-DSRG-MRPT3/cc-pVDZ optimized geometries of singlet and triplet 9,10-anthracyne are shown in Fig.\ \ref{fig:anthracyne}.
All \ce{C-C} bond lengths are consistent with those predicted by density functional theory (DFT) except for the \ce{C_{4a}-C_{9a}} bond of the singlet ground state.\cite{Schuler:2016dv}
Interestingly, the singlet \ce{C_{4a}-C_{9a}} bond length reported in Ref.\ \citenum{Schuler:2016dv} (1.578 \AA) is not only 0.11 {\AA} longer than our prediction (1.466 \AA), but also larger than a typical $sp^3$ hybridized \ce{C-C} \emph{single} bond like in ethane (1.536 \AA{}, from Ref.~\citenum{Herzberg1966}).

\newcolumntype{d}[1]{D{.}{.}{#1}}
\begin{table}[ht!]
\begin{threeparttable}
\centering
\ifpreprint
\renewcommand{\arraystretch}{0.8}
\else
\renewcommand{\arraystretch}{1.25}
\fi
\caption{Adiabatic singlet-triplet splitting ($\Delta_{\rm ST} = E_{\rm T} - E_{\rm S}$, in \kcal) of 9,10-anthracyne computed using various methods with the cc-pVTZ basis set. All values are corrected by the zero-point vibrational energy ($+ 0.264$ \kcal) obtained at the CASSCF(2,2)-DSRG-MRPT3/cc-pVDZ level of theory.}
\label{tab:ST}

\begin{tabular*}{\columnwidth}{@{\extracolsep{\stretch{1}}}l*{5}{d{1.2}}@{}}

\hline
\hline
 & \multicolumn{5}{c}{CAS} \\
 \cline{2-6}
 Method & \multicolumn{1}{c}{(2,2)} & \multicolumn{1}{c}{(4,4)} & \multicolumn{1}{c}{(8,8)} & \multicolumn{1}{c}{(12,12)} & \multicolumn{1}{c}{(16,16)} \\
 \hline
 CASSCF & 0.61 & 2.99 & 4.52 & 2.64 & 2.42 \\
 CASPT2 & 4.08 & 5.47 & 3.24 & 4.78 &  \\
 sc-NEVPT2 & 2.90 & 3.79 & 0.78 & 4.02 & \\
 pc-NEVPT2 & 3.08 & 3.86 & 0.90 & 4.01 & \\
 uDSRG-MRPT2\tnote{a} & 2.15 & 3.82 & 3.20 & 3.72 & 3.66 \\
 DSRG-MRPT2\tnote{a} & 2.58 & 3.89 & 3.21 & 3.86 & 3.78 \\
 DSRG-MRPT2\tnote{b} & 3.19 & 4.00 & 2.52 & 3.96 & 3.85 \\
 uDSRG-MRPT3\tnote{a} & 1.74 & 3.50 & 5.36 & 3.46 & 3.40 \\
 DSRG-MRPT3\tnote{a} & 1.98 & 3.70 & 5.50 & 3.66 & 3.61 \\
 DSRG-MRPT3\tnote{b} & 2.38 & 3.86 & 5.58 & 3.93 & 3.90 \\
 MRCISD & 1.23 & 3.23 & 4.30 & 3.00 & \\
 MRCISD+Q & 1.71 & 3.41 & 4.23 & 3.25 & \\
 Mk-MRCCSD & 4.44 & & \\
 Mk-MRCCSD(T) & 3.98 & & \\
\hline
\hline
\end{tabular*}
\begin{tablenotes}
\item [a] Computed using $s = 0.5$ \sunit.
\item [b] Computed using $s = 1.0$ \sunit.
\end{tablenotes}
\end{threeparttable}
\end{table}

In Table \ref{tab:ST}, we list the singlet-triplet splittings ($\Delta_{\rm ST} = E_{\rm T} - E_{\rm S}$) of 9,10-anthracyne computed using numerous multireference methods and the cc-pVTZ basis set.
All values are adjusted by $+0.264$ \kcal to account for zero-point harmonic vibrational energy corrections computed at the CASSCF(2,2)-DSRG-MRPT3/cc-pVDZ level of theory.
Because the singlet-triplet splitting computed with the cc-pVDZ and cc-pVTZ basis are in very good agreement (within 0.5 \kcal), we report results only for the latter basis set and provide data for the former in the supplementary material.\cite{SI}
Benchmark computations on \textit{p}-benzyne show that Mk-MRCCSD(T) yields an accurate singlet-triplet splitting,\cite{Evangelista:2010cq,Evangelista:2012fo} therefore, in the absence of an experimental value for 9,10-anthracyne, we take the $\Delta_{\rm ST}$ from Mk-MRCCSD(T)  as our reference.

In contrast to the DFT prediction of a triplet ground state,\cite{Schuler:2016dv} all methods reported in Table \ref{tab:ST} favor a singlet ground state.
For the minimal active space, the CASPT2 $\Delta_{\rm ST}$ (4.1 \kcal) is in excellent agreement with our reference value [4.0 \kcal from Mk-MRCCSD(T)], while the remaining methods underestimate the $\Delta_{\rm ST}$ by 1--2 \kcal.
As the active-space size increases, better agreements to the Mk-MRCCSD(T) value are observed for all methods.
For example, when using a CASSCF(16,16) reference, singlet-triplet splittings of DSRG-MRPT2 and -MRPT3 (s = 0.5 \sunit) are 3.8 and 3.6 \kcal, respectively.
For both DSRG-MRPT2 and -MRPT3 approaches, reference relaxation shifts the $\Delta_{\rm ST}$ to higher values, improving the agreement with the Mk-MRCCSD(T) result.
The effect of reference relaxation is largest for the minimal active space, where in the case of the DSRG-MRPT2 ($s = 0.5$ \sunit) it amounts to an increase in the singlet-triplet splitting by 0.4 \kcal.
We also observe that the $\Delta_{\rm ST}$ obtained from DSRG-MRPT2 and -MRPT3 approaches does not depend significantly on the value of the flow variable: as $s$ increases from 0.5 to 1.0 \sunit, the largest variation in $\Delta_{\rm ST}$ is less than 0.4 \kcal.

We note that results for the CAS(8,8) reference are somewhat erratic and they deviate significantly from those obtained with CAS(4,4), CAS(12,12), and CAS(16,16) references.
A plausible reason for this inconsistency is that some of the singlet and triplet semicanonical orbitals of CASSCF(8,8) are considerably different from those obtained with other active spaces (see supplementary material\cite{SI}).

It is instructive to compare the $\Delta_{\rm ST}$ between 9,10-anthracyne and \textit{p}-benzyne.
The $\Delta_{\rm ST}$ of \textit{p}-benzyne is measured to be $3.8 \pm 0.4$ \kcal by photoelectron spectroscopy,\cite{Wenthold:1998cx} and Mk-MRCCSD(T)/cc-pVTZ with a CASSCF(2,2) predicts $\Delta_{\rm ST}$ = 4.45 \kcal.\cite{Evangelista:2012fo}
Therefore, we would expect the experimental $\Delta_{\rm ST}$ of 9,10-anthracyne to fall in the range 2.9--3.7 \kcal.
The predictions of NEVPT2 and DSRG-MRPT2 always fall in this range for all the active spaces considered here, while CASPT2 constantly overestimate the $\Delta_{\rm ST}$ of 9,10-anthracyne.
The DSRG-MRPT3 results are always in line with those of MRCISD, and the minimal active space seems to be insufficient in this case.

\section{Conclusions}
\label{sec:conclusions}

We have introduced the DSRG-MRPT3 scheme, a renormalized third-order multireference perturbation theory derived from the MR-DSRG approach.\cite{Li:2015iz,Li:2016hb,Hannon:2016bh}
By working in a semicanonical basis and using a zeroth-order Hamiltonian [$\hat{H}^{(0)}$] that contains only the diagonal blocks of the Fock matrix, the DSRG-MRPT3 energy may be obtained by a non-iterative procedure that scales as ${\cal O}(N_{\bf C}^2 N_{\bf V}^4)$.
The cost of evaluating the DSRG-MRPT3 energy is reduced by truncating each commutator in the effective Hamiltonian with only one- and two-body operators.\cite{Yanai:2006gi,Yanai:2007ix}
As a result,  the DSRG-MRPT3 energy and amplitude equations require up to three-body density cumulants.
Furthermore, we show that the third-order effective Hamiltonian may be easily formed from the first- and second-order amplitudes, and that it may be diagonalized to obtain a relaxed model space.

We have benchmarked the DSRG-MRPT3 method on the ground-state potential energy curves of \ce{F2}, \ce{H_2O_2}, \ce{C_2H_6} and \ce{N2} and found that, on average, the DSRG-MRPT3 nonparallelism error (3.9 m\Eh) is comparable to that of CASPT3 (5.0 m\Eh) and MRCISD (4.2 m\Eh).
We have also shown that accounting for reference relaxation effects with a one step diagonalization of the effective Hamiltonian does generally improve the quality of DSRG-MRPT2 and -MRPT3 results.
For example, in the case of DSRG-MRPT3, the one-step relaxation approach reduces the average NPE by 1.1 m\Eh.
Since the cost of evaluating the one-step relaxed energy is a small fraction of the cost of an unrelaxed DSRG-MRPT3 computation, we recommend the former approach as the default method.

In our opinion, the DSRG-MRPT3 is best viewed as an economical multireference method that offers a good compromise between accuracy and computational cost.
Compared to more expensive nonperturbative (i.e. coupled cluster like) multireference methods, the DSRG-MRPT3 has the advantage that it does not rely on an iterative procedure and has reduced I/O costs.  In its current implementation, the DSRG-MRPT3 may be routinely applied in computations with a hundred correlated electrons and 500--600 basis functions.
This point is illustrated with computations of the singlet-triplet splitting ($\Delta_{\rm ST}$) of 9,10-anthracyne (C$_{14}$H$_8$).
The DSRG-MRPT3 based on a CASSCF(16,16) reference with $s = 1.0$ \sunit predicts the $\Delta_{\rm ST}$ to be 3.9 \kcal, only 0.1 \kcal smaller than our best estimate from Mk-MRCCSD(T).
In our experience, the CAS(16,16) active space is too large for practical CASPT2 and NEVPT2 computations on a single computer node.
With Cholesky decomposed integrals (using a $10^{-6}$ a.u. threshold), the CAS(16,16) DSRG-MRPT3 energy of 9,10-anthracyne may be computed in about 12.5 hours using 16 threads on two Intel Xeon E5-2650 v2 processors and 128GB of memory.

In conclusion, we have shown that the DSRG framework may be used to formulate a third-order multireference perturbation theory that avoids some of the major limitations of other MRPT approaches and has a favorable accuracy/cost ratio.
These results suggest that it might be worthwhile to explore multi-state generalizations of the DSRG-MRPT2 and DSRG-MRPT3 methods to compute excited state energies and their analytic gradients.
This work also presents additional benchmark results that validate the accuracy of perturbative and nonperturbative computational methods based on the MR-DSRG formalism.

\begin{acknowledgments}
C.L.\ would like to thank Dr.\ Zhi Sun for performing the CASSCF(12,12)-NEVPT2 computations of 9,10-anthracyne.
This work was supported by Emory University and the U.S. Department of Energy under Award No. DE-SC0016004.
\end{acknowledgments}

\appendix

\section{Reference relaxation}
\label{sec:ref_relax}

In this section we provide details of the reference relaxation procedure used in the DSRG-MRPT2/3 and MR-LDSRG(2) methods.
We first introduce the many-body expression of the DSRG transformed Hamiltonian
\begin{align}
\label{eq:Hbar_mb}
    \bar{H}(s) = \bar{H}_{0}(s) + \sum_{k=1}^{n} \bar{H}_{k}(s),
\end{align}
where $\bar{H}_{0}(s)$ is the scalar term obtained by summing the reference energy ($E_0$) and all the fully contracted contributions from $\bar{H}(s)$.  The quantity $\bar{H}_{k}(s)$ contains the $k$-body contributions to $\bar{H}(s)$:
\begin{align}
    \bar{H}_k(s) =  \frac{1}{(k!)^2} \sum_{pqrs\cdots}^{\mathbf{G}} \tens{\bar{H}}{pq\cdots}{rs\cdots} (s) \no{\sqop{pq\cdots}{rs\cdots}}.
\end{align}
The approximated Hamiltonian $\bar{H}_{1,2} (s)$ is thus obtained when setting $n = 2$ in Eq.~\eqref{eq:Hbar_mb}.

In order to solve the eigenvalue problem of Eq.~\eqref{eq:E_relaxed}, we express the transformed Hamiltonian using operators normal ordered with respect to the true vacuum.
Defining the two-particle density matrix $\density{pq}{rs} = \braket{\mref| \cop{p} \cop{q} \aop{s} \aop{r} |\mref}$, the $\bar{H}_{1,2} (s)$ operator is written as [``$(s)$'' is dropped for brevity],\cite{Li:2016hb}
\begin{align}
\label{eq:Hbar_DNO}
    \bar{H}_{1,2} =& \bar{H}_0 - \sum_{uv} \tens{\bar{H}}{u}{v} \density{u}{v} - \sum_{uvxy} \tens{\bar{H}}{uv}{xy} (\frac{1}{4} \density{uv}{xy} - \density{u}{x}\density{v}{y}) \notag\\
    &- \sum_{m} \tens{\bar{H}}{m}{m} + \frac{1}{2} \sum_{mn} \tens{\bar{H}}{mn}{mn} + \sum_{muv} \tens{\bar{H}}{mu}{mv} \density{u}{v} \notag\\
    &+\sum_{pq} \Big[ \tens{\bar{H}}{p}{q} - \sum_{ij} \tens{\bar{H}}{pi}{qj} \density{i}{j}  \Big] \sqop{p}{q} + \frac{1}{4} \sum_{pqrs} \tens{\bar{H}}{pq}{rs} \sqop{pq}{rs}.
\end{align}
The quantities $[\tens{o}{p}{q} = \tens{\bar{H}}{p}{q} - \sum_{ij}\tens{\bar{H}}{pi}{qj} \density{i}{j}]$ and $[\tens{w}{pq}{rs} = \tens{\bar{H}}{pq}{rs}]$ may be considered as MR-DSRG dressed one- and two-electron integrals, respectively.

A common strategy to compute the CASCI energy is to fold the contribution of core orbitals into a scalar term ($c_0$) and the one-body term labeled by all active indices ($\tens{c}{u}{v}$).
Using the dressed integrals ($\tens{o}{p}{q}, \tens{w}{pq}{rs}$), the additional scalar term coming from the core orbitals is given by
\begin{align}
\label{eq:c0_cas}
    c_0 &= \sum_{m}^{\mathbf{C}} \tens{o}{m}{m} + \frac{1}{2} \sum_{mn}^{\mathbf{C}} \tens{w}{mn}{mn} \notag\\
    &= \sum_{m}^{\mathbf{C}} \tens{\bar{H}}{m}{m} - \sum_{m}^{\mathbf{C}}\sum_{uv}^{\mathbf{A}} \tens{\bar{H}}{mu}{mv} \density{u}{v} - \frac{1}{2} \sum_{mn}^{\mathbf{C}} \tens{\bar{H}}{mn}{mn} ,
\end{align}
and the corresponding modified one-body operator is
\begin{align}
    \tens{c}{u}{v} &= \tens{o}{u}{v} + \sum_{m}^{\mathbf{C}} \tens{w}{um}{vm} = \tens{\bar{H}}{u}{v} - \sum_{xy}^{\mathbf{A}} \tens{\bar{H}}{ux}{vy} \density{x}{y} .
\end{align}
Obviously, Eq.~\eqref{eq:c0_cas} cancels the second line of Eq.~\eqref{eq:Hbar_DNO}.
Thus we prove that \emph{only} those elements of $\bar{H}(s)$ labeled by active indices are necessary to obtain the relaxed MR-DSRG energy.
The cost to evaluate $\tens{\bar{H}}{uv}{xy}$ scales as ${\cal O}(N_{\bf V}^{2}N_{\bf A}^{4})$ and is significantly smaller than then the cost to evaluate the DSRG-MRPT3 energy, which scales as ${\cal O}(N_{\bf V}^{4}N_{\bf C}^{2})$.

\section{Other choices of zeroth-order Hamiltonian}
\label{sec:other_H0th}

The DSRG-MRPT3 approach proposed in this work uses a diagonal one-body zeroth-order Hamiltonian.
However, partitioning schemes that include two-body operators may also be used with the DSRG-MRPT3, including the Dyall Hamiltonian\cite{Dyall:1995ct} and the retaining-excitation (RE) Hamiltonian.\cite{Fink:2006gu,Fink:2009gx}
In this section we analyze the implications of including two-body operators in the zeroth-order Hamiltonian, focusing on the computational efficiency and the avoidance of intruders in the DSRG-MRPT3.
If these partitionings led to an intruder-free DSRG-MRPT approach, then one could avoid the $s$-dependence of the energy by taking the limit $s \rightarrow \infty$.
Unfortunately, our analysis shows that even when two-body interactions are included in the zeroth-order Hamiltonian, some form of renormalization is necessary to avoid intruders.

When written in normal-ordered form with respect to the reference wave function, the zeroth-order Hamiltonian of Dyall may be written as:
\begin{align}
    \hat{H}^{(0)}_{\rm Dyall} = E_0 + \hat{F}^{(0)} + \frac{1}{4} \sum_{uvxy}^{\bf A} \tens{v}{uv}{xy,(0)} \no{\sqop{uv}{xy}},
\end{align}
where $\hat{F}^{(0)}$ is defined as in Eq.~\eqref{eq:F0th}.
A first important consequence of the presence of the two-body term is the fact that already at first-order the commutator $[\hat{H}^{(0)}, \hat{A}^{(1)}(s)]$ produces three-body terms.
Consequently, evaluating the second-order energy requires computing of a subset of the three-body operators.
This problem may be avoided by invoking the ``1,2'' operator  approximation, which introduces truncation errors already in the second-order energy.

Another complication resulting from the two-body terms in $\hat{H}^{(0)}$ is the need for an iterative solution of the amplitudes.
Again, we take the zeroth-order Hamiltonian to be $\hat{H}^{(0)}_{\rm Dyall}$ and consider the limit $s \rightarrow \infty$.  The first-order amplitudes corresponding to the promotion two electrons from active to virtual orbitals [$\tens{t}{ef}{uv,(1)} (s)$] are determined by the equation:
\begin{align}
\label{eq:t2_aavv}
    0 =& \tens{v}{ef}{uv,(1)} - \tens{\Delta}{ef}{uv} \tens{t}{ef}{uv,(1)} \notag\\
    &- \frac{1}{2} \sum_{xy}^{\bf A} \tens{v}{xy}{uv,(0)} \tens{t}{ef}{xy,(1)} + \sum_{xyz}^{\bf A} \tens{v}{xy}{uv,(0)} \tens{t}{ef}{xz,(1)} \density{y}{z},
\end{align}
where for clarity we dropped the symbol ``(s)'' from the amplitudes.
In Eq.~\eqref{eq:t2_aavv} the amplitude $\tens{t}{ef}{uv,(1)}$ is coupled to all amplitudes of the form $\tens{t}{ef}{xy,(1)}$ where $x,y \in \mathbf{A}$.
Consequently, the DSRG-MRPT based on the Dyall Hamiltonian requires an iterative solution of a set of linear equations for certain classes of amplitudes.
Nevertheless, the most numerous amplitudes, double excitations from core orbitals to virtual orbitals, $\tens{t}{ef}{mn,(1)}$, may still be computed via a noniterative procedure.

Last, we consider whether the use of Dyall's Hamiltonian in the DSRG-MRPT formalism may avoid the intruder state problem, as it is know to be the case in NEVPT.\cite{Angeli:2001bg,Angeli:2006gf,Angeli:2007by}
To this end, we evaluate the diagonal preconditioner or ``shifted'' denominator corresponding to certain classes of excitations prone to give small denominators and determine if extra terms that arise from Dyall's Hamiltonian may help avoid divergences.
We first go back to Eq.~\eqref{eq:t2_aavv}, and identify the shifted denominator ($\tens{D}{ef}{uv} $) for the $\tens{t}{ef}{uv,(1)}$ amplitudes with the expression:
\begin{equation}
\label{eq:full_preconditioner}
    \tens{D}{ef}{uv} =\epsilon_u + \epsilon_v -\epsilon_e - \epsilon_f +
\frac{1}{2} \sum_{xz}^{\mathbf{A}} [ \gamma^{x}_{u} \gamma^{z}_{v}  v^{u v,(0)}_{x z}
- \eta^{x}_{u} \eta^{z}_{v} v^{u v,(0)}_{x z}]
\end{equation}
The analysis of this expression is simpler in the basis of natural orbitals.  Defining the natural occupation of orbital $\phi^u$ as $n_u = \gamma^{u}_{u}$ and assuming a natural orbital basis, we may simplify Eq.~\eqref{eq:full_preconditioner} to:
\begin{equation}
\label{eq:no_preconditioner}
    \tens{D}{ef}{uv} =\epsilon_u + \epsilon_v -\epsilon_e - \epsilon_f 
     + \frac{1}{2} v^{u v, (0)}_{u v} (1 - n_u - n_v).
\end{equation}
The term $\epsilon_u + \epsilon_v -\epsilon_e - \epsilon_f$ is the standard M{\o}ller--Plesset denominator, and in general it is negative since virtual orbitals are assumed to lie higher in energy than active orbitals.
The term $\frac{1}{2} v^{u v, (0)}_{u v} (1 - n_u - n_v)$ arises from the two-body active part of the Hamiltonian and may be positive (when $n_u + n_v < 1$), negative (when $n_u + n_v > 1$), or zero.
Consequently, the additional terms that arise from the active part of the two-body operator cannot guarantee that $\tens{D}{ef}{uv}\neq 0$.
Excitations most prone to small denominators are singles coupled with a spectator excitation (from active to active orbitals), for example, $\tens{t}{ex}{uv,(1)} (s)$.
In this case we also find that the shifted denominators may accidentally be zero since the contribution of the two-body Hamiltonian does not have a well defined sign.

Next, we consider single excitations.
Interestingly, in this case we find that the Dyall Hamiltonian does not lead to modified energy denominators.
Hence, intruder states may still arise when singles denominators that involve one active orbital  approach zero.
This is in contrast with NEVPT2, where single denominators are shifted by two-body integrals.
This difference arises from the use of many-body conditions in DSRG-MRPT and projective conditions in the case of NEVPT2.

In summary, our analysis shows that the use of Dyall's partitioning of the Hamiltonian cannot avoid the intruder state problem in perturbation theories like the DSRG-MRPT, which are derived from a set of many-body conditions.
In other words, renormalization of small denominators is necessary even when the zeroth-order Hamiltonian contains two-electron terms.
From another perspective, the difficulty in converging the nonperturbative MR-LDSRG(2) equations\cite{Li:2016hb} when $s\rightarrow \infty$, also suggests that using other zeroth-order Hamiltonians with two-body contributions is unlikely solution to the intruder state problem in DSRG multireference perturbation theory.

\bibliography{MRPT3-bib}

\begin{thebibliography}{126}%
\makeatletter
\providecommand \@ifxundefined [1]{%
 \@ifx{#1\undefined}
}%
\providecommand \@ifnum [1]{%
 \ifnum #1\expandafter \@firstoftwo
 \else \expandafter \@secondoftwo
 \fi
}%
\providecommand \@ifx [1]{%
 \ifx #1\expandafter \@firstoftwo
 \else \expandafter \@secondoftwo
 \fi
}%
\providecommand \natexlab [1]{#1}%
\providecommand \enquote  [1]{``#1''}%
\providecommand \bibnamefont  [1]{#1}%
\providecommand \bibfnamefont [1]{#1}%
\providecommand \citenamefont [1]{#1}%
\providecommand \href@noop [0]{\@secondoftwo}%
\providecommand \href [0]{\begingroup \@sanitize@url \@href}%
\providecommand \@href[1]{\@@startlink{#1}\@@href}%
\providecommand \@@href[1]{\endgroup#1\@@endlink}%
\providecommand \@sanitize@url [0]{\catcode `\\12\catcode `\$12\catcode
  `\&12\catcode `\#12\catcode `\^12\catcode `\_12\catcode `\%12\relax}%
\providecommand \@@startlink[1]{}%
\providecommand \@@endlink[0]{}%
\providecommand \url  [0]{\begingroup\@sanitize@url \@url }%
\providecommand \@url [1]{\endgroup\@href {#1}{\urlprefix }}%
\providecommand \urlprefix  [0]{URL }%
\providecommand \Eprint [0]{\href }%
\providecommand \doibase [0]{http://dx.doi.org/}%
\providecommand \selectlanguage [0]{\@gobble}%
\providecommand \bibinfo  [0]{\@secondoftwo}%
\providecommand \bibfield  [0]{\@secondoftwo}%
\providecommand \translation [1]{[#1]}%
\providecommand \BibitemOpen [0]{}%
\providecommand \bibitemStop [0]{}%
\providecommand \bibitemNoStop [0]{.\EOS\space}%
\providecommand \EOS [0]{\spacefactor3000\relax}%
\providecommand \BibitemShut  [1]{\csname bibitem#1\endcsname}%
\let\auto@bib@innerbib\@empty
\bibitem [{\citenamefont {Roos}, \citenamefont {Taylor},\ and\ \citenamefont
  {Siegbahn}(1980)}]{Roos:1980fd}%
  \BibitemOpen
  \bibfield  {author} {\bibinfo {author} {\bibfnamefont {B.~O.}\ \bibnamefont
  {Roos}}, \bibinfo {author} {\bibfnamefont {P.~R.}\ \bibnamefont {Taylor}}, \
  and\ \bibinfo {author} {\bibfnamefont {P.~E.~M.}\ \bibnamefont {Siegbahn}},\
  }\href@noop {} {\bibfield  {journal} {\bibinfo  {journal} {Chem. Phys.}\
  }\textbf {\bibinfo {volume} {48}},\ \bibinfo {pages} {157} (\bibinfo {year}
  {1980})}\BibitemShut {NoStop}%
\bibitem [{\citenamefont {Ruedenberg}\ \emph {et~al.}(1982)\citenamefont
  {Ruedenberg}, \citenamefont {Schmidt}, \citenamefont {Gilbert},\ and\
  \citenamefont {Elbert}}]{Ruedenberg:1982cv}%
  \BibitemOpen
  \bibfield  {author} {\bibinfo {author} {\bibfnamefont {K.}~\bibnamefont
  {Ruedenberg}}, \bibinfo {author} {\bibfnamefont {M.~W.}\ \bibnamefont
  {Schmidt}}, \bibinfo {author} {\bibfnamefont {M.~M.}\ \bibnamefont
  {Gilbert}}, \ and\ \bibinfo {author} {\bibfnamefont {S.~T.}\ \bibnamefont
  {Elbert}},\ }\href@noop {} {\bibfield  {journal} {\bibinfo  {journal} {Chem.
  Phys.}\ }\textbf {\bibinfo {volume} {71}},\ \bibinfo {pages} {41} (\bibinfo
  {year} {1982})}\BibitemShut {NoStop}%
\bibitem [{\citenamefont {Werner}\ and\ \citenamefont
  {Knowles}(1985)}]{Werner:1985dh}%
  \BibitemOpen
  \bibfield  {author} {\bibinfo {author} {\bibfnamefont {H.-J.}\ \bibnamefont
  {Werner}}\ and\ \bibinfo {author} {\bibfnamefont {P.~J.}\ \bibnamefont
  {Knowles}},\ }\href@noop {} {\bibfield  {journal} {\bibinfo  {journal} {J.
  Chem. Phys.}\ }\textbf {\bibinfo {volume} {82}},\ \bibinfo {pages} {5053}
  (\bibinfo {year} {1985})}\BibitemShut {NoStop}%
\bibitem [{\citenamefont {Andersson}\ \emph {et~al.}(1990)\citenamefont
  {Andersson}, \citenamefont {Malmqvist}, \citenamefont {Roos}, \citenamefont
  {Sadlej},\ and\ \citenamefont {Wolinski}}]{Andersson:1990jz}%
  \BibitemOpen
  \bibfield  {author} {\bibinfo {author} {\bibfnamefont {K.}~\bibnamefont
  {Andersson}}, \bibinfo {author} {\bibfnamefont {P.-{\AA}.}\ \bibnamefont
  {Malmqvist}}, \bibinfo {author} {\bibfnamefont {B.~O.}\ \bibnamefont {Roos}},
  \bibinfo {author} {\bibfnamefont {A.~J.}\ \bibnamefont {Sadlej}}, \ and\
  \bibinfo {author} {\bibfnamefont {K.}~\bibnamefont {Wolinski}},\ }\href@noop
  {} {\bibfield  {journal} {\bibinfo  {journal} {J. Phys. Chem.}\ }\textbf
  {\bibinfo {volume} {94}},\ \bibinfo {pages} {5483} (\bibinfo {year}
  {1990})}\BibitemShut {NoStop}%
\bibitem [{\citenamefont {Andersson}, \citenamefont {Malmqvist},\ and\
  \citenamefont {Roos}(1992)}]{Andersson:1992cq}%
  \BibitemOpen
  \bibfield  {author} {\bibinfo {author} {\bibfnamefont {K.}~\bibnamefont
  {Andersson}}, \bibinfo {author} {\bibfnamefont {P.-{\AA}.}\ \bibnamefont
  {Malmqvist}}, \ and\ \bibinfo {author} {\bibfnamefont {B.~O.}\ \bibnamefont
  {Roos}},\ }\href@noop {} {\bibfield  {journal} {\bibinfo  {journal} {J. Chem.
  Phys.}\ }\textbf {\bibinfo {volume} {96}},\ \bibinfo {pages} {1218} (\bibinfo
  {year} {1992})}\BibitemShut {NoStop}%
\bibitem [{\citenamefont {Dyall}(1995)}]{Dyall:1995ct}%
  \BibitemOpen
  \bibfield  {author} {\bibinfo {author} {\bibfnamefont {K.~G.}\ \bibnamefont
  {Dyall}},\ }\href@noop {} {\bibfield  {journal} {\bibinfo  {journal} {J.
  Chem. Phys.}\ }\textbf {\bibinfo {volume} {102}},\ \bibinfo {pages} {4909}
  (\bibinfo {year} {1995})}\BibitemShut {NoStop}%
\bibitem [{\citenamefont {Angeli}\ \emph {et~al.}(2001)\citenamefont {Angeli},
  \citenamefont {Cimiraglia}, \citenamefont {Evangelisti}, \citenamefont
  {Leininger},\ and\ \citenamefont {Malrieu}}]{Angeli:2001bg}%
  \BibitemOpen
  \bibfield  {author} {\bibinfo {author} {\bibfnamefont {C.}~\bibnamefont
  {Angeli}}, \bibinfo {author} {\bibfnamefont {R.}~\bibnamefont {Cimiraglia}},
  \bibinfo {author} {\bibfnamefont {S.}~\bibnamefont {Evangelisti}}, \bibinfo
  {author} {\bibfnamefont {T.}~\bibnamefont {Leininger}}, \ and\ \bibinfo
  {author} {\bibfnamefont {J.~P.}\ \bibnamefont {Malrieu}},\ }\href@noop {}
  {\bibfield  {journal} {\bibinfo  {journal} {J. Chem. Phys.}\ }\textbf
  {\bibinfo {volume} {114}},\ \bibinfo {pages} {10252} (\bibinfo {year}
  {2001})}\BibitemShut {NoStop}%
\bibitem [{\citenamefont {Angeli}, \citenamefont {Cimiraglia},\ and\
  \citenamefont {Malrieu}(2002)}]{Angeli:2002ik}%
  \BibitemOpen
  \bibfield  {author} {\bibinfo {author} {\bibfnamefont {C.}~\bibnamefont
  {Angeli}}, \bibinfo {author} {\bibfnamefont {R.}~\bibnamefont {Cimiraglia}},
  \ and\ \bibinfo {author} {\bibfnamefont {J.~P.}\ \bibnamefont {Malrieu}},\
  }\href@noop {} {\bibfield  {journal} {\bibinfo  {journal} {J. Chem. Phys.}\
  }\textbf {\bibinfo {volume} {117}},\ \bibinfo {pages} {9138} (\bibinfo {year}
  {2002})}\BibitemShut {NoStop}%
\bibitem [{\citenamefont {Angeli}, \citenamefont {Pastore},\ and\ \citenamefont
  {Cimiraglia}(2007)}]{Angeli:2007by}%
  \BibitemOpen
  \bibfield  {author} {\bibinfo {author} {\bibfnamefont {C.}~\bibnamefont
  {Angeli}}, \bibinfo {author} {\bibfnamefont {M.}~\bibnamefont {Pastore}}, \
  and\ \bibinfo {author} {\bibfnamefont {R.}~\bibnamefont {Cimiraglia}},\
  }\href@noop {} {\bibfield  {journal} {\bibinfo  {journal} {Theor. Chem.
  Acc.}\ }\textbf {\bibinfo {volume} {117}},\ \bibinfo {pages} {743} (\bibinfo
  {year} {2007})}\BibitemShut {NoStop}%
\bibitem [{\citenamefont {Evangelisti}, \citenamefont {Daudey},\ and\
  \citenamefont {Malrieu}(1987)}]{Evangelisti:1987fw}%
  \BibitemOpen
  \bibfield  {author} {\bibinfo {author} {\bibfnamefont {S.}~\bibnamefont
  {Evangelisti}}, \bibinfo {author} {\bibfnamefont {J.~P.}\ \bibnamefont
  {Daudey}}, \ and\ \bibinfo {author} {\bibfnamefont {J.~P.}\ \bibnamefont
  {Malrieu}},\ }\href@noop {} {\bibfield  {journal} {\bibinfo  {journal} {Phys.
  Rev. A}\ }\textbf {\bibinfo {volume} {35}},\ \bibinfo {pages} {4930}
  (\bibinfo {year} {1987})}\BibitemShut {NoStop}%
\bibitem [{\citenamefont {Paldus}\ \emph {et~al.}(1993)\citenamefont {Paldus},
  \citenamefont {Piecuch}, \citenamefont {Pylypow},\ and\ \citenamefont
  {Jeziorski}}]{Paldus:1993dx}%
  \BibitemOpen
  \bibfield  {author} {\bibinfo {author} {\bibfnamefont {J.}~\bibnamefont
  {Paldus}}, \bibinfo {author} {\bibfnamefont {P.}~\bibnamefont {Piecuch}},
  \bibinfo {author} {\bibfnamefont {L.}~\bibnamefont {Pylypow}}, \ and\
  \bibinfo {author} {\bibfnamefont {B.}~\bibnamefont {Jeziorski}},\ }\href@noop
  {} {\bibfield  {journal} {\bibinfo  {journal} {Phys. Rev. A}\ }\textbf
  {\bibinfo {volume} {47}},\ \bibinfo {pages} {2738} (\bibinfo {year}
  {1993})}\BibitemShut {NoStop}%
\bibitem [{\citenamefont {Kowalski}\ and\ \citenamefont
  {Piecuch}(2000)}]{Kowalski:2000cj}%
  \BibitemOpen
  \bibfield  {author} {\bibinfo {author} {\bibfnamefont {K.}~\bibnamefont
  {Kowalski}}\ and\ \bibinfo {author} {\bibfnamefont {P.}~\bibnamefont
  {Piecuch}},\ }\href@noop {} {\bibfield  {journal} {\bibinfo  {journal} {Phys.
  Rev. A}\ }\textbf {\bibinfo {volume} {61}},\ \bibinfo {pages} {052506}
  (\bibinfo {year} {2000})}\BibitemShut {NoStop}%
\bibitem [{\citenamefont {Roos}\ and\ \citenamefont
  {Andersson}(1995)}]{Roos:1995jz}%
  \BibitemOpen
  \bibfield  {author} {\bibinfo {author} {\bibfnamefont {B.~O.}\ \bibnamefont
  {Roos}}\ and\ \bibinfo {author} {\bibfnamefont {K.}~\bibnamefont
  {Andersson}},\ }\href@noop {} {\bibfield  {journal} {\bibinfo  {journal}
  {Chem. Phys. Lett.}\ }\textbf {\bibinfo {volume} {245}},\ \bibinfo {pages}
  {215} (\bibinfo {year} {1995})}\BibitemShut {NoStop}%
\bibitem [{\citenamefont {Andersson}(1995)}]{Andersson:1995vv}%
  \BibitemOpen
  \bibfield  {author} {\bibinfo {author} {\bibfnamefont {K.}~\bibnamefont
  {Andersson}},\ }\href@noop {} {\bibfield  {journal} {\bibinfo  {journal}
  {Theor. Chim. Acta}\ }\textbf {\bibinfo {volume} {91}},\ \bibinfo {pages}
  {31} (\bibinfo {year} {1995})}\BibitemShut {NoStop}%
\bibitem [{\citenamefont {Ghigo}, \citenamefont {Roos},\ and\ \citenamefont
  {Malmqvist}(2004)}]{Ghigo:2004fd}%
  \BibitemOpen
  \bibfield  {author} {\bibinfo {author} {\bibfnamefont {G.}~\bibnamefont
  {Ghigo}}, \bibinfo {author} {\bibfnamefont {B.~O.}\ \bibnamefont {Roos}}, \
  and\ \bibinfo {author} {\bibfnamefont {P.-{\AA}.}\ \bibnamefont
  {Malmqvist}},\ }\href@noop {} {\bibfield  {journal} {\bibinfo  {journal}
  {Chem. Phys. Lett.}\ }\textbf {\bibinfo {volume} {396}},\ \bibinfo {pages}
  {142} (\bibinfo {year} {2004})}\BibitemShut {NoStop}%
\bibitem [{\citenamefont {Witek}\ \emph {et~al.}(2002)\citenamefont {Witek},
  \citenamefont {Choe}, \citenamefont {Finley},\ and\ \citenamefont
  {Hirao}}]{Witek:2002jd}%
  \BibitemOpen
  \bibfield  {author} {\bibinfo {author} {\bibfnamefont {H.~A.}\ \bibnamefont
  {Witek}}, \bibinfo {author} {\bibfnamefont {Y.-K.}\ \bibnamefont {Choe}},
  \bibinfo {author} {\bibfnamefont {J.~P.}\ \bibnamefont {Finley}}, \ and\
  \bibinfo {author} {\bibfnamefont {K.}~\bibnamefont {Hirao}},\ }\href@noop {}
  {\bibfield  {journal} {\bibinfo  {journal} {J. Comput. Chem.}\ }\textbf
  {\bibinfo {volume} {23}},\ \bibinfo {pages} {957} (\bibinfo {year}
  {2002})}\BibitemShut {NoStop}%
\bibitem [{\citenamefont {Taube}\ and\ \citenamefont
  {Bartlett}(2009)}]{Taube:2009jz}%
  \BibitemOpen
  \bibfield  {author} {\bibinfo {author} {\bibfnamefont {A.~G.}\ \bibnamefont
  {Taube}}\ and\ \bibinfo {author} {\bibfnamefont {R.~J.}\ \bibnamefont
  {Bartlett}},\ }\href@noop {} {\bibfield  {journal} {\bibinfo  {journal} {J.
  Chem. Phys.}\ }\textbf {\bibinfo {volume} {130}},\ \bibinfo {pages} {144112}
  (\bibinfo {year} {2009})}\BibitemShut {NoStop}%
\bibitem [{\citenamefont {Camacho}, \citenamefont {Witek},\ and\ \citenamefont
  {Yamamoto}(2009)}]{Camacho:2009cr}%
  \BibitemOpen
  \bibfield  {author} {\bibinfo {author} {\bibfnamefont {C.}~\bibnamefont
  {Camacho}}, \bibinfo {author} {\bibfnamefont {H.~A.}\ \bibnamefont {Witek}},
  \ and\ \bibinfo {author} {\bibfnamefont {S.}~\bibnamefont {Yamamoto}},\
  }\href@noop {} {\bibfield  {journal} {\bibinfo  {journal} {J. Comput. Chem.}\
  }\textbf {\bibinfo {volume} {30}},\ \bibinfo {pages} {468} (\bibinfo {year}
  {2009})}\BibitemShut {NoStop}%
\bibitem [{\citenamefont {Forsberg}\ and\ \citenamefont
  {Malmqvist}(1997)}]{Forsberg:1997ke}%
  \BibitemOpen
  \bibfield  {author} {\bibinfo {author} {\bibfnamefont {N.}~\bibnamefont
  {Forsberg}}\ and\ \bibinfo {author} {\bibfnamefont {P.-{\AA}.}\ \bibnamefont
  {Malmqvist}},\ }\href@noop {} {\bibfield  {journal} {\bibinfo  {journal}
  {Chem. Phys. Lett.}\ }\textbf {\bibinfo {volume} {274}},\ \bibinfo {pages}
  {196} (\bibinfo {year} {1997})}\BibitemShut {NoStop}%
\bibitem [{\citenamefont {Roos}\ \emph {et~al.}(1996)\citenamefont {Roos},
  \citenamefont {Andersson}, \citenamefont {F{\"u}lscher}, \citenamefont
  {Serrano-Andr{\'e}s}, \citenamefont {Pierloot}, \citenamefont {Merch{\'a}n},\
  and\ \citenamefont {Molina}}]{Roos:1996ea}%
  \BibitemOpen
  \bibfield  {author} {\bibinfo {author} {\bibfnamefont {B.~O.}\ \bibnamefont
  {Roos}}, \bibinfo {author} {\bibfnamefont {K.}~\bibnamefont {Andersson}},
  \bibinfo {author} {\bibfnamefont {M.~P.}\ \bibnamefont {F{\"u}lscher}},
  \bibinfo {author} {\bibfnamefont {L.}~\bibnamefont {Serrano-Andr{\'e}s}},
  \bibinfo {author} {\bibfnamefont {K.}~\bibnamefont {Pierloot}}, \bibinfo
  {author} {\bibfnamefont {M.}~\bibnamefont {Merch{\'a}n}}, \ and\ \bibinfo
  {author} {\bibfnamefont {V.}~\bibnamefont {Molina}},\ }\href@noop {}
  {\bibfield  {journal} {\bibinfo  {journal} {J. Mol. Struct. (Theochem)}\
  }\textbf {\bibinfo {volume} {388}},\ \bibinfo {pages} {257} (\bibinfo {year}
  {1996})}\BibitemShut {NoStop}%
\bibitem [{\citenamefont {Vancoillie}, \citenamefont {Malmqvist},\ and\
  \citenamefont {Veryazov}(2016)}]{Vancoillie:2016gp}%
  \BibitemOpen
  \bibfield  {author} {\bibinfo {author} {\bibfnamefont {S.}~\bibnamefont
  {Vancoillie}}, \bibinfo {author} {\bibfnamefont {P.-{\AA}.}\ \bibnamefont
  {Malmqvist}}, \ and\ \bibinfo {author} {\bibfnamefont {V.}~\bibnamefont
  {Veryazov}},\ }\href@noop {} {\bibfield  {journal} {\bibinfo  {journal} {J.
  Chem. Theory Comput.}\ }\textbf {\bibinfo {volume} {12}},\ \bibinfo {pages}
  {1647} (\bibinfo {year} {2016})}\BibitemShut {NoStop}%
\bibitem [{\citenamefont {Rintelman}\ \emph {et~al.}(2005)\citenamefont
  {Rintelman}, \citenamefont {Adamovic}, \citenamefont {Varganov},\ and\
  \citenamefont {Gordon}}]{Rintelman:2005fj}%
  \BibitemOpen
  \bibfield  {author} {\bibinfo {author} {\bibfnamefont {J.~M.}\ \bibnamefont
  {Rintelman}}, \bibinfo {author} {\bibfnamefont {I.}~\bibnamefont {Adamovic}},
  \bibinfo {author} {\bibfnamefont {S.}~\bibnamefont {Varganov}}, \ and\
  \bibinfo {author} {\bibfnamefont {M.~S.}\ \bibnamefont {Gordon}},\
  }\href@noop {} {\bibfield  {journal} {\bibinfo  {journal} {J. Chem. Phys.}\
  }\textbf {\bibinfo {volume} {122}},\ \bibinfo {pages} {044105} (\bibinfo
  {year} {2005})}\BibitemShut {NoStop}%
\bibitem [{\citenamefont {Helgaker}, \citenamefont {J{\o}rgensen},\ and\
  \citenamefont {Olsen}(2000)}]{Helgaker2000book}%
  \BibitemOpen
  \bibfield  {author} {\bibinfo {author} {\bibfnamefont {T.}~\bibnamefont
  {Helgaker}}, \bibinfo {author} {\bibfnamefont {P.}~\bibnamefont
  {J{\o}rgensen}}, \ and\ \bibinfo {author} {\bibfnamefont {J.}~\bibnamefont
  {Olsen}},\ }\href@noop {} {\emph {\bibinfo {title} {Molecular
  Electronic-Structure Theory}}}\ (\bibinfo  {publisher} {John Wiley \& Sons},\
  \bibinfo {year} {2000})\BibitemShut {NoStop}%
\bibitem [{\citenamefont {Kutzelnigg}\ and\ \citenamefont
  {Mukherjee}(1997)}]{Kutzelnigg:1997ut}%
  \BibitemOpen
  \bibfield  {author} {\bibinfo {author} {\bibfnamefont {W.}~\bibnamefont
  {Kutzelnigg}}\ and\ \bibinfo {author} {\bibfnamefont {D.}~\bibnamefont
  {Mukherjee}},\ }\href@noop {} {\bibfield  {journal} {\bibinfo  {journal} {J.
  Chem. Phys.}\ }\textbf {\bibinfo {volume} {107}},\ \bibinfo {pages} {432}
  (\bibinfo {year} {1997})}\BibitemShut {NoStop}%
\bibitem [{\citenamefont {Kutzelnigg}\ and\ \citenamefont
  {Mukherjee}(1999)}]{Kutzelnigg:1999cj}%
  \BibitemOpen
  \bibfield  {author} {\bibinfo {author} {\bibfnamefont {W.}~\bibnamefont
  {Kutzelnigg}}\ and\ \bibinfo {author} {\bibfnamefont {D.}~\bibnamefont
  {Mukherjee}},\ }\href@noop {} {\bibfield  {journal} {\bibinfo  {journal} {J.
  Chem. Phys.}\ }\textbf {\bibinfo {volume} {110}},\ \bibinfo {pages} {2800}
  (\bibinfo {year} {1999})}\BibitemShut {NoStop}%
\bibitem [{\citenamefont {Yanai}\ and\ \citenamefont
  {Chan}(2007)}]{Yanai:2007ix}%
  \BibitemOpen
  \bibfield  {author} {\bibinfo {author} {\bibfnamefont {T.}~\bibnamefont
  {Yanai}}\ and\ \bibinfo {author} {\bibfnamefont {G.~K.-L.}\ \bibnamefont
  {Chan}},\ }\href@noop {} {\bibfield  {journal} {\bibinfo  {journal} {J. Chem.
  Phys.}\ }\textbf {\bibinfo {volume} {127}},\ \bibinfo {pages} {104107}
  (\bibinfo {year} {2007})}\BibitemShut {NoStop}%
\bibitem [{\citenamefont {Kurashige}\ and\ \citenamefont
  {Yanai}(2011)}]{Kurashige:2011ck}%
  \BibitemOpen
  \bibfield  {author} {\bibinfo {author} {\bibfnamefont {Y.}~\bibnamefont
  {Kurashige}}\ and\ \bibinfo {author} {\bibfnamefont {T.}~\bibnamefont
  {Yanai}},\ }\href@noop {} {\bibfield  {journal} {\bibinfo  {journal} {J.
  Chem. Phys.}\ }\textbf {\bibinfo {volume} {135}},\ \bibinfo {pages} {094104}
  (\bibinfo {year} {2011})}\BibitemShut {NoStop}%
\bibitem [{\citenamefont {Zgid}\ \emph {et~al.}(2009)\citenamefont {Zgid},
  \citenamefont {Ghosh}, \citenamefont {Neuscamman},\ and\ \citenamefont
  {Chan}}]{Zgid:2009fu}%
  \BibitemOpen
  \bibfield  {author} {\bibinfo {author} {\bibfnamefont {D.}~\bibnamefont
  {Zgid}}, \bibinfo {author} {\bibfnamefont {D.}~\bibnamefont {Ghosh}},
  \bibinfo {author} {\bibfnamefont {E.}~\bibnamefont {Neuscamman}}, \ and\
  \bibinfo {author} {\bibfnamefont {G.~K.-L.}\ \bibnamefont {Chan}},\
  }\href@noop {} {\bibfield  {journal} {\bibinfo  {journal} {J. Chem. Phys.}\
  }\textbf {\bibinfo {volume} {130}},\ \bibinfo {pages} {194107} (\bibinfo
  {year} {2009})}\BibitemShut {NoStop}%
\bibitem [{\citenamefont {Evangelista}(2014)}]{Evangelista:2014kt}%
  \BibitemOpen
  \bibfield  {author} {\bibinfo {author} {\bibfnamefont {F.~A.}\ \bibnamefont
  {Evangelista}},\ }\href@noop {} {\bibfield  {journal} {\bibinfo  {journal}
  {J. Chem. Phys.}\ }\textbf {\bibinfo {volume} {141}},\ \bibinfo {pages}
  {054109} (\bibinfo {year} {2014})}\BibitemShut {NoStop}%
\bibitem [{\citenamefont {G{\l}azek}\ and\ \citenamefont
  {Wilson}(1993)}]{Giazek:1993ik}%
  \BibitemOpen
  \bibfield  {author} {\bibinfo {author} {\bibfnamefont {S.~D.}\ \bibnamefont
  {G{\l}azek}}\ and\ \bibinfo {author} {\bibfnamefont {K.~G.}\ \bibnamefont
  {Wilson}},\ }\href@noop {} {\bibfield  {journal} {\bibinfo  {journal} {Phys.
  Rev. D}\ }\textbf {\bibinfo {volume} {48}},\ \bibinfo {pages} {5863}
  (\bibinfo {year} {1993})}\BibitemShut {NoStop}%
\bibitem [{\citenamefont {Wegner}(2000)}]{Wegner:2000kq}%
  \BibitemOpen
  \bibfield  {author} {\bibinfo {author} {\bibfnamefont {F.}~\bibnamefont
  {Wegner}},\ }in\ \href {\doibase 10.1007/BFb0108350} {\emph {\bibinfo
  {booktitle} {Advances in Solid State Physics 40}}},\ \bibinfo {editor}
  {edited by\ \bibinfo {editor} {\bibfnamefont {B.}~\bibnamefont {Kramer}}}\
  (\bibinfo  {publisher} {Springer Berlin Heidelberg},\ \bibinfo {address}
  {Berlin, Heidelberg},\ \bibinfo {year} {2000})\ pp.\ \bibinfo {pages}
  {133--142}\BibitemShut {NoStop}%
\bibitem [{\citenamefont {Tsukiyama}, \citenamefont {Bogner},\ and\
  \citenamefont {Schwenk}(2011)}]{Tsukiyama:2011eo}%
  \BibitemOpen
  \bibfield  {author} {\bibinfo {author} {\bibfnamefont {K.}~\bibnamefont
  {Tsukiyama}}, \bibinfo {author} {\bibfnamefont {S.~K.}\ \bibnamefont
  {Bogner}}, \ and\ \bibinfo {author} {\bibfnamefont {A.}~\bibnamefont
  {Schwenk}},\ }\href@noop {} {\bibfield  {journal} {\bibinfo  {journal} {Phys.
  Rev. Lett.}\ }\textbf {\bibinfo {volume} {106}},\ \bibinfo {pages} {222502}
  (\bibinfo {year} {2011})}\BibitemShut {NoStop}%
\bibitem [{\citenamefont {Hergert}\ \emph {et~al.}(2013)\citenamefont
  {Hergert}, \citenamefont {Bogner}, \citenamefont {Binder}, \citenamefont
  {Calci}, \citenamefont {Langhammer}, \citenamefont {Roth},\ and\
  \citenamefont {Schwenk}}]{Hergert:2013kf}%
  \BibitemOpen
  \bibfield  {author} {\bibinfo {author} {\bibfnamefont {H.}~\bibnamefont
  {Hergert}}, \bibinfo {author} {\bibfnamefont {S.~K.}\ \bibnamefont {Bogner}},
  \bibinfo {author} {\bibfnamefont {S.}~\bibnamefont {Binder}}, \bibinfo
  {author} {\bibfnamefont {A.}~\bibnamefont {Calci}}, \bibinfo {author}
  {\bibfnamefont {J.}~\bibnamefont {Langhammer}}, \bibinfo {author}
  {\bibfnamefont {R.}~\bibnamefont {Roth}}, \ and\ \bibinfo {author}
  {\bibfnamefont {A.}~\bibnamefont {Schwenk}},\ }\href@noop {} {\bibfield
  {journal} {\bibinfo  {journal} {Phys. Rev. C}\ }\textbf {\bibinfo {volume}
  {87}},\ \bibinfo {pages} {034307} (\bibinfo {year} {2013})}\BibitemShut
  {NoStop}%
\bibitem [{\citenamefont {Hergert}\ \emph {et~al.}(2016)\citenamefont
  {Hergert}, \citenamefont {Bogner}, \citenamefont {Morris}, \citenamefont
  {Schwenk},\ and\ \citenamefont {Tsukiyama}}]{Hergert:2016fd}%
  \BibitemOpen
  \bibfield  {author} {\bibinfo {author} {\bibfnamefont {H.}~\bibnamefont
  {Hergert}}, \bibinfo {author} {\bibfnamefont {S.~K.}\ \bibnamefont {Bogner}},
  \bibinfo {author} {\bibfnamefont {T.~D.}\ \bibnamefont {Morris}}, \bibinfo
  {author} {\bibfnamefont {A.}~\bibnamefont {Schwenk}}, \ and\ \bibinfo
  {author} {\bibfnamefont {K.}~\bibnamefont {Tsukiyama}},\ }\href {\doibase
  http://doi:10.1016/j.physrep.2015.12.007} {\bibfield  {journal} {\bibinfo
  {journal} {Phys. Rep.}\ }\textbf {\bibinfo {volume} {621}},\ \bibinfo {pages}
  {165} (\bibinfo {year} {2016})}\BibitemShut {NoStop}%
\bibitem [{\citenamefont {{\v C}{\'\i}{\v z}ek}(1966)}]{Cizek:1966cy}%
  \BibitemOpen
  \bibfield  {author} {\bibinfo {author} {\bibfnamefont {J.}~\bibnamefont {{\v
  C}{\'\i}{\v z}ek}},\ }\href@noop {} {\bibfield  {journal} {\bibinfo
  {journal} {J. Chem. Phys.}\ }\textbf {\bibinfo {volume} {45}},\ \bibinfo
  {pages} {4256} (\bibinfo {year} {1966})}\BibitemShut {NoStop}%
\bibitem [{\citenamefont {Bartlett}(1981)}]{Bartlett:1981hg}%
  \BibitemOpen
  \bibfield  {author} {\bibinfo {author} {\bibfnamefont {R.~J.}\ \bibnamefont
  {Bartlett}},\ }\href {\doibase 10.1146/annurev.pc.32.100181.002043}
  {\bibfield  {journal} {\bibinfo  {journal} {Annu. Rev. Phys. Chem.}\ }\textbf
  {\bibinfo {volume} {32}},\ \bibinfo {pages} {359} (\bibinfo {year}
  {1981})}\BibitemShut {NoStop}%
\bibitem [{\citenamefont {Crawford}\ and\ \citenamefont
  {Schaefer}(2000)}]{Crawford:2000by}%
  \BibitemOpen
  \bibfield  {author} {\bibinfo {author} {\bibfnamefont {T.~D.}\ \bibnamefont
  {Crawford}}\ and\ \bibinfo {author} {\bibfnamefont {H.~F.}\ \bibnamefont
  {Schaefer}},\ }\href@noop {} {\bibfield  {journal} {\bibinfo  {journal} {Rev.
  Comput. Chem.}\ }\textbf {\bibinfo {volume} {14}},\ \bibinfo {pages} {33}
  (\bibinfo {year} {2000})}\BibitemShut {NoStop}%
\bibitem [{\citenamefont {Bartlett}\ and\ \citenamefont
  {Musia{\l}}(2007)}]{Bartlett:2007kv}%
  \BibitemOpen
  \bibfield  {author} {\bibinfo {author} {\bibfnamefont {R.~J.}\ \bibnamefont
  {Bartlett}}\ and\ \bibinfo {author} {\bibfnamefont {M.}~\bibnamefont
  {Musia{\l}}},\ }\href@noop {} {\bibfield  {journal} {\bibinfo  {journal}
  {Rev. Mod. Phys.}\ }\textbf {\bibinfo {volume} {79}},\ \bibinfo {pages} {291}
  (\bibinfo {year} {2007})}\BibitemShut {NoStop}%
\bibitem [{\citenamefont {Yanai}\ and\ \citenamefont
  {Chan}(2006)}]{Yanai:2006gi}%
  \BibitemOpen
  \bibfield  {author} {\bibinfo {author} {\bibfnamefont {T.}~\bibnamefont
  {Yanai}}\ and\ \bibinfo {author} {\bibfnamefont {G.~K.-L.}\ \bibnamefont
  {Chan}},\ }\href@noop {} {\bibfield  {journal} {\bibinfo  {journal} {J. Chem.
  Phys.}\ }\textbf {\bibinfo {volume} {124}},\ \bibinfo {pages} {194106}
  (\bibinfo {year} {2006})}\BibitemShut {NoStop}%
\bibitem [{\citenamefont {Neuscamman}, \citenamefont {Yanai},\ and\
  \citenamefont {Chan}(2010)}]{Neuscamman:2010ej}%
  \BibitemOpen
  \bibfield  {author} {\bibinfo {author} {\bibfnamefont {E.}~\bibnamefont
  {Neuscamman}}, \bibinfo {author} {\bibfnamefont {T.}~\bibnamefont {Yanai}}, \
  and\ \bibinfo {author} {\bibfnamefont {G.~K.-L.}\ \bibnamefont {Chan}},\
  }\href {\doibase 10.1080/01442351003620540} {\bibfield  {journal} {\bibinfo
  {journal} {Int. Rev. Phys. Chem.}\ }\textbf {\bibinfo {volume} {29}},\
  \bibinfo {pages} {231} (\bibinfo {year} {2010})}\BibitemShut {NoStop}%
\bibitem [{\citenamefont {Kehrein}(2006)}]{Kehrein2010book}%
  \BibitemOpen
  \bibfield  {author} {\bibinfo {author} {\bibfnamefont {S.}~\bibnamefont
  {Kehrein}},\ }\href@noop {} {\emph {\bibinfo {title} {The Flow Equation
  Approach to Many-Particle Systems}}}\ (\bibinfo  {publisher} {Springer Berlin
  Heidelberg},\ \bibinfo {year} {2006})\BibitemShut {NoStop}%
\bibitem [{\citenamefont {Hergert}(2017)}]{Hergert:2016un}%
  \BibitemOpen
  \bibfield  {author} {\bibinfo {author} {\bibfnamefont {H.}~\bibnamefont
  {Hergert}},\ }\href {http://stacks.iop.org/1402-4896/92/i=2/a=023002}
  {\bibfield  {journal} {\bibinfo  {journal} {Phys. Scripta}\ }\textbf
  {\bibinfo {volume} {92}},\ \bibinfo {pages} {023002} (\bibinfo {year}
  {2017})}\BibitemShut {NoStop}%
\bibitem [{\citenamefont {Kutzelnigg}(1982)}]{Kutzelnigg:1982kr}%
  \BibitemOpen
  \bibfield  {author} {\bibinfo {author} {\bibfnamefont {W.}~\bibnamefont
  {Kutzelnigg}},\ }\href@noop {} {\bibfield  {journal} {\bibinfo  {journal} {J.
  Chem. Phys.}\ }\textbf {\bibinfo {volume} {77}},\ \bibinfo {pages} {3081}
  (\bibinfo {year} {1982})}\BibitemShut {NoStop}%
\bibitem [{\citenamefont {Kutzelnigg}\ and\ \citenamefont
  {Koch}(1983)}]{Kutzelnigg:1983dr}%
  \BibitemOpen
  \bibfield  {author} {\bibinfo {author} {\bibfnamefont {W.}~\bibnamefont
  {Kutzelnigg}}\ and\ \bibinfo {author} {\bibfnamefont {S.}~\bibnamefont
  {Koch}},\ }\href@noop {} {\bibfield  {journal} {\bibinfo  {journal} {J. Chem.
  Phys.}\ }\textbf {\bibinfo {volume} {79}},\ \bibinfo {pages} {4315} (\bibinfo
  {year} {1983})}\BibitemShut {NoStop}%
\bibitem [{\citenamefont {Kutzelnigg}(1984)}]{Kutzelnigg:1984eg}%
  \BibitemOpen
  \bibfield  {author} {\bibinfo {author} {\bibfnamefont {W.}~\bibnamefont
  {Kutzelnigg}},\ }\href@noop {} {\bibfield  {journal} {\bibinfo  {journal} {J.
  Chem. Phys.}\ }\textbf {\bibinfo {volume} {80}},\ \bibinfo {pages} {822}
  (\bibinfo {year} {1984})}\BibitemShut {NoStop}%
\bibitem [{\citenamefont {Kutzelnigg}(1985)}]{Kutzelnigg:1985fj}%
  \BibitemOpen
  \bibfield  {author} {\bibinfo {author} {\bibfnamefont {W.}~\bibnamefont
  {Kutzelnigg}},\ }\href@noop {} {\bibfield  {journal} {\bibinfo  {journal} {J.
  Chem. Phys.}\ }\textbf {\bibinfo {volume} {82}},\ \bibinfo {pages} {4166}
  (\bibinfo {year} {1985})}\BibitemShut {NoStop}%
\bibitem [{\citenamefont {Stolarczyk}\ and\ \citenamefont
  {Monkhorst}(1985{\natexlab{a}})}]{Stolarczyk:1985fk}%
  \BibitemOpen
  \bibfield  {author} {\bibinfo {author} {\bibfnamefont {L.~Z.}\ \bibnamefont
  {Stolarczyk}}\ and\ \bibinfo {author} {\bibfnamefont {H.~J.}\ \bibnamefont
  {Monkhorst}},\ }\href@noop {} {\bibfield  {journal} {\bibinfo  {journal}
  {Phys. Rev. A}\ }\textbf {\bibinfo {volume} {32}},\ \bibinfo {pages} {725}
  (\bibinfo {year} {1985}{\natexlab{a}})}\BibitemShut {NoStop}%
\bibitem [{\citenamefont {Stolarczyk}\ and\ \citenamefont
  {Monkhorst}(1985{\natexlab{b}})}]{Stolarczyk:1985ct}%
  \BibitemOpen
  \bibfield  {author} {\bibinfo {author} {\bibfnamefont {L.~Z.}\ \bibnamefont
  {Stolarczyk}}\ and\ \bibinfo {author} {\bibfnamefont {H.~J.}\ \bibnamefont
  {Monkhorst}},\ }\href@noop {} {\bibfield  {journal} {\bibinfo  {journal}
  {Phys. Rev. A}\ }\textbf {\bibinfo {volume} {32}},\ \bibinfo {pages} {743}
  (\bibinfo {year} {1985}{\natexlab{b}})}\BibitemShut {NoStop}%
\bibitem [{\citenamefont {Stolarczyk}\ and\ \citenamefont
  {Monkhorst}(1988{\natexlab{a}})}]{Stolarczyk:1988ci}%
  \BibitemOpen
  \bibfield  {author} {\bibinfo {author} {\bibfnamefont {L.~Z.}\ \bibnamefont
  {Stolarczyk}}\ and\ \bibinfo {author} {\bibfnamefont {H.~J.}\ \bibnamefont
  {Monkhorst}},\ }\href@noop {} {\bibfield  {journal} {\bibinfo  {journal}
  {Phys. Rev. A}\ }\textbf {\bibinfo {volume} {37}},\ \bibinfo {pages} {1908}
  (\bibinfo {year} {1988}{\natexlab{a}})}\BibitemShut {NoStop}%
\bibitem [{\citenamefont {Stolarczyk}\ and\ \citenamefont
  {Monkhorst}(1988{\natexlab{b}})}]{Stolarczyk:1988cv}%
  \BibitemOpen
  \bibfield  {author} {\bibinfo {author} {\bibfnamefont {L.~Z.}\ \bibnamefont
  {Stolarczyk}}\ and\ \bibinfo {author} {\bibfnamefont {H.~J.}\ \bibnamefont
  {Monkhorst}},\ }\href@noop {} {\bibfield  {journal} {\bibinfo  {journal}
  {Phys. Rev. A}\ }\textbf {\bibinfo {volume} {37}},\ \bibinfo {pages} {1926}
  (\bibinfo {year} {1988}{\natexlab{b}})}\BibitemShut {NoStop}%
\bibitem [{\citenamefont {Lindgren}(1978)}]{Lindgren:1978io}%
  \BibitemOpen
  \bibfield  {author} {\bibinfo {author} {\bibfnamefont {I.}~\bibnamefont
  {Lindgren}},\ }\href@noop {} {\bibfield  {journal} {\bibinfo  {journal} {Int.
  J. Quantum Chem.}\ }\textbf {\bibinfo {volume} {14}},\ \bibinfo {pages} {33}
  (\bibinfo {year} {1978})}\BibitemShut {NoStop}%
\bibitem [{\citenamefont {Nooijen}\ and\ \citenamefont
  {Bartlett}(1996)}]{Nooijen:1996hp}%
  \BibitemOpen
  \bibfield  {author} {\bibinfo {author} {\bibfnamefont {M.}~\bibnamefont
  {Nooijen}}\ and\ \bibinfo {author} {\bibfnamefont {R.~J.}\ \bibnamefont
  {Bartlett}},\ }\href@noop {} {\bibfield  {journal} {\bibinfo  {journal} {J.
  Chem. Phys.}\ }\textbf {\bibinfo {volume} {104}},\ \bibinfo {pages} {2652}
  (\bibinfo {year} {1996})}\BibitemShut {NoStop}%
\bibitem [{\citenamefont {Datta}, \citenamefont {Kong},\ and\ \citenamefont
  {Nooijen}(2011)}]{Datta:2011ca}%
  \BibitemOpen
  \bibfield  {author} {\bibinfo {author} {\bibfnamefont {D.}~\bibnamefont
  {Datta}}, \bibinfo {author} {\bibfnamefont {L.}~\bibnamefont {Kong}}, \ and\
  \bibinfo {author} {\bibfnamefont {M.}~\bibnamefont {Nooijen}},\ }\href@noop
  {} {\bibfield  {journal} {\bibinfo  {journal} {J. Chem. Phys.}\ }\textbf
  {\bibinfo {volume} {134}},\ \bibinfo {pages} {214116} (\bibinfo {year}
  {2011})}\BibitemShut {NoStop}%
\bibitem [{\citenamefont {Datta}\ and\ \citenamefont
  {Nooijen}(2012)}]{Datta:2012hu}%
  \BibitemOpen
  \bibfield  {author} {\bibinfo {author} {\bibfnamefont {D.}~\bibnamefont
  {Datta}}\ and\ \bibinfo {author} {\bibfnamefont {M.}~\bibnamefont
  {Nooijen}},\ }\href@noop {} {\bibfield  {journal} {\bibinfo  {journal} {J.
  Chem. Phys.}\ }\textbf {\bibinfo {volume} {137}},\ \bibinfo {pages} {204107}
  (\bibinfo {year} {2012})}\BibitemShut {NoStop}%
\bibitem [{\citenamefont {Li}\ and\ \citenamefont
  {Evangelista}(2015)}]{Li:2015iz}%
  \BibitemOpen
  \bibfield  {author} {\bibinfo {author} {\bibfnamefont {C.}~\bibnamefont
  {Li}}\ and\ \bibinfo {author} {\bibfnamefont {F.~A.}\ \bibnamefont
  {Evangelista}},\ }\href@noop {} {\bibfield  {journal} {\bibinfo  {journal}
  {J. Chem. Theory Comput.}\ }\textbf {\bibinfo {volume} {11}},\ \bibinfo
  {pages} {2097} (\bibinfo {year} {2015})}\BibitemShut {NoStop}%
\bibitem [{\citenamefont {Li}\ and\ \citenamefont
  {Evangelista}(2016)}]{Li:2016hb}%
  \BibitemOpen
  \bibfield  {author} {\bibinfo {author} {\bibfnamefont {C.}~\bibnamefont
  {Li}}\ and\ \bibinfo {author} {\bibfnamefont {F.~A.}\ \bibnamefont
  {Evangelista}},\ }\href@noop {} {\bibfield  {journal} {\bibinfo  {journal}
  {J. Chem. Phys.}\ }\textbf {\bibinfo {volume} {144}},\ \bibinfo {pages}
  {164114} (\bibinfo {year} {2016})}\BibitemShut {NoStop}%
\bibitem [{\citenamefont {Hannon}, \citenamefont {Li},\ and\ \citenamefont
  {Evangelista}(2016)}]{Hannon:2016bh}%
  \BibitemOpen
  \bibfield  {author} {\bibinfo {author} {\bibfnamefont {K.~P.}\ \bibnamefont
  {Hannon}}, \bibinfo {author} {\bibfnamefont {C.}~\bibnamefont {Li}}, \ and\
  \bibinfo {author} {\bibfnamefont {F.~A.}\ \bibnamefont {Evangelista}},\
  }\href@noop {} {\bibfield  {journal} {\bibinfo  {journal} {J. Chem. Phys.}\
  }\textbf {\bibinfo {volume} {144}},\ \bibinfo {pages} {204111} (\bibinfo
  {year} {2016})}\BibitemShut {NoStop}%
\bibitem [{\citenamefont {Mukherjee}(1997)}]{Mukherjee:1997tk}%
  \BibitemOpen
  \bibfield  {author} {\bibinfo {author} {\bibfnamefont {D.}~\bibnamefont
  {Mukherjee}},\ }\href@noop {} {\bibfield  {journal} {\bibinfo  {journal}
  {Chem. Phys. Lett.}\ }\textbf {\bibinfo {volume} {274}},\ \bibinfo {pages}
  {561} (\bibinfo {year} {1997})}\BibitemShut {NoStop}%
\bibitem [{\citenamefont {Mahapatra}\ \emph {et~al.}(1998)\citenamefont
  {Mahapatra}, \citenamefont {Datta}, \citenamefont {Bandyopadhyay},\ and\
  \citenamefont {Mukherjee}}]{Mahapatra:1998cp}%
  \BibitemOpen
  \bibfield  {author} {\bibinfo {author} {\bibfnamefont {U.~S.}\ \bibnamefont
  {Mahapatra}}, \bibinfo {author} {\bibfnamefont {B.}~\bibnamefont {Datta}},
  \bibinfo {author} {\bibfnamefont {B.}~\bibnamefont {Bandyopadhyay}}, \ and\
  \bibinfo {author} {\bibfnamefont {D.}~\bibnamefont {Mukherjee}},\ }\href@noop
  {} {\bibfield  {journal} {\bibinfo  {journal} {Adv. Quantum Chem.}\ }\textbf
  {\bibinfo {volume} {30}},\ \bibinfo {pages} {163} (\bibinfo {year}
  {1998})}\BibitemShut {NoStop}%
\bibitem [{\citenamefont {Shamasundar}(2009)}]{Shamasundar:2009ee}%
  \BibitemOpen
  \bibfield  {author} {\bibinfo {author} {\bibfnamefont {K.~R.}\ \bibnamefont
  {Shamasundar}},\ }\href@noop {} {\bibfield  {journal} {\bibinfo  {journal}
  {J. Chem. Phys.}\ }\textbf {\bibinfo {volume} {131}},\ \bibinfo {pages}
  {174109} (\bibinfo {year} {2009})}\BibitemShut {NoStop}%
\bibitem [{\citenamefont {Kong}, \citenamefont {Nooijen},\ and\ \citenamefont
  {Mukherjee}(2010)}]{Kong:2010kg}%
  \BibitemOpen
  \bibfield  {author} {\bibinfo {author} {\bibfnamefont {L.}~\bibnamefont
  {Kong}}, \bibinfo {author} {\bibfnamefont {M.}~\bibnamefont {Nooijen}}, \
  and\ \bibinfo {author} {\bibfnamefont {D.}~\bibnamefont {Mukherjee}},\
  }\href@noop {} {\bibfield  {journal} {\bibinfo  {journal} {J. Chem. Phys.}\
  }\textbf {\bibinfo {volume} {132}},\ \bibinfo {pages} {234107} (\bibinfo
  {year} {2010})}\BibitemShut {NoStop}%
\bibitem [{\citenamefont {Kutzelnigg}, \citenamefont {Shamasundar},\ and\
  \citenamefont {Mukherjee}(2010)}]{Kutzelnigg:2010iu}%
  \BibitemOpen
  \bibfield  {author} {\bibinfo {author} {\bibfnamefont {W.}~\bibnamefont
  {Kutzelnigg}}, \bibinfo {author} {\bibfnamefont {K.~R.}\ \bibnamefont
  {Shamasundar}}, \ and\ \bibinfo {author} {\bibfnamefont {D.}~\bibnamefont
  {Mukherjee}},\ }\href {\doibase 10.1080/00268970903547926} {\bibfield
  {journal} {\bibinfo  {journal} {Mol. Phys.}\ }\textbf {\bibinfo {volume}
  {108}},\ \bibinfo {pages} {433} (\bibinfo {year} {2010})}\BibitemShut
  {NoStop}%
\bibitem [{\citenamefont {Sinha}, \citenamefont {Maitra},\ and\ \citenamefont
  {Mukherjee}(2013)}]{Sinha:2013dx}%
  \BibitemOpen
  \bibfield  {author} {\bibinfo {author} {\bibfnamefont {D.}~\bibnamefont
  {Sinha}}, \bibinfo {author} {\bibfnamefont {R.}~\bibnamefont {Maitra}}, \
  and\ \bibinfo {author} {\bibfnamefont {D.}~\bibnamefont {Mukherjee}},\
  }\href@noop {} {\bibfield  {journal} {\bibinfo  {journal} {Comput. Theor.
  Chem.}\ }\textbf {\bibinfo {volume} {1003}},\ \bibinfo {pages} {62} (\bibinfo
  {year} {2013})}\BibitemShut {NoStop}%
\bibitem [{\citenamefont {Hanauer}\ and\ \citenamefont
  {K{\"o}hn}(2012)}]{Hanauer:2012bd}%
  \BibitemOpen
  \bibfield  {author} {\bibinfo {author} {\bibfnamefont {M.}~\bibnamefont
  {Hanauer}}\ and\ \bibinfo {author} {\bibfnamefont {A.}~\bibnamefont
  {K{\"o}hn}},\ }\href@noop {} {\bibfield  {journal} {\bibinfo  {journal}
  {Chem. Phys.}\ }\textbf {\bibinfo {volume} {401}},\ \bibinfo {pages} {50}
  (\bibinfo {year} {2012})}\BibitemShut {NoStop}%
\bibitem [{\citenamefont {Wolinski}, \citenamefont {Sellers},\ and\
  \citenamefont {Pulay}(1987)}]{Wolinski:1987jb}%
  \BibitemOpen
  \bibfield  {author} {\bibinfo {author} {\bibfnamefont {K.}~\bibnamefont
  {Wolinski}}, \bibinfo {author} {\bibfnamefont {H.~L.}\ \bibnamefont
  {Sellers}}, \ and\ \bibinfo {author} {\bibfnamefont {P.}~\bibnamefont
  {Pulay}},\ }\href@noop {} {\bibfield  {journal} {\bibinfo  {journal} {Chem.
  Phys. Lett.}\ }\textbf {\bibinfo {volume} {140}},\ \bibinfo {pages} {225}
  (\bibinfo {year} {1987})}\BibitemShut {NoStop}%
\bibitem [{\citenamefont {Werner}(1996)}]{Werner:1996in}%
  \BibitemOpen
  \bibfield  {author} {\bibinfo {author} {\bibfnamefont {H.-J.}\ \bibnamefont
  {Werner}},\ }\href@noop {} {\bibfield  {journal} {\bibinfo  {journal} {Mol.
  Phys.}\ }\textbf {\bibinfo {volume} {89}},\ \bibinfo {pages} {645} (\bibinfo
  {year} {1996})}\BibitemShut {NoStop}%
\bibitem [{\citenamefont {Angeli}\ and\ \citenamefont
  {Cimiraglia}(2002)}]{Angeli:2002ip}%
  \BibitemOpen
  \bibfield  {author} {\bibinfo {author} {\bibfnamefont {C.}~\bibnamefont
  {Angeli}}\ and\ \bibinfo {author} {\bibfnamefont {R.}~\bibnamefont
  {Cimiraglia}},\ }\href@noop {} {\bibfield  {journal} {\bibinfo  {journal}
  {Theor. Chem. Acc.}\ }\textbf {\bibinfo {volume} {107}},\ \bibinfo {pages}
  {313} (\bibinfo {year} {2002})}\BibitemShut {NoStop}%
\bibitem [{\citenamefont {Angeli}\ \emph {et~al.}(2006)\citenamefont {Angeli},
  \citenamefont {Bories}, \citenamefont {Cavallini},\ and\ \citenamefont
  {Cimiraglia}}]{Angeli:2006gf}%
  \BibitemOpen
  \bibfield  {author} {\bibinfo {author} {\bibfnamefont {C.}~\bibnamefont
  {Angeli}}, \bibinfo {author} {\bibfnamefont {B.}~\bibnamefont {Bories}},
  \bibinfo {author} {\bibfnamefont {A.}~\bibnamefont {Cavallini}}, \ and\
  \bibinfo {author} {\bibfnamefont {R.}~\bibnamefont {Cimiraglia}},\
  }\href@noop {} {\bibfield  {journal} {\bibinfo  {journal} {J. Chem. Phys.}\
  }\textbf {\bibinfo {volume} {124}},\ \bibinfo {pages} {054108} (\bibinfo
  {year} {2006})}\BibitemShut {NoStop}%
\bibitem [{\citenamefont {Jiang}, \citenamefont {Khait},\ and\ \citenamefont
  {Hoffmann}(2006)}]{Jiang:2006cn}%
  \BibitemOpen
  \bibfield  {author} {\bibinfo {author} {\bibfnamefont {W.}~\bibnamefont
  {Jiang}}, \bibinfo {author} {\bibfnamefont {Y.~G.}\ \bibnamefont {Khait}}, \
  and\ \bibinfo {author} {\bibfnamefont {M.~R.}\ \bibnamefont {Hoffmann}},\
  }\href@noop {} {\bibfield  {journal} {\bibinfo  {journal} {J. Mol. Struct.
  (Theochem)}\ }\textbf {\bibinfo {volume} {771}},\ \bibinfo {pages} {73}
  (\bibinfo {year} {2006})}\BibitemShut {NoStop}%
\bibitem [{\citenamefont {Khait}, \citenamefont {Jiang},\ and\ \citenamefont
  {Hoffmann}(2009)}]{Khait:2009ew}%
  \BibitemOpen
  \bibfield  {author} {\bibinfo {author} {\bibfnamefont {Y.~G.}\ \bibnamefont
  {Khait}}, \bibinfo {author} {\bibfnamefont {W.}~\bibnamefont {Jiang}}, \ and\
  \bibinfo {author} {\bibfnamefont {M.~R.}\ \bibnamefont {Hoffmann}},\
  }\href@noop {} {\bibfield  {journal} {\bibinfo  {journal} {Int. J. Quantum
  Chem.}\ }\textbf {\bibinfo {volume} {109}},\ \bibinfo {pages} {1855}
  (\bibinfo {year} {2009})}\BibitemShut {NoStop}%
\bibitem [{\citenamefont {Mahapatra}, \citenamefont {Datta},\ and\
  \citenamefont {Mukherjee}(1999{\natexlab{a}})}]{Mahapatra:1999gh}%
  \BibitemOpen
  \bibfield  {author} {\bibinfo {author} {\bibfnamefont {U.~S.}\ \bibnamefont
  {Mahapatra}}, \bibinfo {author} {\bibfnamefont {B.}~\bibnamefont {Datta}}, \
  and\ \bibinfo {author} {\bibfnamefont {D.}~\bibnamefont {Mukherjee}},\
  }\href@noop {} {\bibfield  {journal} {\bibinfo  {journal} {Chem. Phys.
  Lett.}\ }\textbf {\bibinfo {volume} {299}},\ \bibinfo {pages} {42} (\bibinfo
  {year} {1999}{\natexlab{a}})}\BibitemShut {NoStop}%
\bibitem [{\citenamefont {Mahapatra}, \citenamefont {Datta},\ and\
  \citenamefont {Mukherjee}(1998)}]{Mahapatra:1998kj}%
  \BibitemOpen
  \bibfield  {author} {\bibinfo {author} {\bibfnamefont {U.~S.}\ \bibnamefont
  {Mahapatra}}, \bibinfo {author} {\bibfnamefont {B.}~\bibnamefont {Datta}}, \
  and\ \bibinfo {author} {\bibfnamefont {D.}~\bibnamefont {Mukherjee}},\
  }\href@noop {} {\bibfield  {journal} {\bibinfo  {journal} {Mol. Phys.}\
  }\textbf {\bibinfo {volume} {94}},\ \bibinfo {pages} {157} (\bibinfo {year}
  {1998})}\BibitemShut {NoStop}%
\bibitem [{\citenamefont {Sinha~Mahapatra}, \citenamefont {Datta},\ and\
  \citenamefont {Mukherjee}(1999)}]{SinhaMahapatra:1999bp}%
  \BibitemOpen
  \bibfield  {author} {\bibinfo {author} {\bibfnamefont {U.}~\bibnamefont
  {Sinha~Mahapatra}}, \bibinfo {author} {\bibfnamefont {B.}~\bibnamefont
  {Datta}}, \ and\ \bibinfo {author} {\bibfnamefont {D.}~\bibnamefont
  {Mukherjee}},\ }\href@noop {} {\bibfield  {journal} {\bibinfo  {journal} {J.
  Phys. Chem. A}\ }\textbf {\bibinfo {volume} {103}},\ \bibinfo {pages} {1822}
  (\bibinfo {year} {1999})}\BibitemShut {NoStop}%
\bibitem [{\citenamefont {Chaudhuri}\ \emph {et~al.}(2005)\citenamefont
  {Chaudhuri}, \citenamefont {Freed}, \citenamefont {Hose}, \citenamefont
  {Piecuch}, \citenamefont {Kowalski}, \citenamefont {W{\l}och}, \citenamefont
  {Chattopadhyay}, \citenamefont {Mukherjee}, \citenamefont {Rolik},
  \citenamefont {Szabados}, \citenamefont {T{\'o}th},\ and\ \citenamefont
  {Surj{\'a}n}}]{Chaudhuri:2005dt}%
  \BibitemOpen
  \bibfield  {author} {\bibinfo {author} {\bibfnamefont {R.~K.}\ \bibnamefont
  {Chaudhuri}}, \bibinfo {author} {\bibfnamefont {K.~F.}\ \bibnamefont
  {Freed}}, \bibinfo {author} {\bibfnamefont {G.}~\bibnamefont {Hose}},
  \bibinfo {author} {\bibfnamefont {P.}~\bibnamefont {Piecuch}}, \bibinfo
  {author} {\bibfnamefont {K.}~\bibnamefont {Kowalski}}, \bibinfo {author}
  {\bibfnamefont {M.}~\bibnamefont {W{\l}och}}, \bibinfo {author}
  {\bibfnamefont {S.}~\bibnamefont {Chattopadhyay}}, \bibinfo {author}
  {\bibfnamefont {D.}~\bibnamefont {Mukherjee}}, \bibinfo {author}
  {\bibfnamefont {Z.}~\bibnamefont {Rolik}}, \bibinfo {author} {\bibfnamefont
  {{\'A}.}~\bibnamefont {Szabados}}, \bibinfo {author} {\bibfnamefont
  {G.}~\bibnamefont {T{\'o}th}}, \ and\ \bibinfo {author} {\bibfnamefont
  {P.~R.}\ \bibnamefont {Surj{\'a}n}},\ }\href@noop {} {\bibfield  {journal}
  {\bibinfo  {journal} {J. Chem. Phys.}\ }\textbf {\bibinfo {volume} {122}},\
  \bibinfo {pages} {134105} (\bibinfo {year} {2005})}\BibitemShut {NoStop}%
\bibitem [{\citenamefont {Chattopadhyay}\ \emph {et~al.}(2016)\citenamefont
  {Chattopadhyay}, \citenamefont {Chaudhuri}, \citenamefont {Mahapatra},
  \citenamefont {Ghosh},\ and\ \citenamefont {Ray}}]{Chattopadhyay:2016cc}%
  \BibitemOpen
  \bibfield  {author} {\bibinfo {author} {\bibfnamefont {S.}~\bibnamefont
  {Chattopadhyay}}, \bibinfo {author} {\bibfnamefont {R.~K.}\ \bibnamefont
  {Chaudhuri}}, \bibinfo {author} {\bibfnamefont {U.~S.}\ \bibnamefont
  {Mahapatra}}, \bibinfo {author} {\bibfnamefont {A.}~\bibnamefont {Ghosh}}, \
  and\ \bibinfo {author} {\bibfnamefont {S.~S.}\ \bibnamefont {Ray}},\
  }\href@noop {} {\bibfield  {journal} {\bibinfo  {journal} {WIREs: Comput.
  Mol. Sci.}\ }\textbf {\bibinfo {volume} {6}},\ \bibinfo {pages} {266}
  (\bibinfo {year} {2016})}\BibitemShut {NoStop}%
\bibitem [{\citenamefont {Shavitt}\ and\ \citenamefont
  {Redmon}(1980)}]{Shavitt:1980jc}%
  \BibitemOpen
  \bibfield  {author} {\bibinfo {author} {\bibfnamefont {I.}~\bibnamefont
  {Shavitt}}\ and\ \bibinfo {author} {\bibfnamefont {L.~T.}\ \bibnamefont
  {Redmon}},\ }\href@noop {} {\bibfield  {journal} {\bibinfo  {journal} {J.
  Chem. Phys.}\ }\textbf {\bibinfo {volume} {73}},\ \bibinfo {pages} {5711}
  (\bibinfo {year} {1980})}\BibitemShut {NoStop}%
\bibitem [{\citenamefont {Kirtman}(1981)}]{Kirtman:1981kf}%
  \BibitemOpen
  \bibfield  {author} {\bibinfo {author} {\bibfnamefont {B.}~\bibnamefont
  {Kirtman}},\ }\href@noop {} {\bibfield  {journal} {\bibinfo  {journal} {J.
  Chem. Phys.}\ }\textbf {\bibinfo {volume} {75}},\ \bibinfo {pages} {798}
  (\bibinfo {year} {1981})}\BibitemShut {NoStop}%
\bibitem [{\citenamefont {Khait}, \citenamefont {Song},\ and\ \citenamefont
  {Hoffmann}(2002)}]{Khait:2002ca}%
  \BibitemOpen
  \bibfield  {author} {\bibinfo {author} {\bibfnamefont {Y.~G.}\ \bibnamefont
  {Khait}}, \bibinfo {author} {\bibfnamefont {J.}~\bibnamefont {Song}}, \ and\
  \bibinfo {author} {\bibfnamefont {M.~R.}\ \bibnamefont {Hoffmann}},\
  }\href@noop {} {\bibfield  {journal} {\bibinfo  {journal} {J. Chem. Phys.}\
  }\textbf {\bibinfo {volume} {117}},\ \bibinfo {pages} {4133} (\bibinfo {year}
  {2002})}\BibitemShut {NoStop}%
\bibitem [{\citenamefont {Hoffmann}\ \emph {et~al.}(2009)\citenamefont
  {Hoffmann}, \citenamefont {Datta}, \citenamefont {Das}, \citenamefont
  {Mukherjee}, \citenamefont {Szabados}, \citenamefont {Rolik},\ and\
  \citenamefont {Surj{\'a}n}}]{Hoffmann:2009ih}%
  \BibitemOpen
  \bibfield  {author} {\bibinfo {author} {\bibfnamefont {M.~R.}\ \bibnamefont
  {Hoffmann}}, \bibinfo {author} {\bibfnamefont {D.}~\bibnamefont {Datta}},
  \bibinfo {author} {\bibfnamefont {S.}~\bibnamefont {Das}}, \bibinfo {author}
  {\bibfnamefont {D.}~\bibnamefont {Mukherjee}}, \bibinfo {author}
  {\bibfnamefont {A.}~\bibnamefont {Szabados}}, \bibinfo {author}
  {\bibfnamefont {Z.}~\bibnamefont {Rolik}}, \ and\ \bibinfo {author}
  {\bibfnamefont {P.~R.}\ \bibnamefont {Surj{\'a}n}},\ }\href@noop {}
  {\bibfield  {journal} {\bibinfo  {journal} {J. Chem. Phys.}\ }\textbf
  {\bibinfo {volume} {131}},\ \bibinfo {pages} {204104} (\bibinfo {year}
  {2009})}\BibitemShut {NoStop}%
\bibitem [{\citenamefont {Rolik}, \citenamefont {Szabados},\ and\ \citenamefont
  {Surj{\'a}n}(2003)}]{Rolik:2003be}%
  \BibitemOpen
  \bibfield  {author} {\bibinfo {author} {\bibfnamefont {Z.}~\bibnamefont
  {Rolik}}, \bibinfo {author} {\bibfnamefont {{\'A}.}~\bibnamefont {Szabados}},
  \ and\ \bibinfo {author} {\bibfnamefont {P.~R.}\ \bibnamefont {Surj{\'a}n}},\
  }\href@noop {} {\bibfield  {journal} {\bibinfo  {journal} {J. Chem. Phys.}\
  }\textbf {\bibinfo {volume} {119}},\ \bibinfo {pages} {1922} (\bibinfo {year}
  {2003})}\BibitemShut {NoStop}%
\bibitem [{\citenamefont {Szabados}\ \emph {et~al.}(2005)\citenamefont
  {Szabados}, \citenamefont {Rolik}, \citenamefont {T{\'o}th},\ and\
  \citenamefont {Surj{\'a}n}}]{Szabados:2005jk}%
  \BibitemOpen
  \bibfield  {author} {\bibinfo {author} {\bibfnamefont {{\'A}.}~\bibnamefont
  {Szabados}}, \bibinfo {author} {\bibfnamefont {Z.}~\bibnamefont {Rolik}},
  \bibinfo {author} {\bibfnamefont {G.}~\bibnamefont {T{\'o}th}}, \ and\
  \bibinfo {author} {\bibfnamefont {P.~R.}\ \bibnamefont {Surj{\'a}n}},\
  }\href@noop {} {\bibfield  {journal} {\bibinfo  {journal} {J. Chem. Phys.}\
  }\textbf {\bibinfo {volume} {122}},\ \bibinfo {pages} {114104} (\bibinfo
  {year} {2005})}\BibitemShut {NoStop}%
\bibitem [{FOR(2016)}]{FORTE2016}%
  \BibitemOpen
  \href@noop {} {}\bibinfo {howpublished} {Forte, a suite of quantum chemistry
  methods for strongly correlated electrons. For the current version, see
  \url{https://github.com/evangelistalab/forte}} (\bibinfo {year}
  {2016})\BibitemShut {NoStop}%
\bibitem [{\citenamefont {Celani}\ and\ \citenamefont
  {Werner}(2000)}]{Celani:2000cf}%
  \BibitemOpen
  \bibfield  {author} {\bibinfo {author} {\bibfnamefont {P.}~\bibnamefont
  {Celani}}\ and\ \bibinfo {author} {\bibfnamefont {H.-J.}\ \bibnamefont
  {Werner}},\ }\href@noop {} {\bibfield  {journal} {\bibinfo  {journal} {J.
  Chem. Phys.}\ }\textbf {\bibinfo {volume} {112}},\ \bibinfo {pages} {5546}
  (\bibinfo {year} {2000})}\BibitemShut {NoStop}%
\bibitem [{\citenamefont {Knowles}\ and\ \citenamefont
  {Werner}(1988)}]{Knowles:1988hv}%
  \BibitemOpen
  \bibfield  {author} {\bibinfo {author} {\bibfnamefont {P.~J.}\ \bibnamefont
  {Knowles}}\ and\ \bibinfo {author} {\bibfnamefont {H.-J.}\ \bibnamefont
  {Werner}},\ }\href@noop {} {\bibfield  {journal} {\bibinfo  {journal} {Chem.
  Phys. Lett.}\ }\textbf {\bibinfo {volume} {145}},\ \bibinfo {pages} {514}
  (\bibinfo {year} {1988})}\BibitemShut {NoStop}%
\bibitem [{\citenamefont {Werner}\ and\ \citenamefont
  {Knowles}(1988)}]{Werner:1988ku}%
  \BibitemOpen
  \bibfield  {author} {\bibinfo {author} {\bibfnamefont {H.-J.}\ \bibnamefont
  {Werner}}\ and\ \bibinfo {author} {\bibfnamefont {P.~J.}\ \bibnamefont
  {Knowles}},\ }\href@noop {} {\bibfield  {journal} {\bibinfo  {journal} {J.
  Chem. Phys.}\ }\textbf {\bibinfo {volume} {89}},\ \bibinfo {pages} {5803}
  (\bibinfo {year} {1988})}\BibitemShut {NoStop}%
\bibitem [{\citenamefont {Shamasundar}, \citenamefont {Knizia},\ and\
  \citenamefont {Werner}(2011)}]{Shamasundar:2011ew}%
  \BibitemOpen
  \bibfield  {author} {\bibinfo {author} {\bibfnamefont {K.~R.}\ \bibnamefont
  {Shamasundar}}, \bibinfo {author} {\bibfnamefont {G.}~\bibnamefont {Knizia}},
  \ and\ \bibinfo {author} {\bibfnamefont {H.-J.}\ \bibnamefont {Werner}},\
  }\href@noop {} {\bibfield  {journal} {\bibinfo  {journal} {J. Chem. Phys.}\
  }\textbf {\bibinfo {volume} {135}},\ \bibinfo {pages} {054101} (\bibinfo
  {year} {2011})}\BibitemShut {NoStop}%
\bibitem [{\citenamefont {Mahapatra}, \citenamefont {Datta},\ and\
  \citenamefont {Mukherjee}(1999{\natexlab{b}})}]{Mahapatra:1999ev}%
  \BibitemOpen
  \bibfield  {author} {\bibinfo {author} {\bibfnamefont {U.~S.}\ \bibnamefont
  {Mahapatra}}, \bibinfo {author} {\bibfnamefont {B.}~\bibnamefont {Datta}}, \
  and\ \bibinfo {author} {\bibfnamefont {D.}~\bibnamefont {Mukherjee}},\
  }\href@noop {} {\bibfield  {journal} {\bibinfo  {journal} {J. Chem. Phys.}\
  }\textbf {\bibinfo {volume} {110}},\ \bibinfo {pages} {6171} (\bibinfo {year}
  {1999}{\natexlab{b}})}\BibitemShut {NoStop}%
\bibitem [{\citenamefont {Evangelista}, \citenamefont {Allen},\ and\
  \citenamefont {Schaefer}(2006)}]{Evangelista:2006gf}%
  \BibitemOpen
  \bibfield  {author} {\bibinfo {author} {\bibfnamefont {F.~A.}\ \bibnamefont
  {Evangelista}}, \bibinfo {author} {\bibfnamefont {W.~D.}\ \bibnamefont
  {Allen}}, \ and\ \bibinfo {author} {\bibfnamefont {H.~F.}\ \bibnamefont
  {Schaefer}},\ }\href@noop {} {\bibfield  {journal} {\bibinfo  {journal} {J.
  Chem. Phys.}\ }\textbf {\bibinfo {volume} {125}},\ \bibinfo {pages} {154113}
  (\bibinfo {year} {2006})}\BibitemShut {NoStop}%
\bibitem [{\citenamefont {Mahapatra}, \citenamefont {Datta},\ and\
  \citenamefont {Mukherjee}(2010)}]{Mahapatra:2010kj}%
  \BibitemOpen
  \bibfield  {author} {\bibinfo {author} {\bibfnamefont {U.~S.}\ \bibnamefont
  {Mahapatra}}, \bibinfo {author} {\bibfnamefont {B.}~\bibnamefont {Datta}}, \
  and\ \bibinfo {author} {\bibfnamefont {D.}~\bibnamefont {Mukherjee}},\
  }\href@noop {} {\bibfield  {journal} {\bibinfo  {journal} {Mol. Phys.}\
  }\textbf {\bibinfo {volume} {94}},\ \bibinfo {pages} {157} (\bibinfo {year}
  {2010})}\BibitemShut {NoStop}%
\bibitem [{\citenamefont {Evangelista}\ \emph {et~al.}(2010)\citenamefont
  {Evangelista}, \citenamefont {Prochnow}, \citenamefont {Gauss},\ and\
  \citenamefont {Schaefer}}]{Evangelista:2010cq}%
  \BibitemOpen
  \bibfield  {author} {\bibinfo {author} {\bibfnamefont {F.~A.}\ \bibnamefont
  {Evangelista}}, \bibinfo {author} {\bibfnamefont {E.}~\bibnamefont
  {Prochnow}}, \bibinfo {author} {\bibfnamefont {J.}~\bibnamefont {Gauss}}, \
  and\ \bibinfo {author} {\bibfnamefont {H.~F.}\ \bibnamefont {Schaefer}},\
  }\href@noop {} {\bibfield  {journal} {\bibinfo  {journal} {J. Chem. Phys.}\
  }\textbf {\bibinfo {volume} {132}},\ \bibinfo {pages} {074107} (\bibinfo
  {year} {2010})}\BibitemShut {NoStop}%
\bibitem [{\citenamefont {Demel}, \citenamefont {Datta},\ and\ \citenamefont
  {Nooijen}(2013)}]{Demel:2013kz}%
  \BibitemOpen
  \bibfield  {author} {\bibinfo {author} {\bibfnamefont {O.}~\bibnamefont
  {Demel}}, \bibinfo {author} {\bibfnamefont {D.}~\bibnamefont {Datta}}, \ and\
  \bibinfo {author} {\bibfnamefont {M.}~\bibnamefont {Nooijen}},\ }\href@noop
  {} {\bibfield  {journal} {\bibinfo  {journal} {J. Chem. Phys.}\ }\textbf
  {\bibinfo {volume} {138}},\ \bibinfo {pages} {134108} (\bibinfo {year}
  {2013})}\BibitemShut {NoStop}%
\bibitem [{\citenamefont {Evangelista}\ and\ \citenamefont
  {Gauss}(2012)}]{Evangelista:2012hz}%
  \BibitemOpen
  \bibfield  {author} {\bibinfo {author} {\bibfnamefont {F.~A.}\ \bibnamefont
  {Evangelista}}\ and\ \bibinfo {author} {\bibfnamefont {J.}~\bibnamefont
  {Gauss}},\ }\href@noop {} {\bibfield  {journal} {\bibinfo  {journal} {Chem.
  Phys.}\ }\textbf {\bibinfo {volume} {401}},\ \bibinfo {pages} {27} (\bibinfo
  {year} {2012})}\BibitemShut {NoStop}%
\bibitem [{\citenamefont {Fink}(2006)}]{Fink:2006gu}%
  \BibitemOpen
  \bibfield  {author} {\bibinfo {author} {\bibfnamefont {R.~F.}\ \bibnamefont
  {Fink}},\ }\href@noop {} {\bibfield  {journal} {\bibinfo  {journal} {Chem.
  Phys. Lett.}\ }\textbf {\bibinfo {volume} {428}},\ \bibinfo {pages} {461}
  (\bibinfo {year} {2006})}\BibitemShut {NoStop}%
\bibitem [{\citenamefont {Fink}(2009)}]{Fink:2009gx}%
  \BibitemOpen
  \bibfield  {author} {\bibinfo {author} {\bibfnamefont {R.~F.}\ \bibnamefont
  {Fink}},\ }\href@noop {} {\bibfield  {journal} {\bibinfo  {journal} {Chem.
  Phys.}\ }\textbf {\bibinfo {volume} {356}},\ \bibinfo {pages} {39} (\bibinfo
  {year} {2009})}\BibitemShut {NoStop}%
\bibitem [{\citenamefont {Whitten}(1973)}]{Whitten:1973ju}%
  \BibitemOpen
  \bibfield  {author} {\bibinfo {author} {\bibfnamefont {J.~L.}\ \bibnamefont
  {Whitten}},\ }\href@noop {} {\bibfield  {journal} {\bibinfo  {journal} {J.
  Chem. Phys.}\ }\textbf {\bibinfo {volume} {58}},\ \bibinfo {pages} {4496}
  (\bibinfo {year} {1973})}\BibitemShut {NoStop}%
\bibitem [{\citenamefont {Dunlap}, \citenamefont {Connolly},\ and\
  \citenamefont {Sabin}(1979)}]{Dunlap:1979gh}%
  \BibitemOpen
  \bibfield  {author} {\bibinfo {author} {\bibfnamefont {B.~I.}\ \bibnamefont
  {Dunlap}}, \bibinfo {author} {\bibfnamefont {J.~W.~D.}\ \bibnamefont
  {Connolly}}, \ and\ \bibinfo {author} {\bibfnamefont {J.~R.}\ \bibnamefont
  {Sabin}},\ }\href@noop {} {\bibfield  {journal} {\bibinfo  {journal} {J.
  Chem. Phys.}\ }\textbf {\bibinfo {volume} {71}},\ \bibinfo {pages} {3396}
  (\bibinfo {year} {1979})}\BibitemShut {NoStop}%
\bibitem [{\citenamefont {Werner}, \citenamefont {Manby},\ and\ \citenamefont
  {Knowles}(2003)}]{Werner:2003gq}%
  \BibitemOpen
  \bibfield  {author} {\bibinfo {author} {\bibfnamefont {H.-J.}\ \bibnamefont
  {Werner}}, \bibinfo {author} {\bibfnamefont {F.~R.}\ \bibnamefont {Manby}}, \
  and\ \bibinfo {author} {\bibfnamefont {P.~J.}\ \bibnamefont {Knowles}},\
  }\href@noop {} {\bibfield  {journal} {\bibinfo  {journal} {J. Chem. Phys.}\
  }\textbf {\bibinfo {volume} {118}},\ \bibinfo {pages} {8149} (\bibinfo {year}
  {2003})}\BibitemShut {NoStop}%
\bibitem [{\citenamefont {Beebe}\ and\ \citenamefont
  {Linderberg}(1977)}]{Beebe:1977dp}%
  \BibitemOpen
  \bibfield  {author} {\bibinfo {author} {\bibfnamefont {N.~H.~F.}\
  \bibnamefont {Beebe}}\ and\ \bibinfo {author} {\bibfnamefont
  {J.}~\bibnamefont {Linderberg}},\ }\href@noop {} {\bibfield  {journal}
  {\bibinfo  {journal} {Int. J. Quantum Chem.}\ }\textbf {\bibinfo {volume}
  {12}},\ \bibinfo {pages} {683} (\bibinfo {year} {1977})}\BibitemShut
  {NoStop}%
\bibitem [{\citenamefont {Koch}, \citenamefont {S{\'a}nchez~de Mer{\'a}s},\
  and\ \citenamefont {Pedersen}(2003)}]{Koch:2003go}%
  \BibitemOpen
  \bibfield  {author} {\bibinfo {author} {\bibfnamefont {H.}~\bibnamefont
  {Koch}}, \bibinfo {author} {\bibfnamefont {A.}~\bibnamefont {S{\'a}nchez~de
  Mer{\'a}s}}, \ and\ \bibinfo {author} {\bibfnamefont {T.~B.}\ \bibnamefont
  {Pedersen}},\ }\href@noop {} {\bibfield  {journal} {\bibinfo  {journal} {J.
  Chem. Phys.}\ }\textbf {\bibinfo {volume} {118}},\ \bibinfo {pages} {9481}
  (\bibinfo {year} {2003})}\BibitemShut {NoStop}%
\bibitem [{\citenamefont {Aquilante}\ \emph {et~al.}(2008)\citenamefont
  {Aquilante}, \citenamefont {Malmqvist}, \citenamefont {Pedersen},
  \citenamefont {Ghosh},\ and\ \citenamefont {Roos}}]{Aquilante:2008gk}%
  \BibitemOpen
  \bibfield  {author} {\bibinfo {author} {\bibfnamefont {F.}~\bibnamefont
  {Aquilante}}, \bibinfo {author} {\bibfnamefont {P.-{\AA}.}\ \bibnamefont
  {Malmqvist}}, \bibinfo {author} {\bibfnamefont {T.~B.}\ \bibnamefont
  {Pedersen}}, \bibinfo {author} {\bibfnamefont {A.}~\bibnamefont {Ghosh}}, \
  and\ \bibinfo {author} {\bibfnamefont {B.~O.}\ \bibnamefont {Roos}},\
  }\href@noop {} {\bibfield  {journal} {\bibinfo  {journal} {J. Chem. Theory
  Comput.}\ }\textbf {\bibinfo {volume} {4}},\ \bibinfo {pages} {694} (\bibinfo
  {year} {2008})}\BibitemShut {NoStop}%
\bibitem [{\citenamefont {Aquilante}\ \emph {et~al.}(2011)\citenamefont
  {Aquilante}, \citenamefont {Boman}, \citenamefont {Bostr{\"o}m},
  \citenamefont {Koch}, \citenamefont {Lindh}, \citenamefont {de~Mer{\'a}s},\
  and\ \citenamefont {Pedersen}}]{Aquilante:2011fd}%
  \BibitemOpen
  \bibfield  {author} {\bibinfo {author} {\bibfnamefont {F.}~\bibnamefont
  {Aquilante}}, \bibinfo {author} {\bibfnamefont {L.}~\bibnamefont {Boman}},
  \bibinfo {author} {\bibfnamefont {J.}~\bibnamefont {Bostr{\"o}m}}, \bibinfo
  {author} {\bibfnamefont {H.}~\bibnamefont {Koch}}, \bibinfo {author}
  {\bibfnamefont {R.}~\bibnamefont {Lindh}}, \bibinfo {author} {\bibfnamefont
  {A.~S.}\ \bibnamefont {de~Mer{\'a}s}}, \ and\ \bibinfo {author}
  {\bibfnamefont {T.~B.}\ \bibnamefont {Pedersen}},\ }in\ \href {\doibase
  10.1007/978-90-481-2853-2_13} {\emph {\bibinfo {booktitle} {Linear-Scaling
  Techniques in Computational Chemistry and Physics}}},\ \bibinfo {series}
  {Challenges and Advances in Computational Chemistry and Physics},
  Vol.~\bibinfo {volume} {13},\ \bibinfo {editor} {edited by\ \bibinfo {editor}
  {\bibfnamefont {R.}~\bibnamefont {Zalesny}}, \bibinfo {editor} {\bibfnamefont
  {M.~G.}\ \bibnamefont {Papadopoulos}}, \bibinfo {editor} {\bibfnamefont
  {P.~G.}\ \bibnamefont {Mezey}}, \ and\ \bibinfo {editor} {\bibfnamefont
  {J.}~\bibnamefont {Leszczynski}}}\ (\bibinfo  {publisher} {Springer
  Netherlands},\ \bibinfo {year} {2011})\ pp.\ \bibinfo {pages}
  {301--343}\BibitemShut {NoStop}%
\bibitem [{\citenamefont {Weigend}, \citenamefont {Kattannek},\ and\
  \citenamefont {Ahlrichs}(2009)}]{Weigend:2009bh}%
  \BibitemOpen
  \bibfield  {author} {\bibinfo {author} {\bibfnamefont {F.}~\bibnamefont
  {Weigend}}, \bibinfo {author} {\bibfnamefont {M.}~\bibnamefont {Kattannek}},
  \ and\ \bibinfo {author} {\bibfnamefont {R.}~\bibnamefont {Ahlrichs}},\
  }\href@noop {} {\bibfield  {journal} {\bibinfo  {journal} {J. Chem. Phys.}\
  }\textbf {\bibinfo {volume} {130}},\ \bibinfo {pages} {164106} (\bibinfo
  {year} {2009})}\BibitemShut {NoStop}%
\bibitem [{\citenamefont {DePrince}\ and\ \citenamefont
  {Sherrill}(2013)}]{DePrince:2013ki}%
  \BibitemOpen
  \bibfield  {author} {\bibinfo {author} {\bibfnamefont {A.~E.}\ \bibnamefont
  {DePrince}}\ and\ \bibinfo {author} {\bibfnamefont {C.~D.}\ \bibnamefont
  {Sherrill}},\ }\href@noop {} {\bibfield  {journal} {\bibinfo  {journal} {J.
  Chem. Theory Comput.}\ }\textbf {\bibinfo {volume} {9}},\ \bibinfo {pages}
  {2687} (\bibinfo {year} {2013})}\BibitemShut {NoStop}%
\bibitem [{\citenamefont {Schutz}\ and\ \citenamefont
  {Manby}(2003)}]{Schutz:2003jy}%
  \BibitemOpen
  \bibfield  {author} {\bibinfo {author} {\bibfnamefont {M.}~\bibnamefont
  {Schutz}}\ and\ \bibinfo {author} {\bibfnamefont {F.~R.}\ \bibnamefont
  {Manby}},\ }\href@noop {} {\bibfield  {journal} {\bibinfo  {journal} {Phys.
  Chem. Chem. Phys.}\ }\textbf {\bibinfo {volume} {5}},\ \bibinfo {pages}
  {3349} (\bibinfo {year} {2003})}\BibitemShut {NoStop}%
\bibitem [{\citenamefont {Parrish}\ \emph {et~al.}(2014)\citenamefont
  {Parrish}, \citenamefont {Sherrill}, \citenamefont {Hohenstein},
  \citenamefont {Kokkila},\ and\ \citenamefont
  {Mart{\'\i}nez}}]{Parrish:2014ig}%
  \BibitemOpen
  \bibfield  {author} {\bibinfo {author} {\bibfnamefont {R.~M.}\ \bibnamefont
  {Parrish}}, \bibinfo {author} {\bibfnamefont {C.~D.}\ \bibnamefont
  {Sherrill}}, \bibinfo {author} {\bibfnamefont {E.~G.}\ \bibnamefont
  {Hohenstein}}, \bibinfo {author} {\bibfnamefont {S.~I.~L.}\ \bibnamefont
  {Kokkila}}, \ and\ \bibinfo {author} {\bibfnamefont {T.~J.}\ \bibnamefont
  {Mart{\'\i}nez}},\ }\href@noop {} {\bibfield  {journal} {\bibinfo  {journal}
  {J. Chem. Phys.}\ }\textbf {\bibinfo {volume} {140}},\ \bibinfo {pages}
  {181102} (\bibinfo {year} {2014})}\BibitemShut {NoStop}%
\bibitem [{\citenamefont {Turney}\ \emph {et~al.}(2012)\citenamefont {Turney},
  \citenamefont {Simmonett}, \citenamefont {Parrish}, \citenamefont
  {Hohenstein}, \citenamefont {Evangelista}, \citenamefont {Fermann},
  \citenamefont {Mintz}, \citenamefont {Burns}, \citenamefont {Wilke},
  \citenamefont {Abrams}, \citenamefont {Russ}, \citenamefont {Leininger},
  \citenamefont {Janssen}, \citenamefont {Seidl}, \citenamefont {Allen},
  \citenamefont {Schaefer}, \citenamefont {King}, \citenamefont {Valeev},
  \citenamefont {Sherrill},\ and\ \citenamefont {Crawford}}]{PSI4}%
  \BibitemOpen
  \bibfield  {author} {\bibinfo {author} {\bibfnamefont {J.~M.}\ \bibnamefont
  {Turney}}, \bibinfo {author} {\bibfnamefont {A.~C.}\ \bibnamefont
  {Simmonett}}, \bibinfo {author} {\bibfnamefont {R.~M.}\ \bibnamefont
  {Parrish}}, \bibinfo {author} {\bibfnamefont {E.~G.}\ \bibnamefont
  {Hohenstein}}, \bibinfo {author} {\bibfnamefont {F.~A.}\ \bibnamefont
  {Evangelista}}, \bibinfo {author} {\bibfnamefont {J.~T.}\ \bibnamefont
  {Fermann}}, \bibinfo {author} {\bibfnamefont {B.~J.}\ \bibnamefont {Mintz}},
  \bibinfo {author} {\bibfnamefont {L.~A.}\ \bibnamefont {Burns}}, \bibinfo
  {author} {\bibfnamefont {J.~J.}\ \bibnamefont {Wilke}}, \bibinfo {author}
  {\bibfnamefont {M.~L.}\ \bibnamefont {Abrams}}, \bibinfo {author}
  {\bibfnamefont {N.~J.}\ \bibnamefont {Russ}}, \bibinfo {author}
  {\bibfnamefont {M.~L.}\ \bibnamefont {Leininger}}, \bibinfo {author}
  {\bibfnamefont {C.~L.}\ \bibnamefont {Janssen}}, \bibinfo {author}
  {\bibfnamefont {E.~T.}\ \bibnamefont {Seidl}}, \bibinfo {author}
  {\bibfnamefont {W.~D.}\ \bibnamefont {Allen}}, \bibinfo {author}
  {\bibfnamefont {H.~F.}\ \bibnamefont {Schaefer}}, \bibinfo {author}
  {\bibfnamefont {R.~A.}\ \bibnamefont {King}}, \bibinfo {author}
  {\bibfnamefont {E.~F.}\ \bibnamefont {Valeev}}, \bibinfo {author}
  {\bibfnamefont {C.~D.}\ \bibnamefont {Sherrill}}, \ and\ \bibinfo {author}
  {\bibfnamefont {T.~D.}\ \bibnamefont {Crawford}},\ }\href {\doibase
  10.1002/wcms.93} {\bibfield  {journal} {\bibinfo  {journal} {WIREs Comput.
  Mol. Sci.}\ }\textbf {\bibinfo {volume} {2}},\ \bibinfo {pages} {556}
  (\bibinfo {year} {2012})}\BibitemShut {NoStop}%
\bibitem [{AMB(2016)}]{AMBIT2016}%
  \BibitemOpen
  \href@noop {} {}\bibinfo {howpublished} {Ambit is a C++ library for the
  implementation of tensor product calculations through a clean and concise
  user interface, written by Turney,~J. M.; Parrish,~R. M.; Evangelista,~F. A.;
  Smith,~D. G. For the current version, see
  \url{https://github.com/jturney/ambit}} (\bibinfo {year} {2016})\BibitemShut
  {NoStop}%
\bibitem [{\citenamefont {Langhoff}\ and\ \citenamefont
  {Davidson}(1974)}]{Langhoff:1974kz}%
  \BibitemOpen
  \bibfield  {author} {\bibinfo {author} {\bibfnamefont {S.~R.}\ \bibnamefont
  {Langhoff}}\ and\ \bibinfo {author} {\bibfnamefont {E.~R.}\ \bibnamefont
  {Davidson}},\ }\href {\doibase 10.1002/qua.560080106} {\bibfield  {journal}
  {\bibinfo  {journal} {Int. J. Quantum Chem.}\ }\textbf {\bibinfo {volume}
  {8}},\ \bibinfo {pages} {61} (\bibinfo {year} {1974})}\BibitemShut {NoStop}%
\bibitem [{\citenamefont {Bytautas}\ \emph {et~al.}(2007)\citenamefont
  {Bytautas}, \citenamefont {Nagata}, \citenamefont {Gordon},\ and\
  \citenamefont {Ruedenberg}}]{Bytautas:2007fa}%
  \BibitemOpen
  \bibfield  {author} {\bibinfo {author} {\bibfnamefont {L.}~\bibnamefont
  {Bytautas}}, \bibinfo {author} {\bibfnamefont {T.}~\bibnamefont {Nagata}},
  \bibinfo {author} {\bibfnamefont {M.~S.}\ \bibnamefont {Gordon}}, \ and\
  \bibinfo {author} {\bibfnamefont {K.}~\bibnamefont {Ruedenberg}},\
  }\href@noop {} {\bibfield  {journal} {\bibinfo  {journal} {J. Chem. Phys.}\
  }\textbf {\bibinfo {volume} {127}},\ \bibinfo {pages} {164317} (\bibinfo
  {year} {2007})}\BibitemShut {NoStop}%
\bibitem [{\citenamefont {Yang}\ \emph {et~al.}(2013)\citenamefont {Yang},
  \citenamefont {Jalan}, \citenamefont {Green},\ and\ \citenamefont
  {Truhlar}}]{Yang:2013gq}%
  \BibitemOpen
  \bibfield  {author} {\bibinfo {author} {\bibfnamefont {K.~R.}\ \bibnamefont
  {Yang}}, \bibinfo {author} {\bibfnamefont {A.}~\bibnamefont {Jalan}},
  \bibinfo {author} {\bibfnamefont {W.~H.}\ \bibnamefont {Green}}, \ and\
  \bibinfo {author} {\bibfnamefont {D.~G.}\ \bibnamefont {Truhlar}},\
  }\href@noop {} {\bibfield  {journal} {\bibinfo  {journal} {J. Chem. Theory
  Comput.}\ }\textbf {\bibinfo {volume} {9}},\ \bibinfo {pages} {418} (\bibinfo
  {year} {2013})}\BibitemShut {NoStop}%
\bibitem [{\citenamefont {Hirata}\ \emph {et~al.}(2004)\citenamefont {Hirata},
  \citenamefont {Fan}, \citenamefont {Auer}, \citenamefont {Nooijen},\ and\
  \citenamefont {Piecuch}}]{Hirata:2004fm}%
  \BibitemOpen
  \bibfield  {author} {\bibinfo {author} {\bibfnamefont {S.}~\bibnamefont
  {Hirata}}, \bibinfo {author} {\bibfnamefont {P.-D.}\ \bibnamefont {Fan}},
  \bibinfo {author} {\bibfnamefont {A.~A.}\ \bibnamefont {Auer}}, \bibinfo
  {author} {\bibfnamefont {M.}~\bibnamefont {Nooijen}}, \ and\ \bibinfo
  {author} {\bibfnamefont {P.}~\bibnamefont {Piecuch}},\ }\href@noop {}
  {\bibfield  {journal} {\bibinfo  {journal} {J. Chem. Phys.}\ }\textbf
  {\bibinfo {volume} {121}},\ \bibinfo {pages} {12197} (\bibinfo {year}
  {2004})}\BibitemShut {NoStop}%
\bibitem [{\citenamefont {Dunning}(1989)}]{Dunning:1989bx}%
  \BibitemOpen
  \bibfield  {author} {\bibinfo {author} {\bibfnamefont {T.~H.}\ \bibnamefont
  {Dunning}},\ }\href@noop {} {\bibfield  {journal} {\bibinfo  {journal} {J.
  Chem. Phys.}\ }\textbf {\bibinfo {volume} {90}},\ \bibinfo {pages} {1007}
  (\bibinfo {year} {1989})}\BibitemShut {NoStop}%
\bibitem [{che(2016)}]{cheMVP}%
  \BibitemOpen
  \href@noop {} {}\bibinfo {howpublished} {cheMVP is free, open-source software
  designed to make clean, simple molecule drawings suitable for publications
  and presentations. Written by Andrew Simmonett, Justin M. Turney, and H.
  Parker Shelton. For the current version, see
  \url{https://github.com/hpshelton/cheMVP}} (\bibinfo {year}
  {2016})\BibitemShut {NoStop}%
\bibitem [{\citenamefont {Schuler}\ \emph {et~al.}(2016)\citenamefont
  {Schuler}, \citenamefont {Fatayer}, \citenamefont {Mohn}, \citenamefont
  {Moll}, \citenamefont {Pavli{\v c}ek}, \citenamefont {Meyer}, \citenamefont
  {Pe{\~n}a},\ and\ \citenamefont {Gross}}]{Schuler:2016dv}%
  \BibitemOpen
  \bibfield  {author} {\bibinfo {author} {\bibfnamefont {B.}~\bibnamefont
  {Schuler}}, \bibinfo {author} {\bibfnamefont {S.}~\bibnamefont {Fatayer}},
  \bibinfo {author} {\bibfnamefont {F.}~\bibnamefont {Mohn}}, \bibinfo {author}
  {\bibfnamefont {N.}~\bibnamefont {Moll}}, \bibinfo {author} {\bibfnamefont
  {N.}~\bibnamefont {Pavli{\v c}ek}}, \bibinfo {author} {\bibfnamefont
  {G.}~\bibnamefont {Meyer}}, \bibinfo {author} {\bibfnamefont
  {D.}~\bibnamefont {Pe{\~n}a}}, \ and\ \bibinfo {author} {\bibfnamefont
  {L.}~\bibnamefont {Gross}},\ }\href@noop {} {\bibfield  {journal} {\bibinfo
  {journal} {Nat. Chem.}\ }\textbf {\bibinfo {volume} {8}},\ \bibinfo {pages}
  {220} (\bibinfo {year} {2016})}\BibitemShut {NoStop}%
\bibitem [{\citenamefont {Werner}\ \emph {et~al.}(2012)\citenamefont {Werner},
  \citenamefont {Knowles}, \citenamefont {Knizia}, \citenamefont {Manby},\ and\
  \citenamefont {Sch{\"u}tz}}]{MOLPRO-WIREs}%
  \BibitemOpen
  \bibfield  {author} {\bibinfo {author} {\bibfnamefont {H.-J.}\ \bibnamefont
  {Werner}}, \bibinfo {author} {\bibfnamefont {P.~J.}\ \bibnamefont {Knowles}},
  \bibinfo {author} {\bibfnamefont {G.}~\bibnamefont {Knizia}}, \bibinfo
  {author} {\bibfnamefont {F.~R.}\ \bibnamefont {Manby}}, \ and\ \bibinfo
  {author} {\bibfnamefont {M.}~\bibnamefont {Sch{\"u}tz}},\ }\href@noop {}
  {\bibfield  {journal} {\bibinfo  {journal} {WIREs Comput. Mol. Sci.}\
  }\textbf {\bibinfo {volume} {2}},\ \bibinfo {pages} {242} (\bibinfo {year}
  {2012})}\BibitemShut {NoStop}%
\bibitem [{\citenamefont {Werner}\ \emph {et~al.}(2015)\citenamefont {Werner},
  \citenamefont {Knowles}, \citenamefont {Knizia}, \citenamefont {Manby},
  \citenamefont {{Sch\"{u}tz}}, \citenamefont {Celani}, \citenamefont
  {Gy\"orffy}, \citenamefont {Kats}, \citenamefont {Korona}, \citenamefont
  {Lindh}, \citenamefont {Mitrushenkov}, \citenamefont {Rauhut}, \citenamefont
  {Shamasundar}, \citenamefont {Adler}, \citenamefont {Amos}, \citenamefont
  {Bernhardsson}, \citenamefont {Berning}, \citenamefont {Cooper},
  \citenamefont {Deegan}, \citenamefont {Dobbyn}, \citenamefont {Eckert},
  \citenamefont {Goll}, \citenamefont {Hampel}, \citenamefont {Hesselmann},
  \citenamefont {Hetzer}, \citenamefont {Hrenar}, \citenamefont {Jansen},
  \citenamefont {K\"oppl}, \citenamefont {Liu}, \citenamefont {Lloyd},
  \citenamefont {Mata}, \citenamefont {May}, \citenamefont {McNicholas},
  \citenamefont {Meyer}, \citenamefont {Mura}, \citenamefont {Nicklass},
  \citenamefont {O'Neill}, \citenamefont {Palmieri}, \citenamefont {Peng},
  \citenamefont {Pfl\"uger}, \citenamefont {Pitzer}, \citenamefont {Reiher},
  \citenamefont {Shiozaki}, \citenamefont {Stoll}, \citenamefont {Stone},
  \citenamefont {Tarroni}, \citenamefont {Thorsteinsson},\ and\ \citenamefont
  {Wang}}]{MOLPRO2015}%
  \BibitemOpen
  \bibfield  {author} {\bibinfo {author} {\bibfnamefont {H.-J.}\ \bibnamefont
  {Werner}}, \bibinfo {author} {\bibfnamefont {P.~J.}\ \bibnamefont {Knowles}},
  \bibinfo {author} {\bibfnamefont {G.}~\bibnamefont {Knizia}}, \bibinfo
  {author} {\bibfnamefont {F.~R.}\ \bibnamefont {Manby}}, \bibinfo {author}
  {\bibfnamefont {M.}~\bibnamefont {{Sch\"{u}tz}}}, \bibinfo {author}
  {\bibfnamefont {P.}~\bibnamefont {Celani}}, \bibinfo {author} {\bibfnamefont
  {W.}~\bibnamefont {Gy\"orffy}}, \bibinfo {author} {\bibfnamefont
  {D.}~\bibnamefont {Kats}}, \bibinfo {author} {\bibfnamefont {T.}~\bibnamefont
  {Korona}}, \bibinfo {author} {\bibfnamefont {R.}~\bibnamefont {Lindh}},
  \bibinfo {author} {\bibfnamefont {A.}~\bibnamefont {Mitrushenkov}}, \bibinfo
  {author} {\bibfnamefont {G.}~\bibnamefont {Rauhut}}, \bibinfo {author}
  {\bibfnamefont {K.~R.}\ \bibnamefont {Shamasundar}}, \bibinfo {author}
  {\bibfnamefont {T.~B.}\ \bibnamefont {Adler}}, \bibinfo {author}
  {\bibfnamefont {R.~D.}\ \bibnamefont {Amos}}, \bibinfo {author}
  {\bibfnamefont {A.}~\bibnamefont {Bernhardsson}}, \bibinfo {author}
  {\bibfnamefont {A.}~\bibnamefont {Berning}}, \bibinfo {author} {\bibfnamefont
  {D.~L.}\ \bibnamefont {Cooper}}, \bibinfo {author} {\bibfnamefont {M.~J.~O.}\
  \bibnamefont {Deegan}}, \bibinfo {author} {\bibfnamefont {A.~J.}\
  \bibnamefont {Dobbyn}}, \bibinfo {author} {\bibfnamefont {F.}~\bibnamefont
  {Eckert}}, \bibinfo {author} {\bibfnamefont {E.}~\bibnamefont {Goll}},
  \bibinfo {author} {\bibfnamefont {C.}~\bibnamefont {Hampel}}, \bibinfo
  {author} {\bibfnamefont {A.}~\bibnamefont {Hesselmann}}, \bibinfo {author}
  {\bibfnamefont {G.}~\bibnamefont {Hetzer}}, \bibinfo {author} {\bibfnamefont
  {T.}~\bibnamefont {Hrenar}}, \bibinfo {author} {\bibfnamefont
  {G.}~\bibnamefont {Jansen}}, \bibinfo {author} {\bibfnamefont
  {C.}~\bibnamefont {K\"oppl}}, \bibinfo {author} {\bibfnamefont
  {Y.}~\bibnamefont {Liu}}, \bibinfo {author} {\bibfnamefont {A.~W.}\
  \bibnamefont {Lloyd}}, \bibinfo {author} {\bibfnamefont {R.~A.}\ \bibnamefont
  {Mata}}, \bibinfo {author} {\bibfnamefont {A.~J.}\ \bibnamefont {May}},
  \bibinfo {author} {\bibfnamefont {S.~J.}\ \bibnamefont {McNicholas}},
  \bibinfo {author} {\bibfnamefont {W.}~\bibnamefont {Meyer}}, \bibinfo
  {author} {\bibfnamefont {M.~E.}\ \bibnamefont {Mura}}, \bibinfo {author}
  {\bibfnamefont {A.}~\bibnamefont {Nicklass}}, \bibinfo {author}
  {\bibfnamefont {D.~P.}\ \bibnamefont {O'Neill}}, \bibinfo {author}
  {\bibfnamefont {P.}~\bibnamefont {Palmieri}}, \bibinfo {author}
  {\bibfnamefont {D.}~\bibnamefont {Peng}}, \bibinfo {author} {\bibfnamefont
  {K.}~\bibnamefont {Pfl\"uger}}, \bibinfo {author} {\bibfnamefont
  {R.}~\bibnamefont {Pitzer}}, \bibinfo {author} {\bibfnamefont
  {M.}~\bibnamefont {Reiher}}, \bibinfo {author} {\bibfnamefont
  {T.}~\bibnamefont {Shiozaki}}, \bibinfo {author} {\bibfnamefont
  {H.}~\bibnamefont {Stoll}}, \bibinfo {author} {\bibfnamefont {A.~J.}\
  \bibnamefont {Stone}}, \bibinfo {author} {\bibfnamefont {R.}~\bibnamefont
  {Tarroni}}, \bibinfo {author} {\bibfnamefont {T.}~\bibnamefont
  {Thorsteinsson}}, \ and\ \bibinfo {author} {\bibfnamefont {M.}~\bibnamefont
  {Wang}},\ }\href@noop {} {\enquote {\bibinfo {title} {Molpro, version 2015.1,
  a package of $ab$ $initio$ programs},}\ } (\bibinfo {year} {2015}),\ \bibinfo
  {note} {see http://www.molpro.net}\BibitemShut {NoStop}%
\bibitem [{SI()}]{SI}%
  \BibitemOpen
  \href@noop {} {}\bibinfo {howpublished} {See supplemental material at [URL
  will be inserted by AIP] for the optimized geometries of 9,10-anthracyne,
  H$_2$O$_2$ and C$_2$H$_6$; all single-point energies on the potential energy
  curves of F$_2$, H$_2$O$_2$, C$_2$H$_6$, and N$_2$; the singlet-triplet
  splittings of 9,10-anthracyne computed using the cc-pVDZ and cc-pVTZ basis
  sets; and the singlet and triplet CASSCF/cc-pVDZ semicanonical orbitals using
  CAS(2,2), CAS(4,4), CAS(8,8), CAS(12,12), and CAS(16,16) active
  spaces.}\BibitemShut {Stop}%
\bibitem [{\citenamefont {Karton}, \citenamefont {Ruscic},\ and\ \citenamefont
  {Martin}(2007)}]{Karton:2007da}%
  \BibitemOpen
  \bibfield  {author} {\bibinfo {author} {\bibfnamefont {A.}~\bibnamefont
  {Karton}}, \bibinfo {author} {\bibfnamefont {B.}~\bibnamefont {Ruscic}}, \
  and\ \bibinfo {author} {\bibfnamefont {J.~M.~L.}\ \bibnamefont {Martin}},\
  }\href@noop {} {\bibfield  {journal} {\bibinfo  {journal} {J. Mol. Struct.
  (Theochem)}\ }\textbf {\bibinfo {volume} {811}},\ \bibinfo {pages} {345}
  (\bibinfo {year} {2007})}\BibitemShut {NoStop}%
\bibitem [{\citenamefont {Das}, \citenamefont {Mukherjee},\ and\ \citenamefont
  {K{\'a}llay}(2010)}]{Das:2010bj}%
  \BibitemOpen
  \bibfield  {author} {\bibinfo {author} {\bibfnamefont {S.}~\bibnamefont
  {Das}}, \bibinfo {author} {\bibfnamefont {D.}~\bibnamefont {Mukherjee}}, \
  and\ \bibinfo {author} {\bibfnamefont {M.}~\bibnamefont {K{\'a}llay}},\
  }\href@noop {} {\bibfield  {journal} {\bibinfo  {journal} {J. Chem. Phys.}\
  }\textbf {\bibinfo {volume} {132}},\ \bibinfo {pages} {074103} (\bibinfo
  {year} {2010})}\BibitemShut {NoStop}%
\bibitem [{\citenamefont {Kurashige}\ \emph {et~al.}(2014)\citenamefont
  {Kurashige}, \citenamefont {Chalupsky}, \citenamefont {Lan},\ and\
  \citenamefont {Yanai}}]{Kurashige:2014bq}%
  \BibitemOpen
  \bibfield  {author} {\bibinfo {author} {\bibfnamefont {Y.}~\bibnamefont
  {Kurashige}}, \bibinfo {author} {\bibfnamefont {J.}~\bibnamefont
  {Chalupsky}}, \bibinfo {author} {\bibfnamefont {T.~N.}\ \bibnamefont {Lan}},
  \ and\ \bibinfo {author} {\bibfnamefont {T.}~\bibnamefont {Yanai}},\
  }\href@noop {} {\bibfield  {journal} {\bibinfo  {journal} {J. Chem. Phys.}\
  }\textbf {\bibinfo {volume} {141}},\ \bibinfo {pages} {174111} (\bibinfo
  {year} {2014})}\BibitemShut {NoStop}%
\bibitem [{\citenamefont {Fosso-Tande}\ \emph {et~al.}(2016)\citenamefont
  {Fosso-Tande}, \citenamefont {Nguyen}, \citenamefont {Gidofalvi},\ and\
  \citenamefont {DePrince~III}}]{FossoTande:2016hb}%
  \BibitemOpen
  \bibfield  {author} {\bibinfo {author} {\bibfnamefont {J.}~\bibnamefont
  {Fosso-Tande}}, \bibinfo {author} {\bibfnamefont {T.-S.}\ \bibnamefont
  {Nguyen}}, \bibinfo {author} {\bibfnamefont {G.}~\bibnamefont {Gidofalvi}}, \
  and\ \bibinfo {author} {\bibfnamefont {A.~E.}\ \bibnamefont {DePrince~III}},\
  }\href@noop {} {\bibfield  {journal} {\bibinfo  {journal} {J. Chem. Theory
  Comput.}\ }\textbf {\bibinfo {volume} {12}},\ \bibinfo {pages} {2260}
  (\bibinfo {year} {2016})}\BibitemShut {NoStop}%
\bibitem [{\citenamefont {Bofill}\ and\ \citenamefont
  {Pulay}(1989)}]{Bofill:1989dd}%
  \BibitemOpen
  \bibfield  {author} {\bibinfo {author} {\bibfnamefont {J.~M.}\ \bibnamefont
  {Bofill}}\ and\ \bibinfo {author} {\bibfnamefont {P.}~\bibnamefont {Pulay}},\
  }\href@noop {} {\bibfield  {journal} {\bibinfo  {journal} {J. Chem. Phys.}\
  }\textbf {\bibinfo {volume} {90}},\ \bibinfo {pages} {3637} (\bibinfo {year}
  {1989})}\BibitemShut {NoStop}%
\bibitem [{\citenamefont {Keller}\ \emph {et~al.}(2015)\citenamefont {Keller},
  \citenamefont {Boguslawski}, \citenamefont {Janowski}, \citenamefont
  {Reiher},\ and\ \citenamefont {Pulay}}]{Keller:2015ds}%
  \BibitemOpen
  \bibfield  {author} {\bibinfo {author} {\bibfnamefont {S.}~\bibnamefont
  {Keller}}, \bibinfo {author} {\bibfnamefont {K.}~\bibnamefont {Boguslawski}},
  \bibinfo {author} {\bibfnamefont {T.}~\bibnamefont {Janowski}}, \bibinfo
  {author} {\bibfnamefont {M.}~\bibnamefont {Reiher}}, \ and\ \bibinfo {author}
  {\bibfnamefont {P.}~\bibnamefont {Pulay}},\ }\href@noop {} {\bibfield
  {journal} {\bibinfo  {journal} {J. Chem. Phys.}\ }\textbf {\bibinfo {volume}
  {142}},\ \bibinfo {pages} {244104} (\bibinfo {year} {2015})}\BibitemShut
  {NoStop}%
\bibitem [{\citenamefont {Herzberg}(1966)}]{Herzberg1966}%
  \BibitemOpen
  \bibfield  {author} {\bibinfo {author} {\bibfnamefont {G.}~\bibnamefont
  {Herzberg}},\ }\href@noop {} {\emph {\bibinfo {title} {Molecular Spectra and
  Molecular Structure: Electronic Spectra and Electronic Structure of
  Polyatomic Molecules}}}\ (\bibinfo  {publisher} {Van Nostrand},\ \bibinfo
  {year} {1966})\BibitemShut {NoStop}%
\bibitem [{\citenamefont {Evangelista}\ \emph {et~al.}(2012)\citenamefont
  {Evangelista}, \citenamefont {Hanauer}, \citenamefont {K{\"o}hn},\ and\
  \citenamefont {Gauss}}]{Evangelista:2012fo}%
  \BibitemOpen
  \bibfield  {author} {\bibinfo {author} {\bibfnamefont {F.~A.}\ \bibnamefont
  {Evangelista}}, \bibinfo {author} {\bibfnamefont {M.}~\bibnamefont
  {Hanauer}}, \bibinfo {author} {\bibfnamefont {A.}~\bibnamefont {K{\"o}hn}}, \
  and\ \bibinfo {author} {\bibfnamefont {J.}~\bibnamefont {Gauss}},\
  }\href@noop {} {\bibfield  {journal} {\bibinfo  {journal} {J. Chem. Phys.}\
  }\textbf {\bibinfo {volume} {136}},\ \bibinfo {pages} {204108} (\bibinfo
  {year} {2012})}\BibitemShut {NoStop}%
\bibitem [{\citenamefont {Wenthold}, \citenamefont {Squires},\ and\
  \citenamefont {Lineberger}(1998)}]{Wenthold:1998cx}%
  \BibitemOpen
  \bibfield  {author} {\bibinfo {author} {\bibfnamefont {P.~G.}\ \bibnamefont
  {Wenthold}}, \bibinfo {author} {\bibfnamefont {R.~R.}\ \bibnamefont
  {Squires}}, \ and\ \bibinfo {author} {\bibfnamefont {W.~C.}\ \bibnamefont
  {Lineberger}},\ }\href@noop {} {\bibfield  {journal} {\bibinfo  {journal} {J.
  Am. Chem. Soc.}\ }\textbf {\bibinfo {volume} {120}},\ \bibinfo {pages} {5279}
  (\bibinfo {year} {1998})}\BibitemShut {NoStop}%
\end{thebibliography}%

\end{document}